\documentclass[11pt]{article}
\usepackage{jheppub}
\usepackage{graphicx}
\usepackage{color}
\usepackage{verbatim}
\usepackage{subfigure}
\usepackage{acronym}
\usepackage{amsfonts}
\usepackage{slashed}
\usepackage{epsfig}

\usepackage{latexsym,amssymb,amsmath,makeidx,mathrsfs,multirow,lineno,bm,bbm}

\def\tr{\mbox{tr}}

\def\ot{\otimes}

\def\beqn{\begin{eqnarray}}
\def\eeqn{\end{eqnarray}}
\def\beqa{\begin{eqnarray}}
\def\eeqa{\end{eqnarray}}
\def\beqs{\begin{subequations}}
\def\eeqs{\end{subequations}}
\def\beq{\begin{equation}}
\def\eeq{\end{equation}}
\def\ba{\begin{array}}
\def\ea{\end{array}}
\def\dis{\displaystyle}
\def\nn{\nonumber}
\def\f{\frac}
\def\hf{\frac{1}{2}}
\def\[{\left[}
\def\]{\right]}
\def\({\left(}
\def\){\right)}

\def\leqq{\leqslant}
\def\geqq{\geqslant}

\newcommand{\fr}[2]{\mbox{$\frac{\,{#1}\,}{#2}$}}

\def\TeV{\rm TeV}
\def\GeV{\rm GeV}

\def\ifb{\textrm{fb}^{-1}}

\def\gU{\rm U}
\def\gSU{\rm SU}

\def\ga{\gamma}
\def\al{\alpha}
\def\la{\lambda}

\def\dip{\gamma\gamma}
\def\gz{g_0^{}}
\def\gga{g_1^{}}
\def\gb{g_2^{}}
\def\fx{f_1^{}}
\def\fy{f_2^{}}

\def\ca{c_{\alpha}^{}}
\def\sa{s_{\alpha}^{}}

\def\mA{\mathcal{A}}

\def\mL{\mathcal{L}}
\def\mM{\mathcal{M}}

\def\mO{\mathcal{O}}

\def\mR{\mathcal{R}}

\def\MF{M_F^{}}

\title{LHC Higgs Signatures from Extended Electroweak Gauge Symmetry}

\author[a]{Tomohiro Abe,}
\emailAdd{tomohiro\_abe@tsinghua.edu.cn}
\author[a]{~Ning Chen,}
\emailAdd{hep\_nchen@tsinghua.edu.cn}
\author[a,b,c]{~Hong-Jian He\,}
\emailAdd{hjhe@tsinghua.edu.cn}
\affiliation[a]{Institute of Modern Physics and Center for High Energy Physics, \\
Tsinghua University, Beijing 100084, China}
\affiliation[b]{Center for High Energy Physics, Peking University, Beijing 100871, China}
\affiliation[c]{Kavli Institute for Theoretical Physics China, CAS, Beijing 100190, China}
\abstract{
\\[1mm]
We study LHC Higgs signatures from the extended electroweak gauge symmetry
$\,\gSU(2) \ot \gSU(2) \ot \gU(1)$.\, Under this gauge structure, we present
an effective UV completion of the 3-site moose model with ideal fermion delocalization,
which contains two neutral Higgs states $(h,\,H)$ plus three new gauge bosons $(W',\,Z')$.\,
We study the unitarity, and reveal that the exact $E^2$ cancellation
in the longitudinal $V_L^{} V_L^{}$ scattering amplitudes is achieved by
the joint role of exchanging both spin-1 new gauge bosons and spin-0 Higgs bosons.
We identify the lighter Higgs state $\,h\,$ with mass 125\,GeV, and derive the
unitarity bound on the mass of heavier Higgs boson $\,H\,$.\,
The parameter space of this model is highly predictive.
We study the production and decay signals of this 125\,GeV Higgs boson $\,h\,$
at the LHC. We demonstrate that the $h$ Higgs boson can naturally have
enhanced signals in the diphoton channel $\,gg\to h\to\gamma\gamma$\,,\,
while the event rates in the reactions $\,gg\to h\to WW^*$ and $\,gg\to h\to ZZ^*$
are generally suppressed relative to the SM expectation.
Searching the $h$ Higgs boson via the associated production and the vector boson fusions
are also discussed for our model.
We further analyze the LHC signals of the heavier Higgs boson $\,H$\,
as a new physics discriminator from the SM. For wide mass-ranges of $\,H\,$,\,
we derive constraints from the existing LHC searches,
and study the discovery potential of $\,H\,$ at the LHC\,(8\,TeV) and LHC\,(14\,TeV).
}

\keywords{Higgs Physics, Beyond Standard Model 
\\[2mm]
JHEP (2012), in Press [\,arXiv:1207.4103\,]}

\begin{document}
\maketitle

\setcounter{page}{2}



\newpage

\section{\hspace*{-2mm}Introduction}
\label{sec1}

The excellent performance of the LHC running at 7\,TeV (2011)
and 8\,TeV (2012) has allowed both ATLAS and CMS collaborations
to explore substantial mass-ranges of the Higgs boson \cite{Higgs}
in the standard model (SM) \cite{SM}.
During 2011 and the first half of 2012, each collaboration has
collected about $10\,\ifb$ data. The planned LHC running in 2012 is
extended for seven weeks until December\,16, 2012, and will accumulate up
to $20\,\ifb$ data in each detector.
So far ATLAS has excluded the SM Higgs mass-range up to $523$\,GeV at 99\%\,C.L.,
except the small window of $(120.8,\,130.7)$GeV \cite{Atlas2012-7}.
At the meantime, CMS excluded the SM Higgs mass-ranges
of $(110,\,112)$GeV, $(113,\,121.5)$GeV and $(128,\,600)$GeV at 99\%\,C.L.\,\cite{CMS2012-7}.
Both experiments have detected exciting event excesses around the mass window
of $125-126$\,GeV \cite{:2012gk,:2012gu}:
ATLAS reached a $5.9\sigma$ significance at $126.0$\,GeV and
CMS deduced a $5.0\sigma$ signal at $125.3$\,GeV.
These are obtained by combining the most sensitive search channels of $(\gamma\gamma,\,WW^*,\,ZZ^*)$.\,
With the combined data from $7\oplus 8$\,TeV collisions, the ratios of
the observed ATLAS and CMS signals over the SM expectations in the $\,\dip\,$
decay mode are $\,1.8\pm 0.5\,$ and $\,1.5^{+0.6}_{-0.3}\,$,\, respectively.
For the $\,ZZ^*\to 4\ell\,$ decay channel, the experiments have the observed rates
over the SM expectations equal $\,1.2\pm 0.6\,$ [ATLAS] and $\,0.7^{+0.6}_{-0.4}\,$ [CMS].
In the $\,WW^*\to\ell\nu\ell\nu\,$ channel, the observed rates relative to the SM
are $\,1.3\pm 0.5\,$ [ATLAS] and $\,0.6\pm 0.4\,$ [CMS].
For the $b\bar{b}$ and $\tau\bar{\tau}$ final states,
the data still have large statistic errors and the central values fall
in between zero and the SM expectations.
--- It is intriguing that the current observed Higgs signals in the diphoton mode are
found to be always higher than the SM values, which is an indication of possible
new physics in the electroweak symmetry breaking mechanism.

Extending the SM gauge symmetry is a fundamental way to construct new physics beyond the SM.
This is especially motivated by the deconstruction \cite{DC0} approach of the
electroweak symmetry breaking \cite{DC-SU2,3site,tri-3site}, which includes additional
SU(2) and/or U(1) gauge groups.
Such extra gauge groups are also generic in the low energy theories
of many unified models \cite{GUT}.
In this work, we consider the minimal gauge extension of the SM with
an extra $\gSU(2)$ gauge group, which will be called 221 model for abbreviation.
We will focus on its Higgs sector for the desired gauge symmetry breaking
$\,\gSU(2)_0^{} \ot \gSU(2)_1^{} \ot \gU(1)_2^{} \to U(1)_{\textrm{em}}^{}$.\,
This can be viewed as an effective UV completion of the nonlinearly realized
3-site Higgsless moose model\,\cite{3site} by adding two Higgs doublets $\Phi_1$ and $\Phi_2$\,.\,
After the spontaneous symmetry breaking,
two Higgs doublets give rise to six would-be Goldstone bosons
for the mass generation of six gauge bosons $\,(W_0^a,\,W_1^a)\,$,
and leave two neutral Higgs states $\,(h^0,\,H^0)$\,
in the physical particle spectrum.\footnote{Here, two Higgs VEVs are
equal\,\cite{3site} or of the same order of magnitudes,
different from all other types of extended $\gSU(2)\ot\gSU(2)\ot\gU(1)$
models in the literature.}\,
The gauge sector of our 221 model is the same as the 3-site moose model\,\cite{3site}
or the nonlinear BESS model\,\cite{BESS}
and the hidden local symmetry model\,\cite{HLS}.
Our fermion sector follows the 3-site model\,\cite{3site}, where the fermions enjoy the ideal
delocalization\,\cite{iDLF}. This makes the new weak gauge bosons essentially fermiophobic,
so they can be relatively light and significantly below 1\,TeV.
The ideal delocalization minimizes the electroweak corrections to the oblique parameters,
and gives the leading contributions to the triple gauge coupling
of $WWZ$, where $(W,Z)$ denote the light weak gauge bosons.
The LEP-II constraint was found to only put a mild lower limit on the new gauge boson
mass of $W'$ to be about 300\,GeV \cite{3site,tri-3site}; it also escapes from the
current direct searches at the LHC and Tevatron \cite{LHC-TEV}.
The signatures of such new gauge bosons were studied
at the LHC\,(8\,TeV) and LHC\,(14\,TeV), which have good discovery potential
over the wide gauge boson mass-range of $\,250\,\GeV\!-\! 1\,\TeV$ \cite{3site-LHC}.
We also note that realizing the ideal delocalization\,\cite{iDLF} adds
additional vector-like fermions, which will contribute to loop-induced processes
for both Higgs productions and decays (cf.\ Appendix).
The phenomenologies of the pure gauge and fermion sectors of our 221 model are
mainly the same as the 3-site moose model\,\cite{3site}.\footnote{The
$221$ gauge group was considered in very different contexts before,
such as the family non-universality model\,\cite{LM},
the un-unified standard model\,\cite{unUF}, the topflavor seesaw model\,\cite{He:1999vp},
and some recent collider studies\,\cite{Jezo:2012rm}, etc,
but the Higgs sector (including the Higgs potential, mass-spectrum and couplings)
and its collider phenomenology were not addressed. Their fermion sectors also
substantially differ from our current setup under the ideal delocalization \cite{iDLF}.}\,

The goal of the present work is to focus on the full Higgs sector of the 221 model
and its LHC signatures, which were not available before.
In section\,\ref{sec2}, we first set up the model, including the gauge and Yukawa
interactions of the two Higgs doublets, and the complete Higgs potential.
Then, we present the non-standard couplings of Higgs bosons
to all gauge bosons and fermions, including the new gauge bosons $W'/Z'$
and the heavy vector-like fermions $F$.\,
In particular, the coupling strengths between the light Higgs boson $h^0$
and the light weak bosons $W/Z$ are generically reduced relative to those of the SM Higgs boson.
We further study the LHC signals of the heavier Higgs boson $H^0$ whose discovery will
play the key role to discriminate the present model from the SM.
In addition, we analyze the perturbative unitarity condition of our model,
and reveal that the exact $E^2$ cancellation
in the longitudinal $V_L V_L$ scattering amplitudes is achieved by {\it the joint role of
exchanging both spin-1 new gauge bosons $\,(W',Z')$\, and spin-0 Higgs bosons
$\,(h,H)$\,.}\,  Identifying the lighter Higgs boson $\,h^0$ with mass of $125\,\GeV$,
we derive the unitarity bound on the mass of heavier Higgs state $\,H^0$.\,

In section\,\ref{sec3}, we systematically analyze the LHC signatures of
the lighter Higgs boson $h^0$ with mass of $M_h^{}=125\,$GeV.
The dominant gluon-fusion processes are shown to be mostly sensitive
to the Higgs mixing angle $\,\al\,$,\, and the Higgs production cross section
can be enhanced over the SM expectation
due to the new heavy quark contributions of our model.
The sub-dominant processes of vector boson fusion (VBF) and the Higgs associated productions
are also studied, which always predict suppressed signals relative to the SM Higgs boson.
Combining the analyses of both productions and decays, we demonstrate
that our model can generate the observed di-photon signals in proper parameter regions.
On the other hand, we will show that $h^0$ signals from the $WW^*/ZZ^*$ channels
are always suppressed, and are consistent with the current LHC data \cite{Atlas2012-7,CMS2012-7}.
No excess of Higgs signals come from either the VBF process
or the vector boson associated production in our model.
In parallel, we analyze the LHC signatures of the heavier Higgs boson $H^0$
as a new physics discriminator from the SM in section\,\ref{sec4}.
Due to the suppressed couplings of $H^0$ with the SM particles,
the discovery potential of $H^0$ is quite challenging,
but it can be further probed at the LHC\,(8\,TeV)
with more data collected by the end this year, and at the LHC\,(14\,TeV).
We also study an interesting case
with $h^0$ and $H^0$ being (nearly) degenerate around mass $125-126$\,GeV.
To further discover the new Higgs boson $H^0$ will be crucial for discriminating
our 221 model from the SM.

We conclude in section\,\ref{sec5}. In addition,
we provide two appendices\,\ref{appA} and \ref{appB} to summarize the partial decay widths
for the Higgs bosons $h^0$ and $H^0$ in the 221 model, respectively.


\vspace*{3mm}

\section{\hspace*{-2mm}Spontaneous Symmetry Breaking and Unitarity}
\label{sec2}

In this section we will study the spontaneous symmetry breaking and the unitarity
of the 221 model. We present the complete Higgs potential, and derive the Higgs mass-spectrum
and their mixing. Afterwards, we analyze all Higgs couplings with the gauge bosons and fermions,
as well as gauge boson self-couplings and gauge-fermion couplings. Finally, we study the
unitarity of the present model. We reveal that the exact $E^2$ cancellation in the
longitudinal $V_L V_L$ scattering amplitudes is achieved
by the joint role of exchanging spin-1 new gauge bosons $W'/Z'$ and spin-0 Higgs bosons.
Identifying the lighter Higgs state $\,h^0\,$ to have mass 125\,GeV, we derive the
unitarity bound on the mass of heavier Higgs boson $\,H^0$.\,
We find that the parameter space of this model is highly predictive.

\vspace*{2.5mm}
\subsection{\hspace*{-2mm}Structure of the 221 Model}
\vspace*{2mm}
\label{sec2.1}

We consider a minimal gauge extension of the SM with an extra SU(2) gauge group,
so the full electroweak gauge symmetry
$\,{\cal G}=\gSU(2)_0 \otimes\gSU(2)_1 \otimes \gU(1)_2$\,
at some high energy scale will be spontaneously broken down to $\gU(1)_{\rm em}$ of QED.
The product gauge groups $\,\gSU(2)_0\otimes\gSU(2)_1\otimes\gU(1)_2$\,
have the gauge couplings, $g_0^{}$, $g_1^{}$, and $g_2^{}$, respectively.
Since we are constructing an effective UV completion
of the 3-site Higgsless nonlinear moose model\,\cite{3site},
we will focus on the distinct parameter space
$\,g_1 \gg (\gz,\,\gb )$\, for the present study, which differs from most other
SU(2) extensions in the literature\,\cite{LM,unUF,He:1999vp,Jezo:2012rm}.
With $\,g_1 \gg (\gz,\,\gb )$,\, the $\,\gSU(2)_0 \otimes \gU(1)_2$\, are approximately reduced to
the SM electroweak gauge group $\,\gSU(2)_L \otimes\gU(1)_Y$,\,
while the $\gSU(2)_1$  becomes the additional new gauge symmetry.
We introduce two Higgs doublets $\Phi_1^{}$ and $\Phi_2^{}$, transforming as
({\bf 2},\,{\bf 2},\,{\bf 0}) and ({\bf 1},\,{\bf 2},\,$\f{1}{2}$)
under ${\cal G}$, respectively.
They develop two vacuum expectation values (VEVs), $\,\fx$\, and $\fy$\,,\,
from minimizing the Higgs potential.
The gauge symmetry breaking takes in the following sequences,
\beqs
\label{eq:break}
\beqn
\label{eq:break1}
\gSU(2)_0\otimes\gSU(2)_1
&~\xrightarrow{\langle\Phi_1\rangle\neq0}~&
\gSU(2)_V\,,
\\
\label{eq:break2}
\gSU(2)_1\otimes\gU(1)_2
&~\xrightarrow{\langle\Phi_2\rangle\neq 0}~&
\gU(1)_V\,,
\eeqn
\eeqs
where the subscript ``$_V$" represents the diagonal part of the product groups.
Thus, the Lagrangian for the gauge and Higgs sectors can be written as,
\beqn
\mL &=&-\frac{1}{4} \sum_{a = 1}^{3} W^{a}_{0 \mu \nu} W^{a \mu \nu}_{0}-\frac{1}{4}
\sum_{a = 1}^{3} W^{a}_{1 \mu \nu} W^{a \mu \nu}_{1}
-\frac{1}{4} B_{2 \mu \nu} B^{\mu \nu}_{2}
\nonumber\\
 && +\sum_{j=1,2}\tr\Big[(D_\mu \Phi_j)^\dagger (D^\mu \Phi_j)\Big]-V(\Phi_1, \Phi_2)
\,,
\label{eq:lagrangian}
\eeqn
where $W^{a \mu \nu}_{0}$, $W^{a \mu \nu}_{1}$, and $B^{\mu \nu}_{2}$
are the field strengths of the $\gSU(2)_0$, $\gSU(2)_1$ and $\gU(1)_2$
gauge symmetries, respectively.
According to our assignments for $\Phi_1^{}$ and $\Phi_2^{}$,
we can write down the general Higgs potential $V(\Phi_1, \Phi_2)$
under the 221 gauge group ${\cal G}$\,,
\beqn
 V(\Phi_1^{}, \Phi_2^{}) &\,=\,&
 \frac{1}{2}\lambda_1^{}  \left[ \tr(\Phi_1^\dagger \Phi_1^{}) -\dfrac{f_1^2}{2} \right]^2
  +\frac{1}{2}\lambda_2^{}\left[ \tr(\Phi_2^\dagger \Phi_2^{}) -\dfrac{f_2^2}{2} \right]^2
  \nonumber\\
&& +\lambda_{12}^{}
  \left[ \tr(\Phi_1^\dagger \Phi_1^{}) -\dfrac{f_1^2}{2} \right]
  \left[ \tr(\Phi_2^\dagger \Phi_2^{}) -\dfrac{f_2^2}{2} \right] .
\label{eq:V}
\eeqn
Unlike the 3-site model\,\cite{3site}, we construct $\Phi_1^{}$ and $\Phi_2^{}$
in the linear realization with physical Higgs bosons.
Hence, our 221 model is renormalizable
and provides an effective UV completion of the 3-site model.
Here we use the matrix representation for the Higgs fields,
\beqn
 \Phi_j^{} ~=~
 \frac{1}{2} \left( f_j^{} + h_j^{} + i \tau^a \pi_j^a \right),
 \label{eq:Phi}
\eeqn
with the Pauli matrices $\tau^a$ and two Higgs VEVs $\,f_{j}^{}\in (\fx,\,\fy)$\,.\,
In this representation, the covariant derivatives of $\Phi_j^{}$ are given by
\beqs
\beqn
 D_{\mu} \Phi_1^{}
 &=& \partial_{\mu} \Phi_1^{} + i g_0^{} \frac{\tau^{a}}{2} W_{0 \mu}^{a} \Phi_1^{}
- i g_1^{} \Phi_1^{}\frac{\tau^{a}}{2} W_{1 \mu}^{a} \,,
\\
 D_{\mu} \Phi_2^{}
 &=& \partial_{\mu} \Phi_2^{} + i g_1^{} \frac{\tau^{a}}{2} W_{1 \mu}^{a} \Phi_2^{}
- i g_2^{}  \Phi_2^{}\frac{\tau^{3}}{2} B_{2 \mu}  \,.
\eeqn
\eeqs

Besides of the SM gauge boson $(W,\, Z,\, \gamma)$,
there are three new gauge bosons $(W',\, Z')$ in the gauge sector.
Under the gauge symmetry breaking (\ref{eq:break}),
all massive gauge bosons acquire their masses as follows,
\beqs
\label{eq:GaugeMass}
\beqn
m_W^2 &=& \frac{1}{4}
\frac{g_0^2f_2^2}{\,1 \!+\! r^2\,}\Big[1-\frac{x^2}{(1\!+\!r^2)^2}+\mO(x^4) \Big] \,,
\label{eq:MWmass}
\\
m_Z^2\, &=&
\frac{1}{4c^2}\frac{g_0^2f_2^2}{\,1 \!+\! r^2\,}
\Big[1-\frac{\,(c^2-r^2 s^2)^2x^2\,}{\,c^2(1\!+\!r^2)^2\,}+\mO(x^4) \Big] \,,
\label{eq:MZmass}
\\
M_{W'}^2 &=&
\frac{1}{4} g_1^2 f_1^2 (1 \!+\! r^2)
\Big[1+\frac{x^2}{(1\!+\!r^2)^2}+\mO(x^4)\Big] \,,
\label{eq:MWpmass}
\\
M_{Z'}^2 &=&
\frac{1}{4} g_1^2 f_1^2 (1 \!+\! r^2)
\Big[1+\frac{1+r^4 (s^2/c^2)}{\,(1\!+\!r^2)^2\,}x^2+\mO(x^4)\Big] \,,
\label{eq:MZpmass}
\eeqn
\eeqs
where $\,x^2 \equiv g_{0}^2/g_{1}^2 \ll 1$\, is a small parameter under
$\,g_1^{2} \gg g_0^{2}, g_2^{2}$,\, and $\,r\equiv f_2^{}/f_1^{}\,$ is the VEV ratio.
In (\ref{eq:MZmass}) and (\ref{eq:MZpmass}), we have adopted the notation,
$\,(c,\,s)\equiv (\cos\theta,\,\sin\theta)\,$,\, with $\,s/c\equiv g_2^{}/g_0^{}\,$.\,
From (\ref{eq:MWmass}) and (\ref{eq:MWpmass}), we can reexpress the small expansion parameter $x$ as,
\beqa
x &=& \f{\,1\!+\!r^2\,}{r}\f{m_W^{}}{M_{W'}^{}}
\[1 + \f{m_W^2}{\,r^2M_{W'}^2\,} + \mO\(\!\f{m_W^4}{M_{W'}^4}\!\)\] .
\eeqa
The formulae given in this subsection hold for general VEVs $(\fx,\,\fy)$,
which extend the gauge-sector results of the 3-site model
under $\,\fx =\fy\,$ \cite{3site}.  We note that the $W'$ and $Z'$
are nearly degenerate and their masses differ only
at the $\mO(x^2)$.\,
The two Higgs VEVs $(f_1^{}, f_2^{})$ are connected to the Fermi constant
$\,G_F^{}\,$ via
\beqn
\label{eq:vev}
\frac{1}{f_1^2}+\frac{1}{f_2^2} ~=~ \frac{1}{v^2}\,,
\eeqn
where
\beqn
\label{eq:vev2}
v^2 &\,=\,& v_0^2
\left[1-\frac{2r^2}{\,(1+r^2)^2\,}x^2
+\frac{\,(3+r^2)r^2\,}{\,(1+r^2)^4\,}x^4+\mO(x^6)\right]\,,
\eeqn
and $\,v_0^2 \equiv (\sqrt{2}G_F^{})^{-1}\,$ \cite{3site}.
Thus, we have $\,f_2^{}= v\sqrt{1\!+\!r^2\,}\,$.\,
So, from the formulae (\ref{eq:MWmass})-(\ref{eq:MZmass}),
it is clear that under $\,g_1^2 \gg g_0^2, g_2^2$\,,\,
we can approximately identify $\,g_0^{}$\, and $\,g_2^{}$\, as the SM electroweak gauge couplings.
It can be further shown that the angle $\,\theta =\arctan (g_2^{}/g_0^{})$\,
approximates the Weinberg angle $\,\theta_W^{}$\, of the SM up to $\,\mO(x^2)\,$ corrections.
In addition, the mass-eigenstates are related to the gauge eigenstates as follows,
\beqs
\beqn
 W_{\mu} &\,\simeq\,& W^{\pm}_{0 \mu} + \frac{x}{\,1 + r^2\,} W^{\pm}_{1 \mu} \,,
\label{eq:Wwave}
\\[1.5mm]
 W'_{\mu} &\,\simeq\,& W^{\pm}_{1 \mu} - \frac{x}{\,1 + r^2\,}W^{\pm}_{0 \mu} \,,
\\[1.5mm]
 Z_{\mu} &\,\simeq\,& \(c W^3_{0 \mu}-s B_{2\mu}\)
 + x\f{\,c^2\!-\! s^2r^2\,}{\,c(1+r^2)\,} W^{3}_{1 \mu} \,,
\\[1.5mm]
 Z'_{\mu} &\,\simeq\,& W^{3}_{1 \mu}- \frac{x}{\,1+r^2\,}W^{3}_{0 \mu}
 - x\frac{sr^2}{\,1+r^2\,}B_{2\mu} \,,
\label{eq:Zpwave}
\eeqn
\eeqs
with $\,\mO(x^3)$\, corrections neglected. The massless photon field is exactly expressed as,
\beqn
 A_{\mu} ~=~
 \frac{e}{g_0}W^{3}_{0 \mu} +\frac{e}{g_1}W^{3}_{1 \mu}+\frac{e}{g_2} B_{2 \mu} \,,
\eeqn
where the QED gauge coupling $\,e$\, of the unbroken $U(1)_{\rm em}$ is
inferred from the gauge couplings of the group $\,{\cal G}$\,,
\beqn
\label{eq:ee}
 \frac{1}{e^2} ~=~ \frac{1}{g_0^2}+ \frac{1}{g_1^2}+\frac{1}{g_2^2} \,.
\eeqn
From the expressions (\ref{eq:Wwave})-(\ref{eq:Zpwave}),
we see that $W^{a}_{0 \mu}$ and $B_{2 \mu}$ are approximately the gauge bosons
of the SM gauge group $\gSU(2)_L\times \gU(1)_Y$,\,
while $W^{a}_{1 \mu}$ are mainly the new gauge bosons $\,W^{\prime a}_{\mu}$\,.\,
Due to the mixing between the lighter and heavier gauge bosons,
the $WWZ$ coupling becomes different from the SM value. The LEP-II limit on
this coupling translates into a lower bound on the $W'$ mass, around
300\,GeV \cite{3site,tri-3site}.

\vspace*{2.5mm}
\subsection{\hspace*{-2mm}Higgs Sector of the Model}
\label{sec2.2}

Minimizing the Higgs potential (\ref{eq:V}), we find that
$\,\left<\Phi_1^{}\right>=\hf f_1^{}\,\mathbb{I}_{2\times 2}\,$ and
$\,\left<\Phi_2^{}\right>=\hf f_2^{}\,\mathbb{I}_{2\times 2}$\,,\,
under the following conditions,
\beqa
\label{eq:cond1-lambda}
\la_1^{},\la_2^{} ~>~ 0\,, ~~~~~~
\la_1^{}\la_2^{} ~\neq~ \la_{12}^2 \,.
\eeqa
After spontaneous symmetry breaking,
six would-be Nambu-Goldstones $(\pi_{1}^a,\,\pi_2^a)$ are eaten
by the gauge bosons $(W_1^a,\,W_2^a)$ which compose the longitudinal
components of the lighter mass-eigenstates $(W,\,Z)$ and the heavier ones $(W',\,Z')$.\,
Only two physical CP-even neutral scalars $(h_1^{}, h_2^{})$ survive.
This is very different from the conventional two-Higgs-doublet model
which contains five physical Higgs states.
The mass matrix for the CP-even states $(h_1^{},\, h_2^{})$ can be generally expressed as,
\begin{equation}
\label{eq:h1h2-massterm}
-\frac{1}{2}
\left(\begin{array}{cc}
 h_1 & h_2 \\
\end{array}
\right)
  \left(
    \begin{array}{cc}
      \lambda_1 f_1^2 & \lambda_{12} f_1 f_2 \\[1.5mm]
      \lambda_{12} f_1 f_2 & \lambda_2 f_2^2
    \end{array}
  \right)
\left(\begin{array}{c}
 h_1\\ h_2\end{array} \right),
\end{equation}
which is diagonalized by the orthogonal transformation,
\beqa
  \left(
    \begin{array}{rr}
      c_{\alpha}^{}  ~& -s_{\alpha}^{} \\[1.5mm]
      s_{\alpha}^{}  ~&  c_{\alpha}^{}
    \end{array}
  \right)\!
  \left(
    \begin{array}{cc}
      \lambda_1 f_1^2 & \lambda_{12} f_1 f_2 \\[1.5mm]
      \lambda_{12} f_1 f_2 & \lambda_2 f_2^2
    \end{array}
  \right)
  \left(
    \begin{array}{rr}
      c_{\alpha} &~\, s_{\alpha} \\[1.5mm]
     -s_{\alpha} &~\, c_{\alpha}
    \end{array}
  \right)
  \,=\,
  \left(
    \begin{array}{cc}
      M_h^2 & 0 \\[1.5mm]
      0  & M_H^2
    \end{array}
  \right),
\label{eq:HiggsMass}
\eeqa
where mass-eigenstates $(h,\, H)$ are connected to the gauge-eigenstates $(h_1,\, h_2)$
via
\begin{equation}
\label{eq:h1h2-massR}
  \left(
    \begin{array}{c}
      h \\[1.5mm]
      H
    \end{array}
  \right)
  =
  \left(
    \begin{array}{rr}
      c_{\alpha}^{} &~ -s_{\alpha}^{} \\[1.5mm]
      s_{\alpha}^{} &~  c_{\alpha}^{}
    \end{array}
  \right)
  \left(
    \begin{array}{c}
      h_1 \\[1.5mm]
      h_2
    \end{array}
  \right).
\end{equation}
Thus we can derive the mass-eigenvalues $M_{h, H}^2$ and the mixing angle $\alpha$\,,
\beqs
\label{eq:HiggsM-alpha}
\beqn
\label{eq:HiggsMass}
M_{h, H}^2 &\,=\,&
\frac{1}{2}\Big[(\lambda_1 f_1^2+\lambda_2 f_2^2) \mp
\sqrt{(\lambda_1 f_1^2 - \lambda_2 f_2^2)^2 + 4\lambda_{12}^2 f_1^2 f_2^2\,}\,\Big],
\\[2mm]
\sin 2\alpha &\,=\,&
\frac{ 2 \lambda_{12}  }
{\sqrt{(\lambda_1^{} r^{-1} - \lambda_2^{} r)^2 + 4\lambda_{12}^2\,}\,} \,,
\label{eq:mixing}
\eeqn
\eeqs
where $\,r = f_2^{}/f_1^{}$\, is the VEV ratio introduced earlier.
In (\ref{eq:HiggsMass}), the positivity condition of the lighter Higgs mass
$\,M_h^2 > 0\,$ further requires,
\beqa
\label{eq:cond2-lambda}
\la_1^{}\la_2^{} ~>~ \la_{12}^2 \,.
\eeqa
which also holds the second inequality of (\ref{eq:cond1-lambda}).
We note that the conditions (\ref{eq:cond1-lambda}) and
(\ref{eq:cond2-lambda}) ensure the tree-level vacuum stability
of the Higgs potential.

Finally, inspecting  Eq.\,(\ref{eq:V}) we see that the Higgs potential contains five
parameters in all,  namely, $\,(\la_1^{},\,\la_2^{},\,\la_{12}^{})$\, and $\,(\fx,\,\fy)$\,.\,
Imposing the new LHC data of $\,M_h\simeq 125\,$GeV \cite{Atlas2012-7,CMS2012-7} and
the VEV condition (\ref{eq:vev}) with the Fermi constant $G_F^{}$,\,
we are left with three input parameters of $(M_H^{},\,\al,\,r)$.\, Here
the VEV ratio $\,r=1\,$ is the default of the 3-site model\,\cite{3site}
for collective symmetry breaking. In the current study, we also set
$\,r=1\,$ as our default, but we will further explore wider parameter space with
$\,r=\mO(1)\,$. Hence, from the VEV condition (\ref{eq:vev}) with $G_F^{}$,
and given the ratio $\,r\,$ and the lighter Higgs mass
$\,M_h\simeq 125\,$GeV \cite{Atlas2012-7,CMS2012-7},\,
we only have two free parameters in the Higgs sector,
which can be expressed as the
Higgs mixing angle $\,\al\,$ and the heavier Higgs mass $\,M_H^{}\,$.\,
In the gauge sector, the three gauge couplings $(\gz,\,\gga,\,\gb)$
can be fixed by inputting the fine-structure constant
$\,\al_{\rm em}^{}=e^2/4\pi\,$ via (\ref{eq:ee})
together with the gauge boson masses $(m_W^{},\,M_W')$
via (\ref{eq:GaugeMass}), i.e.,
$(\al_{\rm em}^{},\,m_W^{},\,M_W')$, or equivalently,
$(\al_{\rm em}^{},\,m_Z^{},\,M_W')$.

In the following sections, we will perform the LHC analyses based on the
two major input parameters  $(\al,\,M_H^{})$.\,
The other secondary input parameters include the VEV ratio $\,r=\mO(1)\,$,
the heavier gauge boson mass $\,M_W'\gtrsim 300\,$GeV,
and the heavier fermion mass $\,M_F^{}\gtrsim 1.8\,$TeV,
as will be summarized in Eq.\,(\ref{eq:Pranges}).
We will first identify the proper parameter range of the Higgs mixing angle $\,\al$\,
which gives rise to the observed $\,\dip\,$ signals
of the light Higgs boson $\,h\,$ around $125$\,GeV.
Then, we further predict the signals
for the heavier Higgs boson $H$ via $WW$ and $ZZ$ channels over
the wide mass-ranges of $\,M_H^{}=130-600\,$GeV,
as a new physics discriminator from the SM.

\vspace*{2.5mm}
\subsection{\hspace*{-2mm}Fermion Sector of the Model}
\label{sec2.3}

The fermion sector contains the left-handed Weyl fermions $\Psi_{0L}$,
the right-handed Weyl fermions $\Psi_{2R}^{}=(\Psi^u_{2R}, \Psi^d_{2R})$,\,
and the vector-like fermions $\Psi_1^{}=(\Psi_{1L}, \Psi_{1R})$.\,
Same as the 3-site model\,\cite{3site}, these vector-like fermions $\Psi_1$ are necessary for
the ideal delocalization condition and providing compatible results
for the precision electroweak measurement \cite{iDLF}.
In addition, the vector-like fermions $\Psi_1$ can naturally have
gauge-invariant renormalizable mass-term.
The representations for fermions are summarized in Table\,1. 

With these, we can write down the Higgs Yukawa interactions and fermion mass-terms
in this model,
\beqn
\mL_{\rm Y}^{} &\,=\,&
-\sum_{i,j}\left[ y_{1 ij}^{}\overline{\Psi}_{0L}^i \Phi_1^{}\Psi_{1R}^j
+  \overline{\Psi}_{1L}^i \Phi_2^{} y_{2 ij}^{} \Psi_{2R}^j
 + \overline{\Psi}_{1L}^i M_{ij}^{}\Psi_{1R}^j+ \textrm{h.c.}\right] ,
\label{eq:lagrangian-Yukawa-sec}
\eeqn
where $i,j=1, 2, 3$ are the generational indices. $\,M_{ij}^{}\,$ represents a Dirac mass matrix,
and $\,y_{2 ij} = {\rm diag}\( y_{2}^{u},\, y_{2}^{d}\)_{ij}^{}$\,.\,
In general, $y_{1ij}^{}$ and $M_{ij}^{}$ are not flavor-blind.
For consistency with the FCNC constraints on the off-diagonal components
of $y_{1ij}^{}$ and $M_{ij}^{}$ \cite{Abe:2011sv},
we will set $y_{1ij}^{}$ and $M_{ij}$ to be flavor-diagonal,
i.e., $\,y_{1ij}^{} = y_1^{} \delta_{ij}^{}$\, and
$\,M_{ij}^{} = \MF \delta_{ij}^{}$.\,
With this, all flavor structures in this model are embedded in $y_{2u}^{}$ and $y_{2d}^{}$.
We also consider  $\,\MF \gg y_1^{} f_1^{}\,$ and
$\,\MF \gg (y_{2}^{u},\,y_2^d) f_2^{}\,$.\,
This makes the non-SM fermion masses much heavier than the SM fermion masses,
and makes $\Psi_{0L}^{}$ and $\Psi_{2R}^{}$ approximately behave like the SM fermions.
It is also consistent with the fact
$\,\gSU(2)_0 \times \gU(1)_2\approx\gSU(2)_L \times \gU(1)_Y$\,,\,
as we commented earlier.
For simplicity, we will take the same $\MF$ and $y_1^{}$ values in both quark and lepton sectors
for the present study.
The mass-eigenstates for the quark sector are expressed as,
up to $\mO(\epsilon_L^2, \epsilon_R^2)$ corrections,
\beqs
\label{eq:qQ}
\beqn
q_{L}^{} &\simeq & - \Psi_{0L}^{} + \epsilon_L^{} \Psi_{1L}^{} \,,
\qquad
q_{R}^{} \simeq - \epsilon_{R}^{} \Psi_{1R}^{} + \Psi_{2R}^{} \,,
\\[1.5mm]
Q_{L}^{} &\simeq & -\epsilon_L^{} \Psi_{0L}^{} - \Psi_{1L}^{}\,,
\qquad Q_{R}^{} \simeq \Psi_{1R}^{} +  \epsilon_{R}^{} \Psi_{2R}^{}\,,
\eeqn
\eeqs
where $\,q_{L, R}^{}\,$ denote the SM quarks,
and $Q_{L, R}^{}$ represent the corresponding heavy fermions.
The relations for lepton sector are expressed in a similar manner.
Eq.\,(\ref{eq:qQ}) contains two small parameters,
$\,\epsilon_L^{} = y_{1}^{} f_1^{} /(2 \MF) \ll 1$,\, and
$\,\epsilon_R^{} = y_{2}^{} f_2^{} /(2 \MF) \ll 1$\,.\,
It is clear that $\Psi_{L0}^{}$ and $\Psi_{R2}^{}$
are approximately the SM fermions.
Up to $\,\mO(\epsilon_L^2)$\, and $\,\mO(\epsilon_R^2)$\, corrections,
the mass-eigenvalues for the SM fermions and the heavy fermions are given by,
\beqn
 m_{q}^{} \,\simeq\, \epsilon_L^{} \epsilon_R^{} \MF\,,
&~~~~&
 M_Q^{} \,\simeq\, \MF \,.
\label{eq:KKfermiMass}
\eeqn
The light fermion mass-term is proportional to
$\,\epsilon_R^{}$\,, or $\,y_{2}^{}$\,,\,
which has all flavor structures and mass hierarchies embedded.

\begin{table}[t]
\label{tab:fermion}
\begin{center}
\caption{Assignments for fermions under the gauge group of the present model.
In the fourth and fifth columns, the $\gU(1)_2$ charges and $\gSU(3)_c$ representations
are shown for the quarks (without parentheses) and leptons (in parentheses), respectively.}
\vspace*{3mm}
\begin{tabular}{c||cccc}\hline\hline
&&&& \\[-3.5mm]
Fermions & $\gSU(2)_0$ & $\gSU(2)_1$ & $\gU(1)_2$ & $\gSU(3)_c$\\
\hline
&&&& \\[-3.5mm]
$\Psi_{0L}$ & $\bf 2$  & $\bf 1$ & $\frac{1}{6}$
	     $\left(-\frac{1}{2}\right)$ & $\bf 3$ ($\bf 1$) \\[1mm]
\hline
$\Psi_{1L}$ & $\bf 1$ & $\bf 2$ & $\frac{1}{6}$
	     $\left(-\frac{1}{2}\right)$ & $\bf 3$ ($\bf 1$)\\[1mm]
$\Psi_{1R}$ & $\bf 1$ & $\bf 2$ & $\frac{1}{6}$
	     $\left(-\frac{1}{2}\right)$ & $\bf 3$ ($\bf 1$)\\[1mm]
\hline
$\Psi^u_{2R}$ & $\bf 1$ & $\bf 1$ & $~~\f{2}{3}~(0)$ & $\bf 3$ ($\bf 1$)\\[1mm]
$\Psi^d_{2R}$ & $\bf 1$ & $\bf 1$ & $-\f{1}{3}~(-1)$ & $\bf 3$ ($\bf 1$)\\[-3.5mm]
&&&&\\
\hline\hline
\end{tabular}
\end{center}
\end{table}

The couplings between the SM gauge boson and the SM fermions,
such as $\,g_{Zf_Lf_L}^{}\,$,\, are different from those in the SM
due to the mixing among the SM particles and extra heavy particles.
This leads to nonzero electroweak parameter $S$ \cite{Peskin-S} at the tree-level.
Such contributions to the $S$ parameter can be reduced to zero
by adjusting $\epsilon_L^{}$ as follows \cite{3site},
\beqn
 \epsilon_L^{} ~=~
 \frac{m_W^{}}{\,M_{W'}^{}\,}\(1+r^{-2}\)^{\hf}\!\left[1+\mO(x^2)\right] ,
\label{eq:idealDelocal}
\eeqn
via the ideal delocalization\,\cite{iDLF}\footnote{It is also possible
to make $S$ small enough by considering large $W'$ mass of $\mO(1)$\,TeV or above.
We do not pursue this possibility for the current construction.},\,
where $\,r\equiv \fy /\fx\,$.\,
To keep $\epsilon_L^{}$ small, it is necessary to control the ratio
$\,f_2^{}/f_1^{}$.\,
For the phenomenological analyses, we will take the natural parameter range,
$\,
\hf \lesssim f_2^{}/f_1^{} \lesssim 2\,.
$\,
The heavy fermion mass-parameter $\MF$ is constrained by the $T$ parameter, which
gives a lower limit, $\,\MF > 1.8\,\TeV$\, \cite{3site,Abe:2011sv}.

As a summary of the model setup, we list the parameter sets and the relevant
ranges consistent with the low energy precision constraints,
\beqn
\label{eq:Pranges}
&& \hf \,\lesssim\, f_2^{}/f_1^{} \,\lesssim\, 2\,,~~~~~ \alpha\in [0, \pi)\,,
\nonumber\\[1.5mm]
&& \MF \,\gtrsim\, 1.8\,\TeV\,,~~~~~  M_{W'}^{} \,\gtrsim\, 300\,\GeV\,.
\eeqn
As we will demonstrate in the next two sections,
the predictions of Higgs signal are mostly sensitive to the mixing angle $\,\al\,$.\,
The other model parameters varying within the ranges of (\ref{eq:Pranges})
only yield sub-leading contributions.

\vspace*{2.5mm}
\subsection{\hspace*{-2mm}Couplings of Higgs Bosons, Gauge Bosons and Fermions}
\label{sec2.4}

In this subsection, we present the Higgs boson couplings with gauge bosons
and fermions, as well as the gauge boson couplings with fermions.
These will be needed for our analyses of Higgs productions and decays
in the next two sections.

\subsubsection{\hspace*{-2mm}Gauge and Yukawa Couplings of Higgs Bosons}
\label{sec2.4.1}


We analyze the gauge and Yukawa couplings of the CP-even Higgs bosons
$\,h^0\,$ and $\,H^0\,$,\, respectively. For the light weak gauge bosons $\,V(=W,Z)$,\,
the triple Higgs-gauge couplings $\,G_{hVV}^{}\,$ and $\,G_{HVV}^{}\,$
take the following forms up to $\mO(x^2)$ corrections,
\beqs
\label{eq:VVh}
\beqn
\label{eq:G-VVh/H}
G_{hVV}^{} \,&=&\,  G_{hVV}^{\rm SM}\xi_{hVV}^{} \,, ~~~~~~~
G_{HVV}^{} \,=\,  G_{hVV}^{\rm SM}\xi_{HVV}^{} \,,
\\[2.5mm]
\label{eq:xi-VVh/H}
\xi_{hWW}^{} \,&=&\, \frac{\,r^3\ca -\sa}{\,(1 + r^2)^{3/2}\,}
+\frac{\,r^2[-3\sa+(r^3-2r)\ca ]\,}{\,(1+r^2)^{7/2}\,}x^2 \,,
\\[2.5mm]
\xi_{hZZ}^{} \,&=&\, \frac{\,r^3\ca\!-\!\sa}{\,(1\!+\!r^2)^{3/2}\,}
+\frac{\,r^2[(-3\!+\!s_W^2(3\!+\!2r))\sa+((r^3\!-\!2r)\!+\!s_W^2r(r^2\!+\!2))\ca]\,}
      {\,c_W^2 (1+r^2)^{7/2}\,}x^2 ,~~~~~~~~~
\\[2.5mm]
\label{eq:xi-hVV}
\xi_{HWW}^{} &=& \frac{\,r^3\sa +\ca}{\,(1 + r^2)^{3/2}\,}
+\frac{\,r^2[3\ca+(r^3-2r)\sa]\,}{\,(1+r^2)^{7/2}\,}x^2 \,,
\\[2.5mm]
\xi_{HZZ}^{} &=&
\frac{\,r^3\sa\!+\!\ca}{\,(1\!+\!r^2)^{3/2}\,}
+\frac{\,r^2[(3\!-\!s_W^2(3\!+\!2r))\ca+((r^3\!-\!2r)\!+\!s_W^2r(r^2\!+\!2))\sa]\,}
      {\,c_W^2 (1+r^2)^{7/2}\,}x^2 ,~~~~~~~~~
\eeqn
\eeqs
with $\,G_{hVV}^{\rm SM}=2 M_V^2/v_0^{}$\, for $\,V=W,Z$\,.

An important feature of the Higgs-gauge-boson triple coupling $\,G_{hVV}^{}$\,
is that it vanishes when the Higgs mixing angle $\,\al\,$ takes a special value,
$\,\tan\al \simeq r^3\,$,\, where $\,r\equiv f_2^{}/f_1^{}\,$.\,
In this case, the light Higgs boson $h$ becomes gaugephobic regarding
its triple couplings. The corresponding collider phenomenology
of $\,h^0$\, would deviate from the SM Higgs boson substantially.
Similarly, for the special mixing angle $\,\tan\al \simeq -1/r^3\,$,\,
the $G_{HVV}^{}$ coupling vanishes, and thus $H$ Higgs state becomes gaugephobic so long as the triple coupling is concerned.
We note that with the generic choices of two Higgs VEVs, $\,r=f_2/f_1 \sim \mO(1)$\,,\,
$\,G_{hVV}^{}\,$ and $\,G_{HVV}^{}\,$ couplings are always smaller than
$G_{hVV}^{\rm SM}$ for any mixing angle $\,\al$\,.\,

The Yukawa couplings between the Higgs bosons and the SM fermions are also modified
relative to the SM values. We derive these couplings in following form
up to $\mO(\epsilon_L^2)$ and $\mO(\epsilon_R^2)$ corrections,
\beqs
\label{eq:tth/H}
\beqn
\label{eq:G-tth/H}
&& G_{h f f}^{} ~=~ \frac{m_f^{}}{v_0^{}}\xi_{hff}^{}\,,~~~~~~~
   G_{H f f}^{} ~=~ \frac{m_f^{}}{v_0^{}}\xi_{Hff}^{}\,,
\\[2mm]
\label{eq:xi-tth/H}
&& \xi_{hff}^{} ~=~  \frac{\,r\ca - \sa\,}{\sqrt{1+r^2\,}\,}-\frac{r(r\sa+\ca)}{(1+r^2)^{5/2}}x^2
+\frac{\sa\sqrt{1+r^2}}{x^2}\frac{m_f^2}{M_F^2}
\,,\\[2mm]
\label{eq:xi-tth}
&& \xi_{Hff}^{} ~=~ \frac{\,r\sa + \ca\,}{\sqrt{1+r^2\,}\,}-\frac{r(-r\ca+\sa)}{(1+r^2)^{5/2}}x^2
-\frac{\ca\sqrt{1+r^2}}{x^2}\frac{m_f^2}{M_F^2} \,,
\label{eq:xi-ttH}
\eeqn
\eeqs
We see that the Yukawa couplings for both $h^0$ and $H^0$ are always smaller
than the corresponding SM Yukawa couplings, i.e.,
\,$|\xi_{hff}^{}|\leqq 1$\, and \,$|\xi_{Hff}^{}|\leqq 1$\,.\,
Moreover, the Yukawa couplings $\xi_{hff}^{}$ or $\xi_{Hff}$
vanishes under the special value of mixing angle $\,\tan\alpha=r\,$
or $\,\tan\alpha =-1/r$\,,\, and thus $h$ or $H$ becomes fermiophobic.

 We also analyze the additional Higgs couplings with the new gauge bosons $\,V'(=W',Z')$\,
 and new fermions $\,F$\,.\,
 The new particles $V'$ and $F$ will appear
 in the loop-induced Higgs boson productions and decays.
 At the leading order, we derive these additional new couplings of the Higgs boson
 $h^0$ as follows,
\beqs
\label{eq:hVV-hFF}
\beqn
&& \hspace*{-16mm}
G_{hVV'}^{} =\, \frac{2m_V^{} M_{V'}^{}}{v_0^{}}\xi_{hVV'}^{}\,,
\qquad~~
\xi_{hVV'} = -\frac{\,r(\sa \!+ r\ca )\,}{\,(1+r^2)^{3/2}\,}\,,
\label{eq:gWWph}
\\
&& \hspace*{-16mm}
G_{hV'V'}^{} =\, \frac{2M_{V'}^2}{v_0^{}}\xi_{hV'V'}^{} \,,
\qquad~~
\xi_{hV'V'}^{} =\, \frac{\,r(\ca \!- r\sa)\,}{\,(1+r^2)^{3/2}\,}\,,
\label{eq:gWpWph}
\\
&& \hspace*{-16mm}
G_{h\bar{f}_L^{} F_R}^{} =\, \frac{\,M_F}{v_0^{}}\xi_{h\bar{f}_L^{}F_R^{}}\,,
\qquad~~
\xi_{h\bar{f}_L^{} F_R^{}} = \frac{\ca r}{\,1\!+\!r^2\,}x \,,
\label{eq:gfFh}
\\
&& \hspace*{-16mm}
G_{h\bar{F}_L^{} f_R^{}} =\, \frac{m_f^{}}{v_0^{}}
\xi_{h\bar{F}_L^{}f_R^{}} \,,
\qquad~~
\xi_{h\bar{F}_L^{}f_R^{}} = \frac{\,\sa}{\,x\,} \,,
\label{eq:gFfh}
\\
&& \hspace*{-16mm}
G_{h\bar{F}_L^{}F_R^{}} =\, \frac{M_F}{v_0^{}}\xi_{hFF}^{}\,,
\qquad~~
\xi_{hFF}^{} = \f{\,\ca r x^2}{\,(1\!+\!r^2)^{3/2}\,}
- \frac{\,\sa\sqrt{1\!+\!r^2}\,}{x^2}\frac{\,m_f^2\,}{\,M_F^2\,}
\label{eq:fFFh}
\,,
\eeqn
\eeqs
where $M_F^{}$ denote the heavy fermion masses.
Due to the ideal delocalization result (\ref{eq:idealDelocal}),
the Higgs couplings to the new heavy fermions depend on the gauge boson masses.
These couplings enter the loop-induced processes for the Higgs boson productions and decays,
and their effects will be taken into account in the following numerical analysis.

In parallel, we derive couplings of the heavier Higgs boson $H^0$
with new gauge bosons $V'$ and new fermions $F$,\,
which are obtained by making the replacement of
$\,(c_{\alpha}^{},\,s_{\alpha}^{})\to (s_{\alpha}^{},\,-c_{\alpha}^{})$\,,
\beqs
\beqn
&& \hspace*{-16mm}
G_{HVV'}^{} =\, \frac{2m_V^{} M_{V'}^{}}{v_0^{}}\xi_{HVV'}^{}\,,
\qquad~~
\xi_{HVV'} = -\frac{\,r(r\sa -\ca)\,}{\,(1+r^2)^{3/2}\,}\,,
\label{eq:gWWpH}
\eeqa
\beqa
&& \hspace*{-16mm}
G_{HV'V'}^{} =\, \frac{2M_{V'}^2}{v_0^{}}\xi_{HV'V'}^{} \,,
\qquad~~
\xi_{HV'V'}^{} = \frac{\,r(\sa + r\ca)\,}{\,(1+r^2)^{3/2}\,}\,,
\label{eq:gWpWpH}
\\
&& \hspace*{-16mm}
G_{H\bar{f}_L^{} F_R}^{} = \frac{\,M_F}{v_0^{}}\xi_{H\bar{f}_L^{}F_R^{}}\,,
\qquad~~
\xi_{H\bar{f}_L^{} F_R^{}} = \frac{\sa r}{\,1\!+\!r^2\,}x \,,
\label{eq:gfFH}
\\
&& \hspace*{-16mm}
G_{H\bar{F}_L^{} f_R^{}} = \frac{m_f^{}}{v_0^{}}
\xi_{H\bar{F}_L^{}f_R^{}} \,,
\qquad~~
\xi_{H\bar{F}_L^{}f_R^{}} = -\frac{\,\ca}{\,x\,} \,,
\label{eq:gFfH}
\\
&& \hspace*{-16mm}
G_{H\bar{F}_L^{}F_R^{}} =\, \frac{M_F}{v_0^{}}\xi_{HFF}^{}\,,
\qquad~~
\xi_{HFF}^{} = \f{\,\sa r x^2}{\,(1\!+\!r^2)^{3/2}\,}
+ \frac{\,\ca\sqrt{1\!+\!r^2}\,}{x^2}\frac{\,m_f^2\,}{\,M_F^2\,} \,.
\label{eq:fFFH}
\eeqn
\eeqs

\vspace*{3mm}

\subsubsection{\hspace*{-2mm}Gauge Boson Self-Couplings and Gauge-Fermion Couplings}
\label{sec2.4.2}

Besides the above gauge and Yukawa couplings with Higgs bosons,
some of the gauge boson self-couplings and gauge-fermion couplings
invoke both the SM fields and the extra heavy states.
They will contribute to the unitarity sum rules (Sec.\,2.5) and
the loop-induced Higgs decay channels (Appendix).

We first summarize the cubic gauge couplings of ${VVV'}$ and $VV'V'$
with extra $V'\,(=W',Z')$ gauge bosons up to the $\mO(x^2)$ corrections,
\beqs
\label{eq:VVV}
\beqn
&&
G_{WW\gamma}^{}\,=\, G_{W'W'\gamma} \,=\, e\,,
\qquad
G_{WW'\gamma}\,=\,0\,,
\\[1.5mm]
&&
G_{WWZ'} \,= -\frac{e}{s_W}\frac{r^2\,x}{\,(1\!+\!r^2)^2\,} \,,
\qquad
G_{WW'Z} \,=\, \frac{e}{s_W^{} c_W^{}}
\frac{r^2\,x}{\,(1\!+\!r^2)^2\,}\,,
\label{eq:VVVpCoup}
\\[1.5mm]
&&
G_{WW'Z'}^{} \,=\,
\frac{e}{s_W^{}}\frac{1}{\,1\!+\!r^2\,}\,,
\qquad
G_{W'W'Z} \,=\,
\frac{e}{s_W^{}c_W^{}}\frac{\,c_W^2 \!-\! s_W^2 r^2\,}{\,1\!+\!r^2\,}\,.
\eeqn
\eeqs

For quartic gauge couplings, we include both $VVVV$ couplings
and those involving the $\,\gamma\gamma$\, or $\,Z\gamma$\,,\,
as relevant to our phenomenological analysis,
\beqs
\beqn
&& \hspace*{-8mm}
G_{WWWW}^{}\,=\,\frac{e^2}{s_W^2}\left[1-\frac{1-2r^2}{(1+r^2)^2}x^2\right],
~~~~~
G_{WWZZ}^{}\,=\,\frac{e^2}{s_W^2}\left[c_W^2-\frac{c_W^2-2r^2}{(1+r^2)^2}x^2\right],~~~~
\\[2mm]
&& \hspace*{-8mm}
G_{WW\gamma\gamma}^{}\,=\,G_{W'W'\gamma\gamma}=e^2\,,
~~~~~
G_{WW'\gamma\gamma}^{} \,=\,0\,,
\\[1.5mm]
&& \hspace*{-8mm}
G_{WWZ\gamma}^{} \,=\, eG_{WWZ}^{} \,,
~~~~~
G_{WW'Z\gamma}^{} \,=\, eG_{WW'Z}^{} \,,
~~~~~
G_{W'W'Z\gamma} \,=\, eG_{W'W'Z}^{} \,.~~~~~~
\eeqn
\eeqs

Finally, we derive all relevant couplings of the light weak gauge bosons $W$ or $Z$
with the fermions. For the $W$ couplings to the fermion pairs, we have,
\beqs
\beqn
\label{eq:Wff-L}
&&\hspace*{-12mm}
G_{W \bar{f'}_L f_L}=\,\frac{e}{s_W}
\,,
\\ 
&&\hspace*{-12mm}
G_{W \bar{F}_L f_L} =\,
G_{W \bar{f}_L F_L}=\,
\frac{e}{s_W}\frac{r^2}{(1+r^2)^{3/2}} x
\,,
\\ 
&&\hspace*{-12mm}
G_{W \overline{F}_L F_L}=\,
\frac{e}{s_W}\frac{1}{1 + r^2}
\,,
\label{eq:Wff-R}
\eeqa
\beqa
&&\hspace*{-12mm}
G_{W \bar{f}_R f_R}=\,
\frac{e}{s_W}
\frac{m_u m_d}{M_F^2}
\frac{1}{x^2}
\,,
\\ 
&&\hspace*{-12mm}
G_{W \overline{U}_R d_R}
=\frac{e}{s_W}
\frac{1}{(1 + r^2)^{1/2}}
\frac{m_d}{M_F}\frac{1}{x}
\,,
\\ 
&&\hspace*{-12mm}
G_{W \overline{u}_R D_R}=\,
\frac{e}{s_W}\frac{1}{(1 + r^2)^{1/2}}\frac{m_u}{M_F}\frac{1}{x}
\,,
\\ 
&&\hspace*{-12mm}
G_{W \overline{F}_R F_R}=\,
\frac{e}{s_W}\frac{1}{1 + r^2} \,.
\eeqn
\eeqs
Then, we present the fermion couplings with weak gauge boson $\,Z\,$,\,
which are also relevant for the decay analysis of $\,h^0\to Z\gamma$\,,
\beqs
\label{eq:Vff}
\beqn
\label{eq:Zff-L}
&&
G_{Z\bar{f}_L^{}f_L^{}}^{} =\,
\frac{e}{s_W^{} c_W^{}}\Big(T_{3f}^{}-Q_f^{} s_W^2\Big)\,,
\\
&&
\label{eq:Zff-R}
G_{Z\bar{f}_Rf_R} =\,
\frac{e}{s_W^{} c_W^{}}
\Big(\f{T_{3f}^{}}{x^2}\frac{\,m_f^2\,}{M_F^2} - Q_f^{} s_W^2\Big)\,,
\\[1.5mm]
&&
G_{Z\bar{F}_L^{}F_L^{}} =\,
G_{Z\bar{F}_R^{}F_R^{}} =\,
\frac{e}{s_W^{} c_W^{}}
\Big(\frac{T_{3f}^{}}{\,1+r^2\,}-Q_f^{} s_W^2\Big)\,,
\\[1.5mm]
&&
G_{Z\bar{F}_L^{}f_L^{}} =\, G_{Z\bar{f}_L^{}F_L^{}}
=\, \frac{e}{s_W^{} c_W^{}}\frac{T_{3f}^{}\,r^2\,x\,}{(1\!+\!r^2)^{3/2}\,}
\,,
\\[1.5mm]
&&
G_{Z\bar{F}_R^{}f_R^{}} =\,
G_{Z\bar{f}_R^{}F_R^{}} =\,
\frac{e}{s_W^{} c_W^{}}
\frac{T_{3f}^{}\,r}{\,(1\!+\!r^2)^{1/2}\,x\,}
\frac{m_f^{}}{\,M_F^{}} \,,
\eeqn
\eeqs
where $\,T_{3f}^{}=\pm\hf\,$ are isospins
for the up-type and down-type fermions, respectively.

\vspace*{4mm}
\subsection{\hspace*{-2mm}Unitarity: Roles of Spin-1 Gauge Bosons versus Spin-0 Higgs Bosons}
\label{sec2.5}

In the conventional SM, a single physical Higgs boson plays the key role
to ensure exact $E^2$ cancellation in the amplitudes of
longitudinal weak boson scattering $\,V_L^{}V_L^{}\to V_L^{}V_L^{}$\,
at high energies \cite{SMuni}.
The present 221 model has two spin-0 neutral Higgs bosons $(h^0,\, H^0)$ and three new
spin-1 gauge bosons $(W',\,Z')$,\,
which will cooperate together to ensure the exact $E^2$ cancellation
in the $\,V_L^{}V_L^{}\,$ scattering, so long as the masses of $h^0/H^0$ and $W'/Z'$
are below about 1\,TeV.
Hence, it differs from either the conventional SM, or the 5d Higgsless models\,\cite{HL}
in which the extra longitudinal Kaluza-Klein (KK) gauge bosons alone ensure the
$E^2$ cancellation\,\cite{HLuni}.\footnote{It also differs from the recent
unitarization proposal \cite{He:2011jv} via spontaneous dimensional reduction (SDR),
where the $\,V_L^{}V_L^{}\,$ scattering cross sections
get unitarized through the reduced phase-space under SDR at high energies.}\,
We will show that even for the special parameter choice of
$\,\tan\al = r^3\,$ (which corresponds to vanishing $hVV$ coupling and thus
a gaugephobic Higgs boson $\,h^0$\,), the perturbative unitarity is preserved
by the heavier Higgs boson $H^0$ together with $\,W'/Z'$\,.\,
This conceptually differs from the conventional two-Higgs-doublet models
under the SM gauge group as there is no extra gauge bosons $W'/Z'$
to participate in the $E^2$ cancellation and unitarity; it also differs
from all other extra $\gSU(2)$ models in the
literature\,\cite{LM,unUF,He:1999vp,Jezo:2012rm} with a large splitting between
the two Higgs VEVs which make the $W'/Z'$ masses significantly above $1\,$TeV and thus
irrelevant to the unitarization of $\,V_L^{}V_L^{}\,$ scattering.

We analyze the perturbative unitarity of the 221 model in this subsection.
For the purpose of unitarity analysis, it is sufficient to set the $\gU(1)_2$
gauge coupling $\,g_2^{} =0\,$,\, so there is no photon-related
mixings. We perform a unitary gauge computation of the
$\,V_L^{}V_L^{}\to V_L^{}V_L^{}$\, amplitude in the present model.
Requiring the $E^2$ cancellation, we derive the following sum rule on
the quartic couplings ($G_{VVVV}^{}$) and triple couplings
($G_{VVV}^{}$, $G_{V'VV}^{}$, $G_{hVV}^{}$, $G_{HVV}^{}$),
\beqa
\label{eq:SR}
G_{VVVV}^{} -\f{3}{4}G_{VVV}^2 ~=~
\f{3}{4}\f{M_{V'}^2}{m_V^2}G_{V'VV}^2 +
\f{\,G_{hVV}^2\!+G_{HVV}^2\,}{4m_V^2} \,,
\eeqa
where $V=W,Z$. This sum rule can be readily extended to more general cases with
multiple new gauge bosons $\,V_n^{}(=V',V'',V''',\cdots)$\, and multiple Higgs bosons
$\,H_n^{}(=h,H,H',H'',\cdots)$.\, Thus, the two terms on the right-hand side (RHS)
of (\ref{eq:SR}) will be replaced by the sums over $\,V_n^{}$\, and $\,H_n^{}$\,
in general,
\beqa
\label{eq:SR-2}
G_{VVVV}^{} -\f{3}{4}G_{VVV}^2 ~=~
\sum_n\f{3}{4}\f{M_{V_n}^2}{m_V^2}G_{VVV_n}^2 +
\sum_n\f{\,G_{VVH_n}^2\,}{4m_V^2} \,.
\eeqa

Let us inspect the sum rule (\ref{eq:SR}) further.
The left-hand side (LHS) of (\ref{eq:SR}) is the coefficient of the
non-canceled $E^2$ terms of the $V_L$-amplitude from
the pure Yang-Mills gauge-interactions involving
only the light gauge bosons $W/Z$; while the first term on
the RHS of (\ref{eq:SR}) is the coefficient of
the contributions given by the $V'$-exchanges ($V' =W',Z'$),
and the second term on the RHS comes from the ($h,H$) Higgs-exchanges.
The net coefficient of the $E^2$ term equals the difference between the
two sides of (\ref{eq:SR}).
Hence, the exact $E^2$ cancellation is ensured
by holding the rum rule (\ref{eq:SR}).
It is now evident that {\it the 221 model achieves the
exact $E^2$ cancellation by the joint role of exchanging
both spin-1 new gauge bosons and spin-0 Higgs bosons.}
Hence, this joint unitarization mechanism differs from
either the conventional unitarization of the SM (with Higgs-exchange alone)
or the 5d Higgsless unitarization (with exchanges of $W'/Z'$-like gauge KK-modes alone).

We have explicitly verified the validity of (\ref{eq:SR}) in this model.
Due to the Higgs boson mass-diagonalization and rotation in
(\ref{eq:h1h2-massR})-(\ref{eq:HiggsM-alpha}), we note that the Higgs mass-eigenstate
couplings $\,G_{hVV}^{}$\, and $\,G_{HVV}^{}\,$ in (\ref{eq:SR}) are connected to
their corresponding weak-eigenstate couplings
$\,G_{h_1^{}VV}^{}$\, and $\,G_{h_2^{}VV}^{}\,$,
\beqs
\label{eq:hHVV-h1h2VV}
\beqa
G_{hVV}^{} &=& \ca G_{h_1^{}VV}^{} - \sa G_{h_2^{}VV}^{} \,,
\\
G_{HVV}^{} &=& \sa G_{h_1^{}VV}^{} + \ca G_{h_2^{}VV}^{} \,,
\eeqa
\eeqs
where the weak-eigenstate couplings
$\,G_{h_1^{}VV}^{}$\, and $\,G_{h_2^{}VV}^{}\,$ are independent of the
Higgs mixing angle $\,\al$\,.\, From (\ref{eq:hHVV-h1h2VV}), we readily deduce the
sum of two squared Higgs couplings in the last term of (\ref{eq:SR}),
\beqa
G_{hVV}^2 + G_{HVV}^2 ~=~ G_{h_1^{}VV}^2 + G_{h_2^{}VV}^2\,,
\eeqa
which does not depend on the mixing angle $\,\al$\,.\,
From the $hVV$ and $HVV$ couplings in (\ref{eq:VVh}), we can
explicitly compute the sum,
\beqa
\f{\,G_{hVV}^2\!+G_{HVV}^2\,}{(G_{hVV}^{\rm SM})^2}
~=~ \xi_{hVV}^2 + \xi_{HVV}^2
~=~ \f{1+r^6}{\,(1+r^2)^3\,} + \frac{2r^2(r^4-3r^2+3)}{(1+r^2)^4}x^2+\mO(x^4)\,,
~~~~~~~
\eeqa
which invokes only the VEV ratio $\,r=\fy/\fx\,$ and the gauge coupling ratio
$\,x=g_0^{}/g_1^{}\,$.\,
This means that for $h^0$ in the gaugephobic limit  $\,\xi_{hVV}^{}=0\,$
(under $\,\tan\al =r^3\,$), the role of the Higgs-exchange in the unitarity sum rule
(\ref{eq:SR}) is played by the heavier Higgs boson $H^0$ alone.
Vice versa, for $H^0$ in gaugephobic limit of $\,\xi_{HVV}^{}=0\,$
(under $\,\tan\al =-1/r^3\,$), it is the ligher Higgs boson $h^0$ that
plays the full role of the Higgs-exchange in the unitarity sum rule
(\ref{eq:SR}).

For illustration, we explicitly analyze two longitudinal weak boson scattering processes,
namely, $\,W_L^{}W_L^{}\to \frac{1}{\sqrt{2}}Z_L^{}Z_L^{}\,$
and $\,W_L^{}Z_L^{}\to W_L^{}Z_L^{}\,$.\,
The scattering amplitudes of these two processes include the following types
of contributions,
\beqs
\beqn
\mM(W_L^{}W_L^{}\to \frac{1}{\sqrt{2}}Z_L^{}Z_L^{})
&\,=\,&
\mM_{\rm ct}^{}+\mM_{W/W'}^{(t)}+\mM_{W/W'}^{(u)}+\mM_{h/H}^{(s)} \,,
\label{eq:WWZZamp}
\\[2mm]
\mM(W_L^{}Z_L^{}\to W_L^{}Z_L^{})
&\,=\,&
\mM_{\rm ct}^{}+\mM_{W/W'}^{(s)}+\mM_{W/W'}^{(t)}+\mM_{h/H}^{(t)} \,,
\label{eq:WZWZamp}
\eeqn
\eeqs
where $\,\mM_{\rm ct}^{}\,$ denotes the contact interaction amplitude, and
$\,\mM_{X}^{(j)}$\, stands for the exchange of particle $X$ via $j$-channel
($j=s,t,u$). In the above,
the $W'$-exchange enters the $t/u$-channel ($s/t$-channel) of the first (second) process
and the $(h,H)$-exchanges appear in the $s$-channel ($t$-channel) of the first (second)
process.  The corresponding $s$-wave amplitudes are computed from,
\beqn
a_0^{}(s) ~=~
\frac{1}{32\pi}\int_{-1}^{1}\!d(\cos\theta)\,\mM(s,\,\theta) \,,
\eeqn
where $\,\sqrt{s}\,$ is the scattering energy in the center-of-mass frame and
$\,\theta\,$ is the scattering angle.

\begin{figure}[t]
\begin{center}
\includegraphics[width=7.6cm,height=6.9cm]{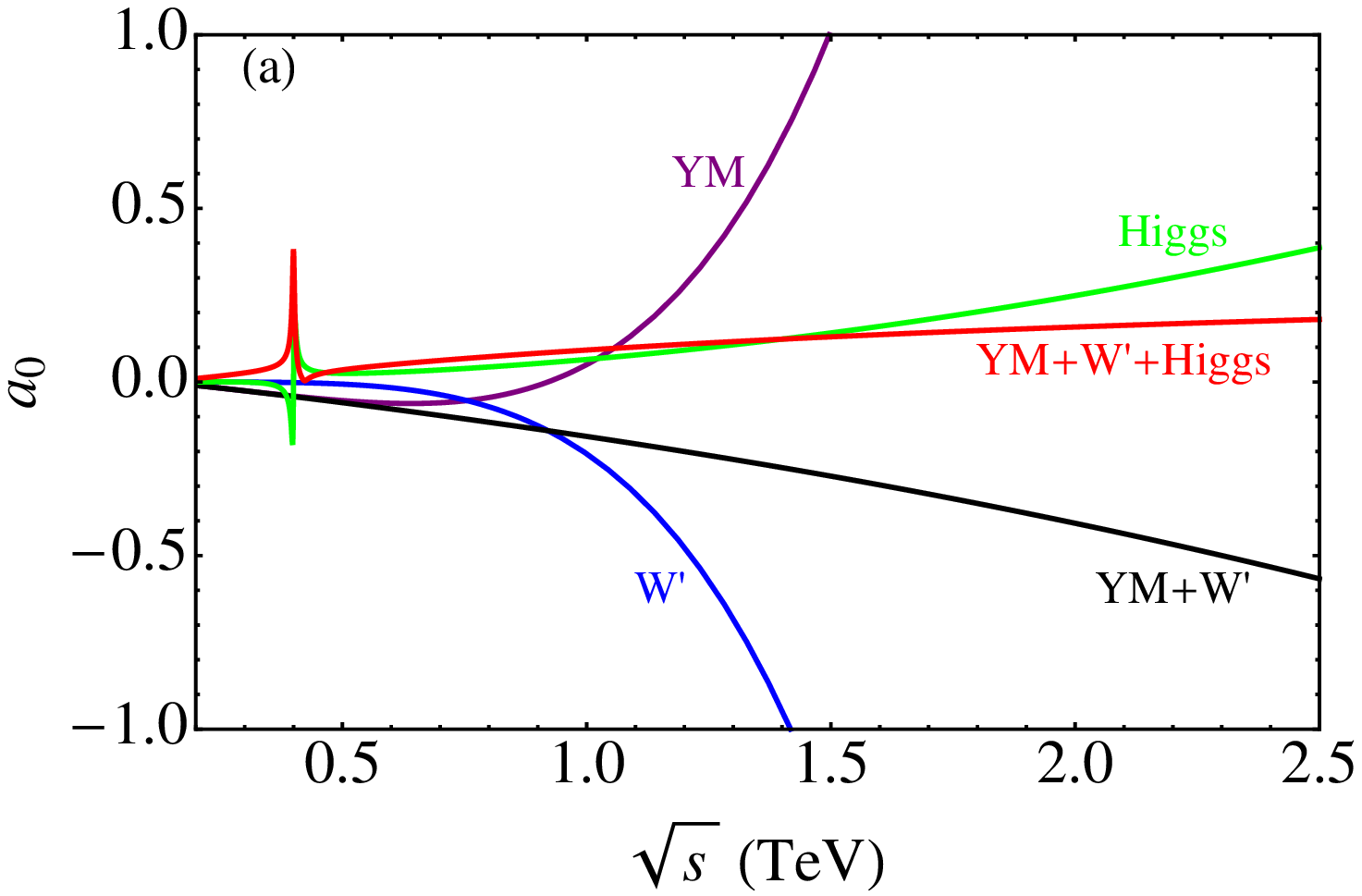}
\includegraphics[width=7.6cm,height=6.9cm]{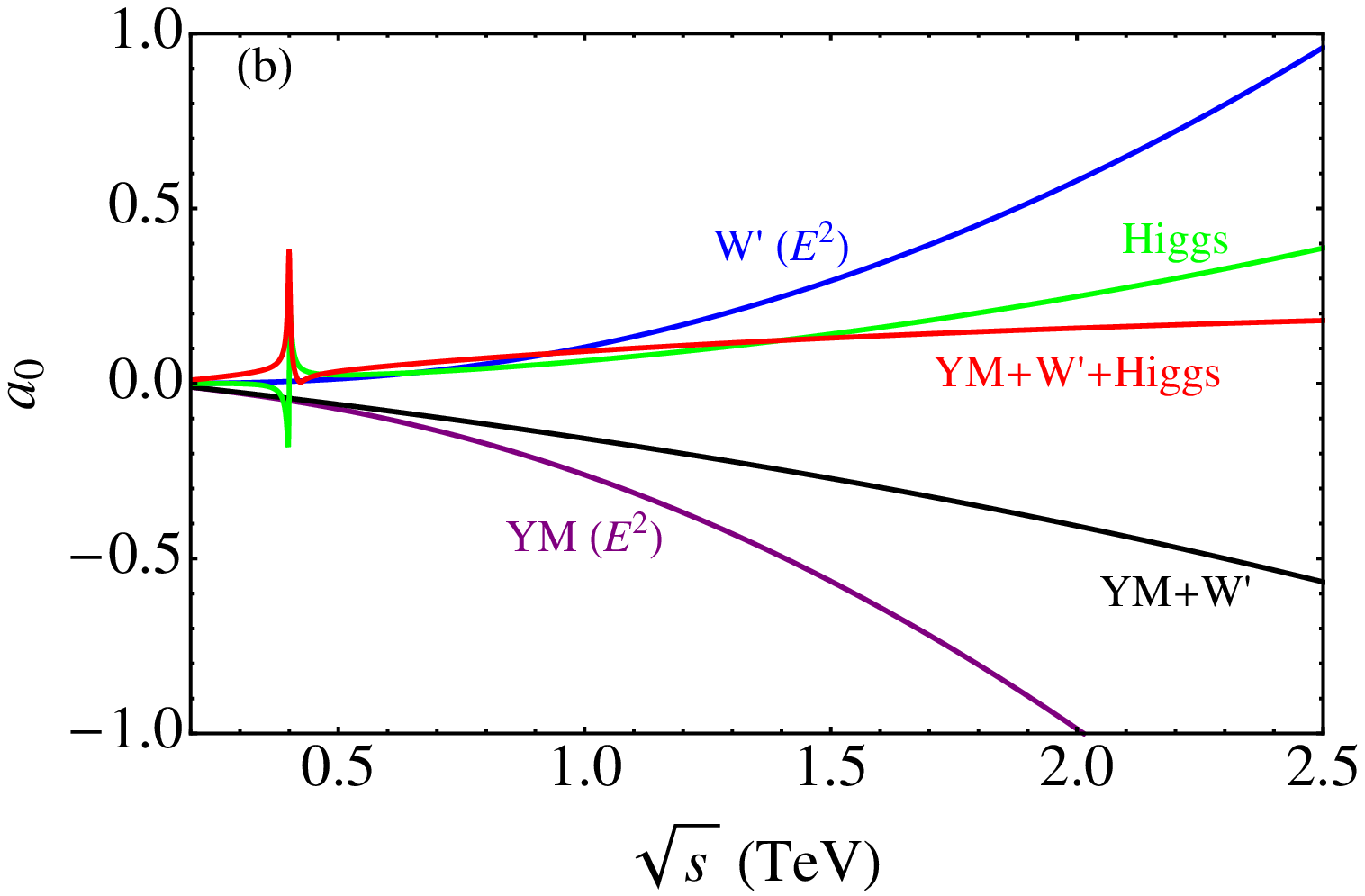}
\vspace*{-2mm}
\caption{The $s$-wave amplitude $\,a_0^{}(W_L^{}W_L^{}\to \frac{1}{\sqrt{2}}Z_L^{}Z_L^{})$\,
as a function of scattering energy $\sqrt{s}$\,.\,
In each plot, we have separately plotted the individual contributions
from the pure Yang-Mills (YM) in purple curves (including contact terms
plus $W$-exchange), the $W'$-exchange (blue curve), the $h/H$-exchanges
(green curve), and the total sum of all three types of contributions
(red curve, marked by YM$+W'+$Higgs).
The sum of all gauge contributions is shown by the black curve (marked by YM$+W'$).
The curves for individual contributions are computed from the corresponding parts
in \,Re$(a_0^{})\,$,\, and only the total sum (red curve)
is given in terms of $\,|a_0^{}|$\,.\, Furthermore, in plot-(b) we only show the separate
Yang-Mills and $W'$-exchange contributions for $E^2$ (and subleading) terms,
with the asymptotic $E^4$ terms subtracted out, which are marked by
YM($E^2$) and $W'(E^2)$, respectively.
Both plots have the parameter inputs,
$\,f_1^{}=f_2^{}$\,,\, $(M_h^{},\,M_H^{})=(125,\,400)$GeV,\,
$M_{W'}^{}=500\,\GeV$, and $\,M_F^{}=2.5\,\TeV$.
}
\label{fig:wwzzamp}
\end{center}
\end{figure}

In Fig.\,\ref{fig:wwzzamp}, we plot the $s$-wave amplitude
$\,a_0^{}(W_L^{}W_L^{}\to \frac{1}{\sqrt{2}}Z_L^{}Z_L^{})$\,
as a function of scattering energy $\sqrt{s}$\,.\,
In plot-(a), we separately display the individual contributions
from the pure Yang-Mills (YM) in purple curve (including contact terms
plus the $W$-exchange contribution), the $W'$-exchange (blue curve), the $h/H$-exchanges
(green curve), and the total sum of all three types of contributions
(red curve, marked by YM$+W'+$Higgs).
We note that both YM and $W'$-exchange contributions contain $E^4$ and
$E^2$ terms in their amplitudes, where the $E^4$ terms exactly cancel
between them at high energies. Thus the summation of the YM and $W'$-exchange
given by the black curve (marked by YM$+W'$) contains only the remaining $E^2$
term in the gauge part, which is negative and will cancel against
the $E^2$ term of Higgs-exchange amplitude (green curve).
Hence, we find that the sum of all three types of contributions
(given by red curve and marked by YM$+W'+$Higgs),
behaves like $\,\mO(E^0)\,$ in high energy,
as required by the unitarity of this renormalizable 221 model.
In this plot, we compute each individual contribution from the corresponding part
in \,Re$(a_0^{})\,$,\, and only the total sum (red curve) is given in terms of $\,|a_0^{}|$\,.\,
For Fig.\,\ref{fig:wwzzamp}(b), the notations are the same as
plot-(a), except that we only show the separate Yang-Mills
and $W'$-exchange contributions for $E^2$ (and subleading) terms,
where we have subtracted out the asymptotic $E^4$ terms.
These two contributions are marked by YM($E^2$) and $W'(E^2)$, respectively.
From this plot, we see that the $W'(E^2)$ contribution is positive
(blue curve) and much larger than the positive contribution of Higgs-exchanges (green curve)
which is at most of $\,\mO(E^2)\,$.\, The sum of Yang-Mills and $W'$-exchange (shown in
black curve) is negative and its magnitude is larger than Higgs-exchange amplitude
(due to the non-$E^2$ subleading terms). Hence, after the $E^2$ cancellation between
YM$+W'$ and Higgs contributions, the final total amplitude of
YM$+W'+$Higgs (red curve) comes from the remaining non-$E^2$ subleading terms in
YM$+W'$ contributions. In both plots, the total $s$-wave amplitude $|a_0^{}|$
is the same, shown as red curves, where the $s$-channel resonance of the heavier
Higgs boson $H^0$ (with a $400$\,GeV mass) explicitly shows up.
In Fig.\,\ref{fig:wwzzamp}(a)-(b), we have set the inputs,
$\,f_1^{}=f_2^{}$\,,\, $(M_h^{},\,M_H^{})=(125,\,400)$ GeV,\,
$M_{W'}^{}=500\,\GeV$, and $\,M_F^{}=2.5\,\TeV$.

\begin{figure}[t]
\begin{center}
\includegraphics[width=7.7cm,height=7cm]{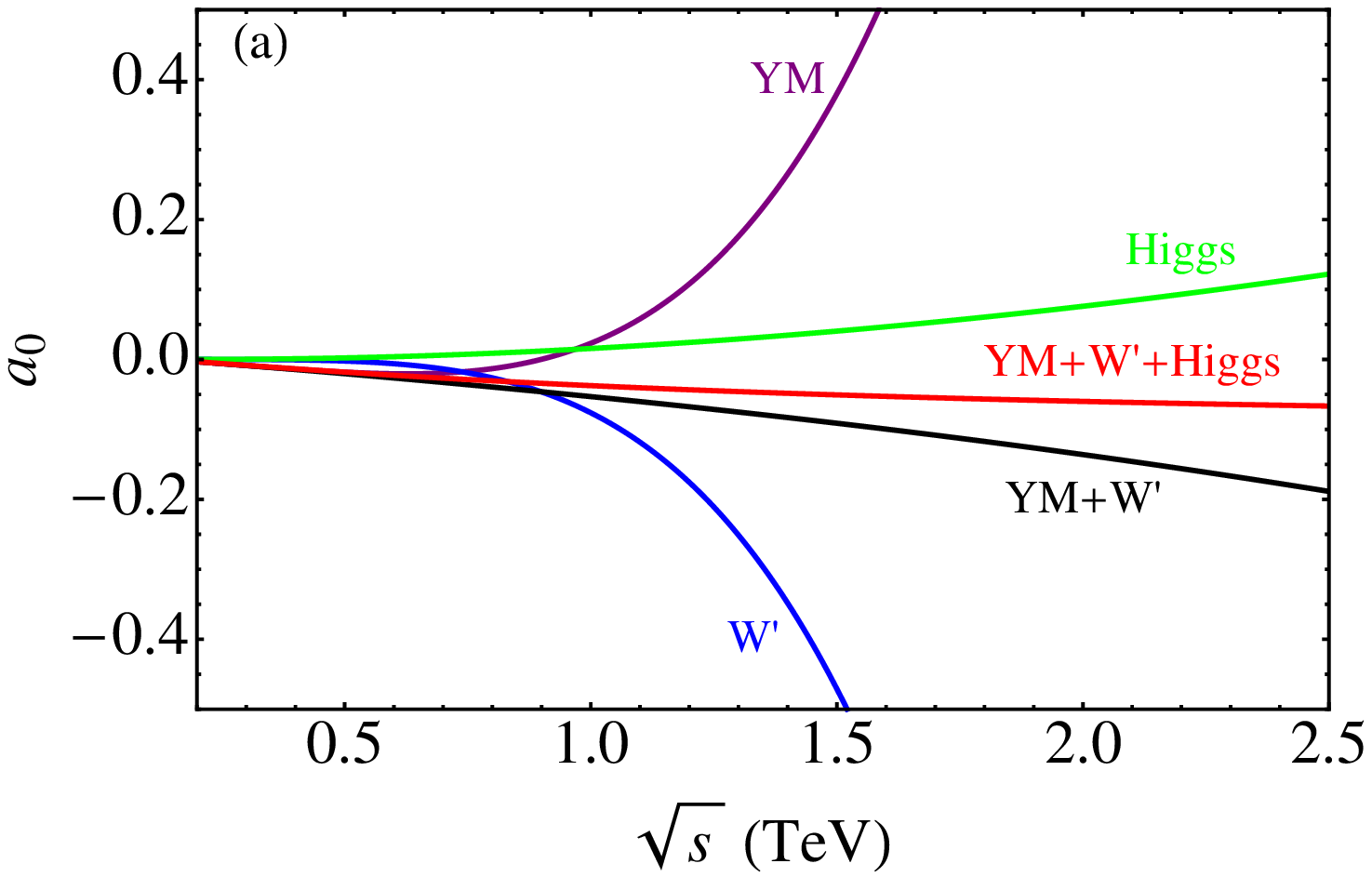}
\includegraphics[width=7.7cm,height=7cm]{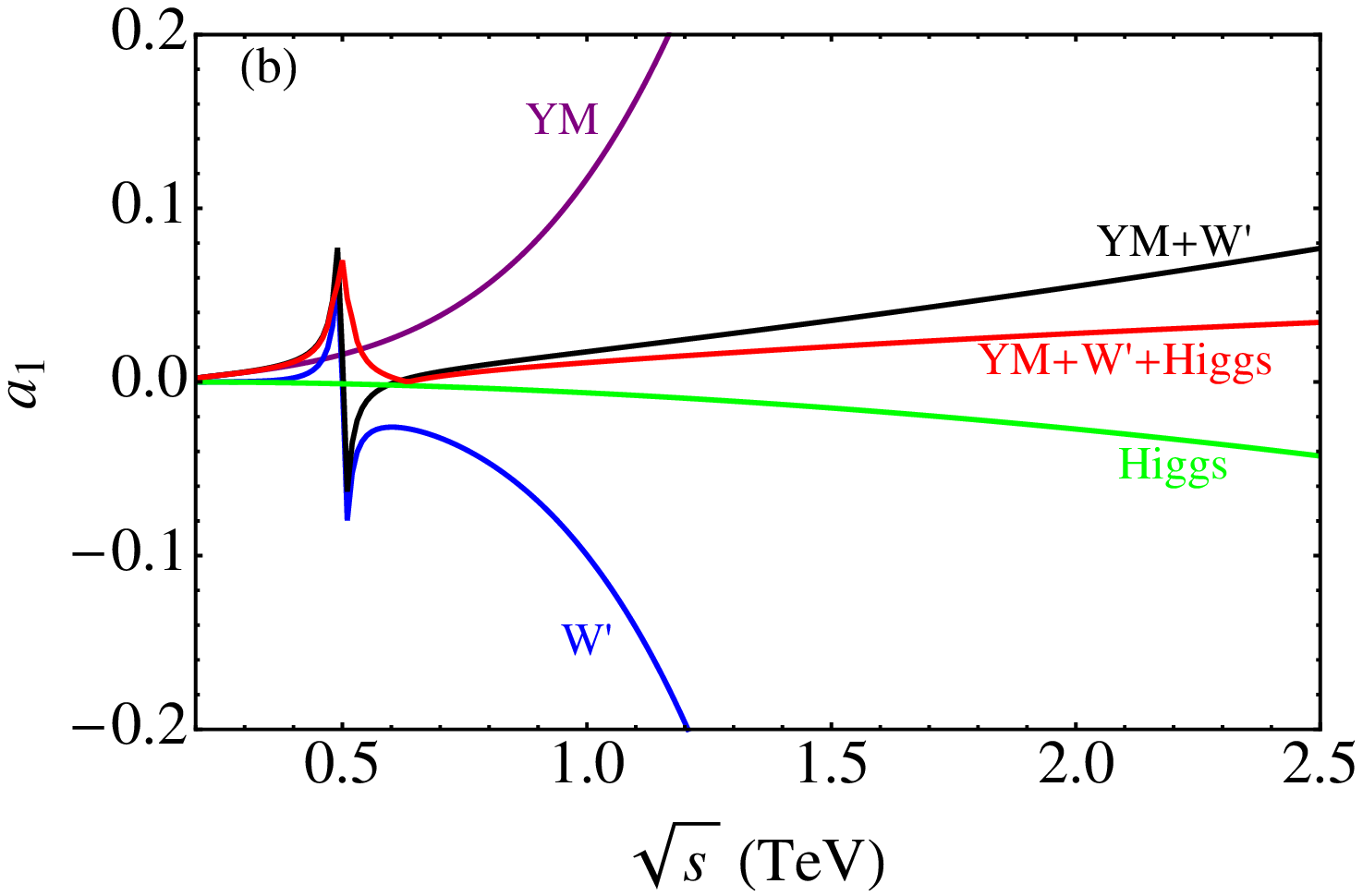}
\vspace*{-2mm}
\caption{The $s$-wave amplitudes [plot-(a)] and the $p$-wave amplitudes [plot-(b)]
for $\,W_L^{}Z_L^{}\to W_L^{}Z_L^{}$\, scattering as functions of the
scattering energy $\sqrt{s}$\,.\,
In each panel, we have separately plotted the individual contributions
from the pure Yang-Mills (YM) in purple curves (including contact terms
plus $W$-exchange), the $W'$-exchange (blue curve), the $h/H$-exchanges
(green curve), and the total sum of all three types of contributions
(red curve, marked by YM$+W'+$Higgs).
The sum of all gauge contributions is shown by the black curve (marked by YM$+W'$).
The curves for individual contributions are computed from the corresponding parts
in \,Re$(a_0^{})\,$.\, The total sum (red curve) in plot-(a) is given in terms of
\,Re$(a_0^{})\,$, but in plot-(b) is computed for $\,|a_0^{}|$\,.\,
The parameter inputs are,
$\,f_1^{}=f_2^{}$\,,\, $(M_h^{},\,M_H^{})=(125,\,400)$ GeV,\,
$M_{W'}^{}=500\,\GeV$, and $\,M_F^{}=2.5\,\TeV$.
}
\label{fig:wzwzamp}
\end{center}
\end{figure}

In Fig.\,\ref{fig:wzwzamp}, we further analyze the longitudinal scattering
process of $\,W_L^{}Z_L^{}\to W_L^{}Z_L^{}\,$,\, where both $s$-wave and $p$-wave amplitudes
are shown as functions of the scattering energy $\sqrt{s}$ in plot-(a) and plot-(b),\,
respectively.  In both plots, we show the individual contributions from the
Yang-Mills interactions (including contact terms plus $W$-exchange, marked by YM in
purple curve), the $W'$-exchange (marked by $W'$ in blue curve) and the Higgs-exchanges
(marked by Higgs in green curve). The sum of YM and $W$-exchange contributions is
depicted by the black curve (marked by YM$+W'$), where the $E^4$ terms are exactly
canceled out between them, and only the $E^2$ (and subleading) terms are contained
in the black curve. Then, we see that the two $E^2$ contributions in the negative
YM$+W'$ amplitude (black curve) and positive Higgs amplitude (green curve) cancel
each other exactly at high energies, while the remaining terms have the asymptotic
behavior of $\,\mO(E^0)\,$,\, as shown in the red curve (marked by YM$+W'+$Higgs),
which is negative since it is dominated by the YM$+W'$ contribution.
Here all curves (including the red curve) are computed
from the corresponding parts of the real $s$-wave amplitude
\,Re$(a_0^{})\,$.\, We choose the same parameter inputs of
VEVs, Higgs masses, $W'$ mass and heavy fermion masses as in Fig.\,\ref{fig:wwzzamp}.
For Fig.\,\ref{fig:wzwzamp}(b), we compute the $p$-wave amplitude,
\beqn
a_1^{}(s) ~=~
\frac{1}{32\pi}\int_{-1}^{1}\!d(\cos\theta)\,\cos\theta\,\mM(s,\,\theta) \,,
\eeqn
for the same process $\,W_L^{}Z_L^{}\to W_L^{}Z_L^{}\,$.\,
Here all the notations and inputs are the same as those in Fig.\,\ref{fig:wwzzamp}(b),
expect that we are computing the $p$-wave for $\,W_L^{}Z_L^{}\to W_L^{}Z_L^{}\,$ scattering.
Different from Fig.\,\ref{fig:wzwzamp}(a) and Fig.\,\ref{fig:wwzzamp} for
the $s$-wave amplitudes, we see from Fig.\,\ref{fig:wzwzamp}(b) that
the Higgs-exchange amplitude becomes negative and the YM$+W'$ contribution is postive;
and they cancel each other for the asymptotic $E^2$ terms. In consequence, the total sum
of the $p$-wave amplitude is shown in the red curve (marked by YM$+W'+$Higgs), which
has good high energy behavior of $\,\mO(E^0)\,$.\,
In the plot-(b), we find that the $s$-channel spin-1 resonance of $W'$ gauge boson
shows up in the $p$-wave amplitude at 500\,GeV, as expected.

\vspace*{2mm}

Next, we analyze the unitarity bound on the heavy Higgs mass.
In the above we have demonstrated how the $E^4$ and $E^2$ terms are separately
canceled out in the longitudinal scattering amplitudes at high energies,
and the crucial role played by the $W'/Z'$-exchanges and Higgs-exchanges
together in the $E^2$ cancellation to ensure the unitarity of the model.
The remaining terms in the scattering amplitudes are of $\,\mO(E^0)\,$
and are dominated by the Higgs self-couplings related to the Higgs boson masses.
Given the light Higgs boson mass of $\,M_h=125\,$GeV as observed
by the LHC \cite{Atlas2012-7,CMS2012-7}, we can further derive
an upper bound on the mass of heavier Higgs state $\,H^0\,$.\,
For this purpose, it is extremely convenient to use the equivalence theorem
and analyze the corresponding Goldstone boson scattering amplitudes at high energies.
Here the external in/out states (either gauge bosons or Goldstone bosons)
can be treated as massless, and only the quartic contact interactions
among the Goldstone and Higgs bosons can contribute to the leading
scattering amplitudes at $\,\mO(E^0)\,$.
This means to take the limit $\,g_0^{},g_2^{}\simeq 0\,$.\,
We also note that the other gauge coupling $g_1^{}$ of $\gSU(2)_1$ may be sizable
since $(W',Z')$ gauge bosons are relatively heavy.
Thus, the scattering amplitudes via the $(W',Z')$-exchanges
may be comparable to the contact interactions.
In the following, we shall estimate the unitarity bounds
on the mass of the heavier Higgs state
$\,H^0\,$ both without and with the $(W',Z')$ contributions.

\begin{figure}[htbp]
\begin{center}
\includegraphics[width=7.5cm,height=6.5cm]{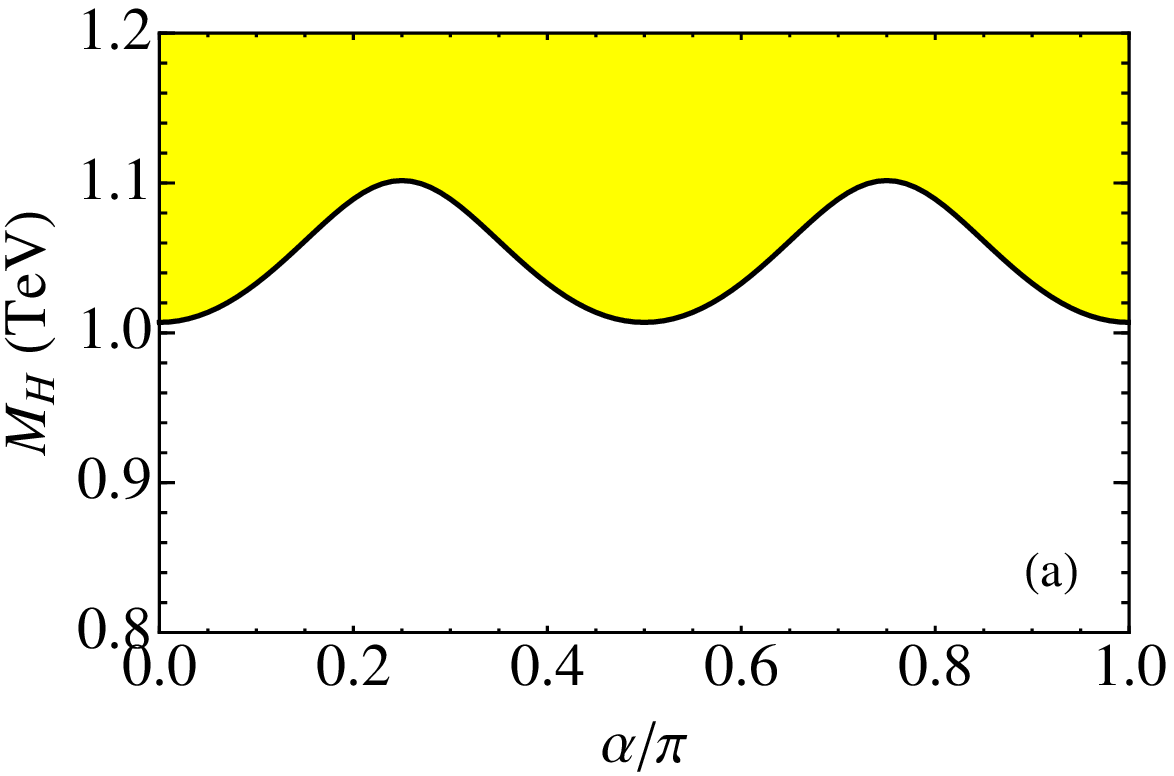}
\includegraphics[width=7.5cm,height=6.5cm]{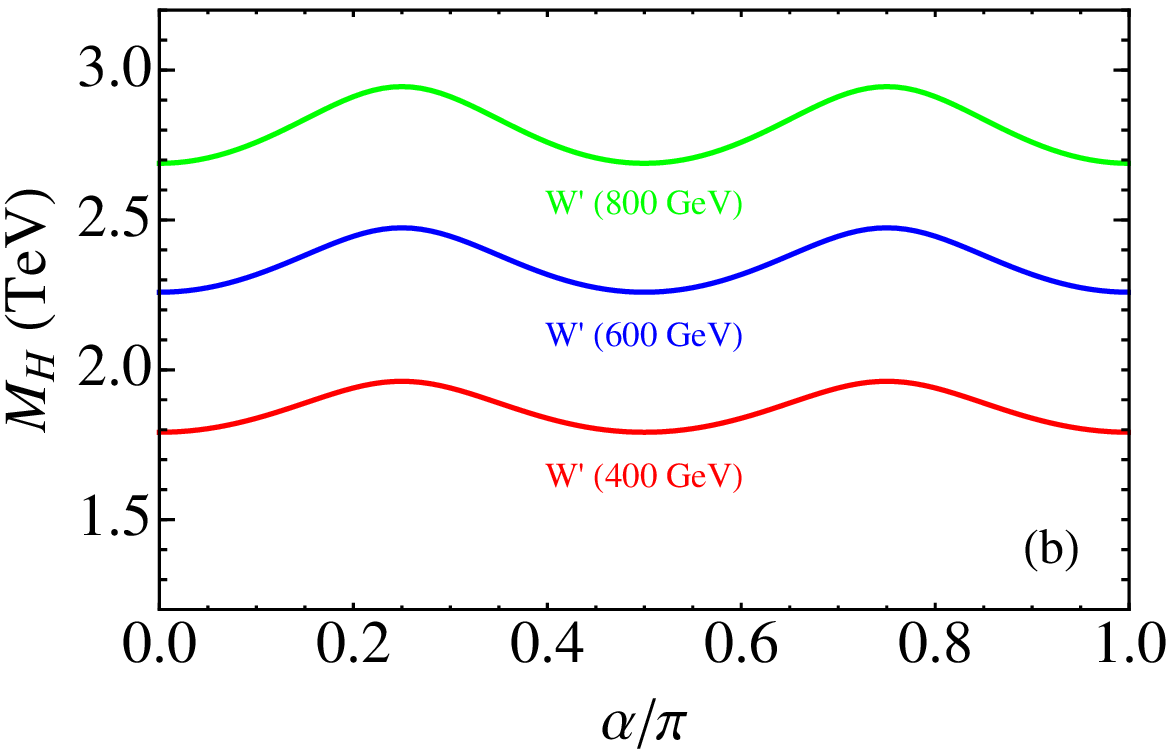}
\vspace*{-2mm}
\caption{Unitarity bound on the heavier Higgs boson mass $\,M_H^{}\,$
as a function of the Higgs mixing angle $\,\alpha$\,.\,
We present the uniarity bounds by taking the contact-term contributions
alone in plot-(a), and by including both contact-term and $W'/Z'$-exchanges
in plot-(b) for three inputs of $M_{W'}$.\,
The yellow region in plot-(a) is excluded, while in plot-(b) the region
above each curve is excluded for the given input of $M_{W'}$.
In both plots, we set the lighter Higgs boson mass $\,M_h^{}=125\,\GeV$\,
and the Higgs VEV ratio $\,r=1\,$. }
\label{fig:MH}
\end{center}
\end{figure}

We note that the Higgs potential (\ref{eq:V})
enjoys a global $\,O(4)\otimes O(4)\,$ symmetry,
under which two Higgs doublets $\Phi_1^{}$ and $\Phi_2$
transform as ({\bf 4}, {\bf 1}) and ({\bf 1}, {\bf 4}), respectively.
Hence we perform the coupled channel analysis by defining
the normalized singlet two-body states under $\,O(4)\otimes O(4)\,$,
\beqn
\label{eq:O4singlets}
|X_j^{}\rangle ~=~
\frac{1}{\sqrt{8}\,}\Big( |h_j^{} h_j^{}\rangle + |\vec{\pi}_j^2 \rangle \Big)\,,
\eeqn
where $\,j=1,2\,$.\,
Thus, we compute the coupled-channel scattering amplitudes of
$\,|X_j^{}\rangle\to |X_{j'}^{}\rangle\,$ in terms of the $2\times 2$ matrix,
\beqn
\label{eq:amp-matrix}
{\cal T} ~=~
\left(\begin{array}{cc}
\langle X_1|T|X_1\rangle ~&~ \langle X_1|T|X_2\rangle
\\[2mm]
\langle X_2|T|X_1\rangle ~&~ \langle X_2|T|X_2\rangle
\end{array}  \right) ,
\eeqn
where we derive the $\,\mO(E^0)\,$ leading contributions to its elements as follows,
\beqa\label{eq:Tmatrix_elem1}
\langle X_j^{}|T|X_j^{}\rangle ~=~{-3\lambda_j^{}}\,,
~~~~~~
\langle X_1^{}|T|X_2^{}\rangle ~=~ \langle X_2^{}|T|X_1^{}\rangle
~=~ {-2\lambda_{12}^{}} \,.
\eeqa
Diagonalizing the matrix (\ref{eq:amp-matrix}),
we derive the maximal eigenvalue for the corresponding $s$-wave amplitude
from the scalar contact interactions,
\beqn
\label{eq:a-max}
a_0^{\max}[\textrm{contact}] ~&=&~
{-}\frac{1}{32\pi}
\left[ 3(\lambda_1+\lambda_2)+\sqrt{9(\lambda_1-\lambda_2)^2+16\lambda_{12}^2}
\right] ,
\eeqn
which is a function of the three quartic Higgs self-couplings
$\,(\lambda_1\,,\lambda_2\,,\lambda_{12})\,$,\,
which is expected to give the best bound on the Higgs mass
under the $s$-wave unitarity condition,
$\,|{\rm Re}\, a_0^{\max}| < \frac{1}{2}$\,.\,
As we noted in Sec.\,\ref{sec2.2}, fixing the two VEVs, the three Higgs self-couplings $\,(\lambda_1^{}\,,\lambda_2^{}\,,\lambda_{12}^{})$ can be equivalently expressed
in terms of the two Higgs masses and one mixing angle, \,$(M_h^{},\,M_H^{},\,\alpha)$.\,
Hence, inputting the observed lighter Higgs boson mass $\,M_h=125\,$GeV
and Higgs VEV ratio $\,r=1$,\, we can derive the upper bound on the
heavier Higgs boson mass $\,M_H^{}\,$ as a function of the Higgs mixing angle $\,\al\,$.\,
We present our findings in Fig.\,\ref{fig:MH}(a). We see that to respect
the unitarity over the full range of mixing angle $\,\alpha$\,
imposes the following bound on the mass of the Higgs boson $H^0$,
\beqa
M_H ~\leqq~ 1.0\,\TeV\,.
\eeqa
Furthermore, Fig.\,\ref{fig:MH}(a) shows that this unitarity limit
could become weaker for some ranges of $\,\al\,$,
but we find that the bound is only slightly weakened to $\,M_H\leqq 1.10\,$TeV.

Including the $W'/Z'$-exchanges for the coupled channel amplitudes
under the limit of $\,g_0^{},g_2^{}\simeq 0$,\, we extend
the elements of the $S$-matrix (\ref{eq:amp-matrix}) as follows,
\beqs
\beqn
\label{eq:Tmatrix_elems}
&& \langle X_j^{}|T|X_j^{}\rangle ~=~{-3\lambda_j^{}+(\delta_a+\delta_b)}\,,
\\[2mm]
&& \langle X_1^{}|T|X_2^{}\rangle ~=~ \langle X_2^{}|T|X_1^{}\rangle
~=~ {-2\lambda_{12}^{}+\delta_a} \,,
\\[2mm]
&& \delta_a^{} ~=~\frac{g_1^2}{8}\frac{s\cos\theta}{s-M_{W'}^2}\,,
\\[2mm]
&&
\delta_b^{} ~=~\frac{\,g_1^2\,}{16}
\left[\frac{7(3+\cos\theta)}{(1-\cos\theta)+2M_{W'}^2/s}
      +\frac{5(3-\cos\theta)}{(1+\cos\theta)+2M_{W'}^2/s}
       \right]\!. ~~~~~~~~~
\eeqn
\eeqs
Accordingly, we find the modified maximal eigenvalue
for the $s$-wave amplitude,
\beqs
\beqn
\label{eq:a-max2}
\widehat{a}_0^{\max} &~=~& a_0^{\max}[\textrm{contact}]\,+\,\frac{1}{16\pi}\Delta\,,
\\[2mm]
\Delta &~=~&
\frac{\,3g_1^2\,}{4}\left[\!\(2+\frac{M_{W'}^2}{s}\)
\log\!\(1+\frac{s}{M_{W'}^2}\)-1 \right]\!,
\eeqn
\eeqs
where $a_0^{\max}[\textrm{contact}]$ is the contribution of contact terms as in (\ref{eq:a-max}),
and the $\Delta$ term arises from the $W'/Z'$-exchanges.
We note that the $W'/Z'$-exchanges partially cancel the contact term contributions.
Imposing the $s$-wave unitarity condition of
$\,|{\rm Re}\,\widehat{a}_0^{\max}|<\frac{1}{2}$
thus further relaxes the $M_H$ upper bound as compared to the contact-terms alone
in Fig.\,\ref{fig:MH}(a). This is explicitly displayed in Fig.\,\ref{fig:MH}(b) with
the center-of-mass energy being $\,\sqrt{s}=5\,\TeV$ for three sample inputs of
$\,M_{W'}=(400,\,600,\,800)$GeV and $\,r=1$\,.\,
Accordingly, we infer the mass bounds on the Higgs boson $H^0$ over the full range
of mixing angle $\alpha$\,,
\beqa
M_{W'}=400\,\GeV\!:~~~~M_H ~\leqq~ 1.80\,\TeV,
\nonumber\\[2mm]
M_{W'}=600\,\GeV\!:~~~~M_H ~\leqq~ 2.26\,\TeV,
\\[2mm]
M_{W'}=800\,\GeV\!:~~~~M_H ~\leqq~ 2.69\,\TeV.
\nn
\eeqa


\vspace*{3mm}
\section{\hspace*{-2mm}LHC Signatures of the Lighter Higgs Boson $\mathbf{h}^{\mathbf{0}}$}
\label{sec3}

In this section, we study the production and decays of the light Higgs boson $h^0$
at the LHC, based upon its non-SM couplings presented in the previous section.

\vspace*{2mm}
\subsection{\hspace*{-2mm}Decay Branching Fractions of h$^0$}
\label{sec:br}

There are three types of the decay processes for the light Higgs boson $h^0$
in the current 221 model.  This includes:
(i) decays into the SM fermions;
(ii) decays into $WW^*$ or $ZZ^*$; and
(iii) the loop-induced decays into $\,\gamma\gamma$\,,\, $\gamma Z$\,,\, or $\,gg$\,.\,
Due to the non-SM couplings of $h^0$ in Sec.\,\ref{sec2.4},
the partial decay widths of $h^0$ differ from the SM values\,\cite{Djouadi:1997yw}
and are explicitly given in the Appendix\,\ref{appA}.
In particular, the $W'$-loop induced contributions can enhance the
$\ga\ga$ and $Z\ga$ partial widths over the proper parameter space of our
model.\footnote{Very recently, the $W'$-loop contribution to the diphoton channel was also
studied for different models\,\cite{Ian}; and possible enhancements of the diphoton rate
for some composite Higgs models and analysis of their $\pi\pi$ scattering channels
were discussed in \cite{compositeH-LHC}.}\,

\begin{figure}[t]
\begin{center}
\includegraphics[width=7.5cm,height=6.5cm]{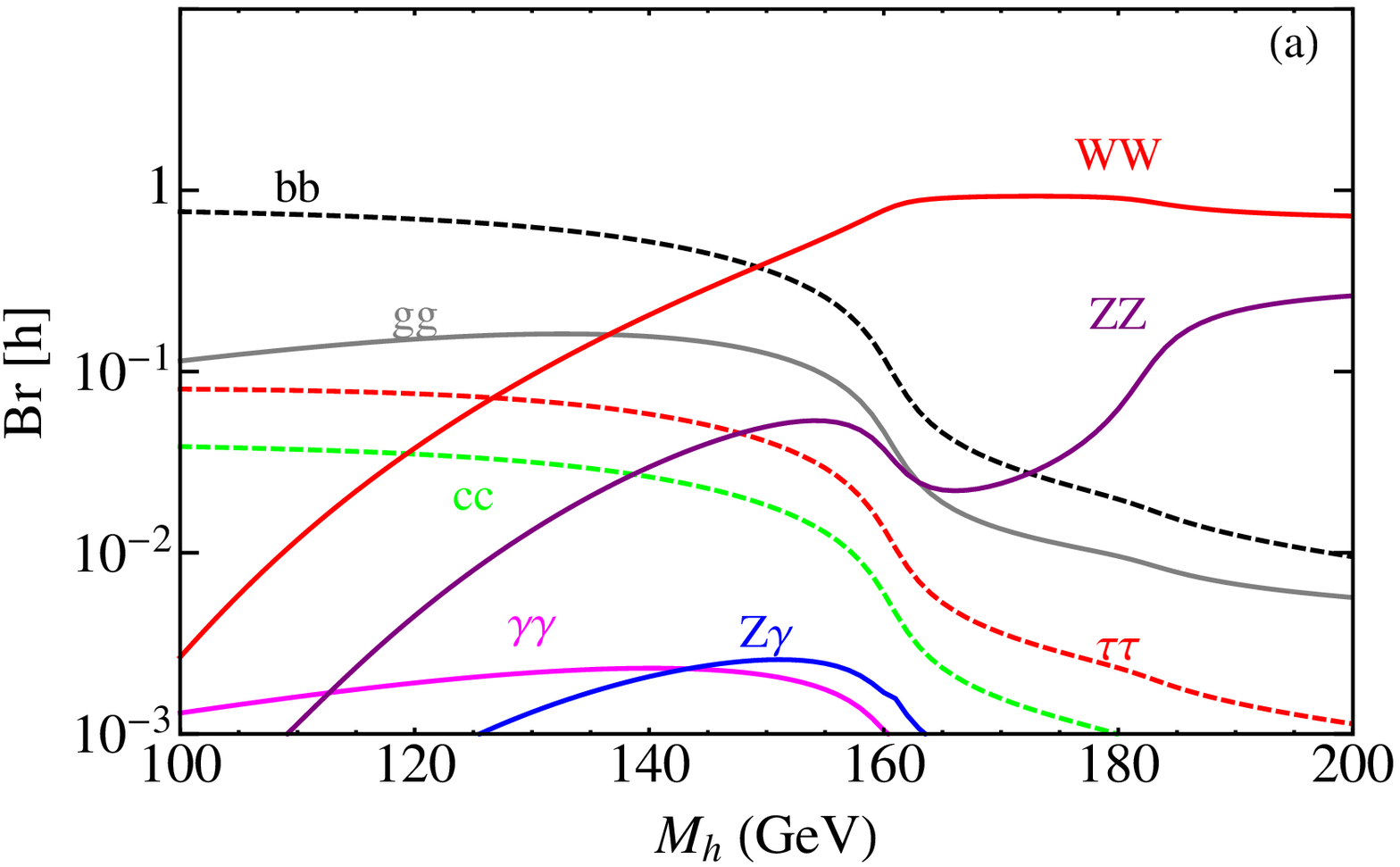}
\includegraphics[width=7.5cm,height=6.5cm]{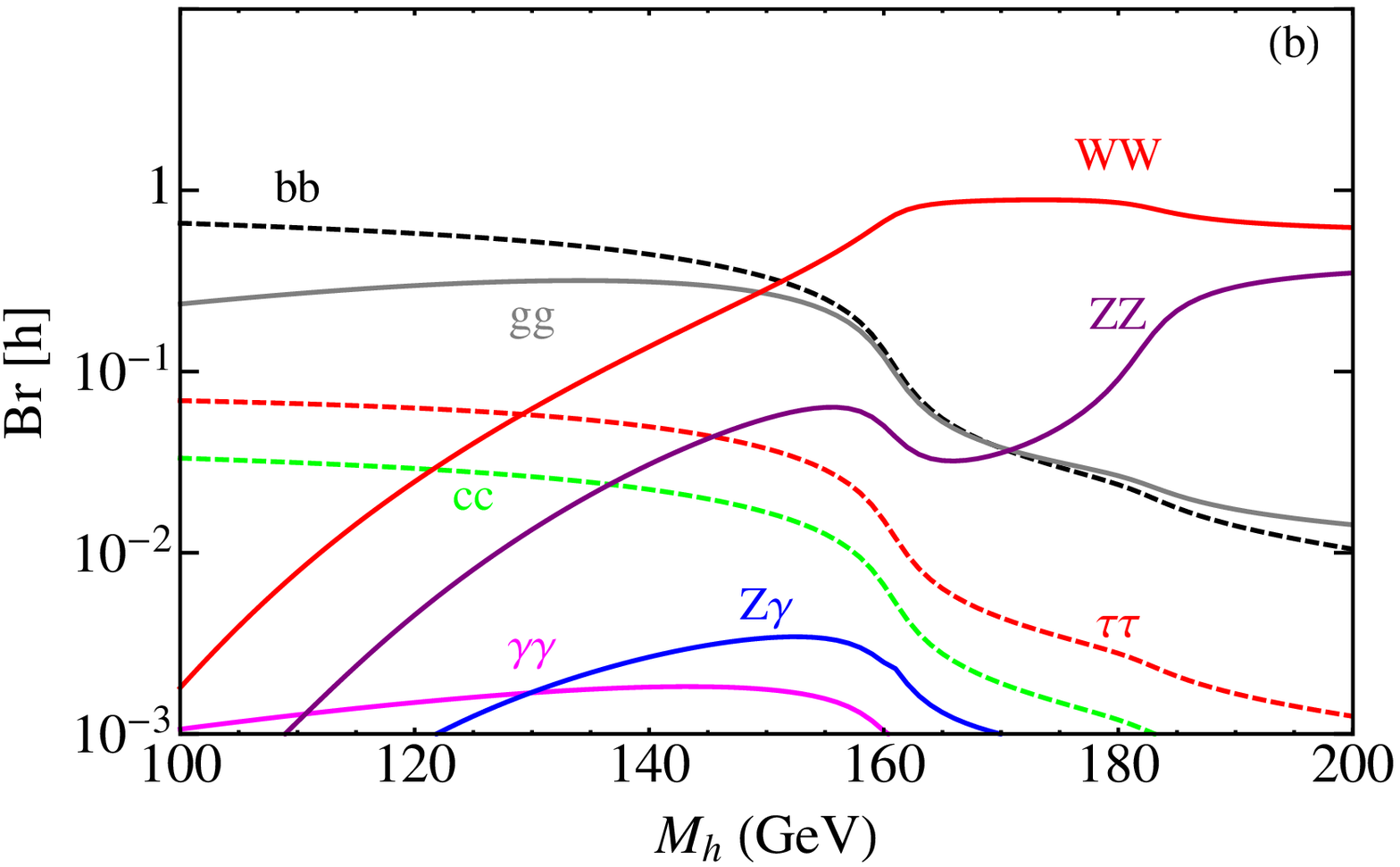}
\vspace*{-2mm}
\caption{Decay branching fractions of the light Higgs boson $h^0$ over the mass-range
$\,100<\,M_h<\,200$\,GeV. We have input $\,f_1^{} = f_2^{}$\,
and $\,(M_{W'},\,M_F^{})=(0.6,\,2.5)$\,TeV.\,
The Higgs mixing angle is taken to be $\,\alpha=0.8\,\pi$\,
in plot-(a) and $\,\alpha=0.2\pi$\, in plot-(b), respectively.}
\label{fig:Br3site}
\end{center}
\end{figure}
\begin{figure}
\begin{center}
\vspace*{-2mm}
\includegraphics[width=7.8cm,height=6.4cm]{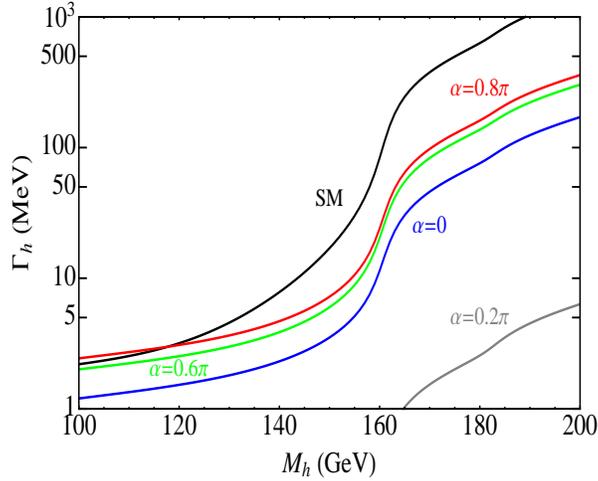}
\vspace*{-2mm}
\caption{Total decay width of the light CP-even Higgs boson $h^0$
with $\,\alpha=0.8\pi$ (red),\, $\alpha=0.6\pi$ (green),
$\,\alpha =0\,$ (blue) and $\,\alpha=0.2\pi$ (grey), over the mass-range
$\,100-200$\,GeV. For comparison, the total width of the SM Higgs boson
is depicted in the same mass-range (black curve).
We set the other inputs as, $\,f_1^{} = f_2^{}$\, and
$\,(M_{W'}^{},\,M_F^{})=(0.6,\,2.5)$\,TeV.}
\vspace*{-3mm}
\label{fig:hwid}
\end{center}
\end{figure}

From these, we compute the decay branching fractions of $h^0$ into the SM final states.
In Fig.\,\ref{fig:Br3site}, we plot $h^0$ branching ratios over the mass range
of $\,100\,\GeV < M_h < 200\,\GeV$.\,
For illustration, we take $\,f_1^{} = f_2^{}$,\,
$\,M_{W'}=600$\,\GeV, and $\,M_F = 2.5\,$TeV, with different values of $\,\alpha=0.2\,\pi$\, and $\,\alpha=0.8\,\pi$,\, respectively.
For comparison, we show the total decay widths for our $h^0$ Higgs boson and
the SM Higgs boson in Fig.\,\ref{fig:hwid}.  For the lighter $h^0$,\, we plot the total
width of $h^0$ for four values of the mixing parameter $\,\al =0\,,0.2\,\pi,\,0.6\,\pi,\,0.8\,\pi\,$,\,
respectively. We see that for $\,\alpha=0.8\,\pi$\, and $\,\alpha=0.6\,\pi$,\,
the total widths vary very little, while they both become visibly smaller than the SM values
for the Higgs mass above 120\,GeV.
Besides, Fig.\,\ref{fig:hwid} shows that our $h^0$ width with $\,\alpha =0.2\,\pi$\,
is substantially lower than the SM.
We have further verified that the main features of Fig.\,\ref{fig:hwid}
still remain by varying the inputs of the VEV ratio $\,f_2^{}/f_1^{}$\,
and heavy masses $(M_{W'}^{},\,M_F^{})$\,.

\vspace*{3mm}
\subsection{\hspace*{-2mm}Signatures of \,h$^0$ via Gluon Fusion}
\label{sec3.2}

\begin{figure}[b]
\begin{center}
\hspace*{-4mm}
\includegraphics[width=7.9cm,height=7.5cm]{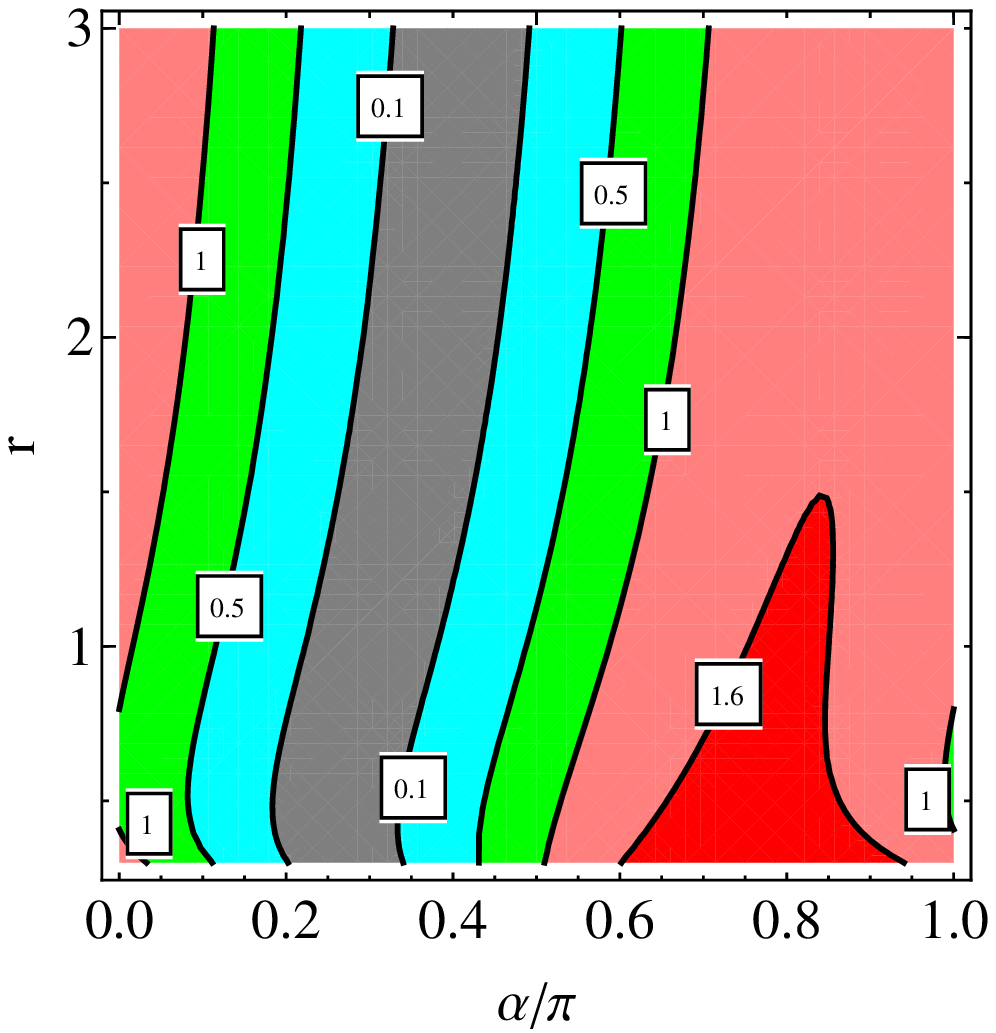}
\hspace*{-3mm}
\includegraphics[width=7.9cm,height=7.5cm]{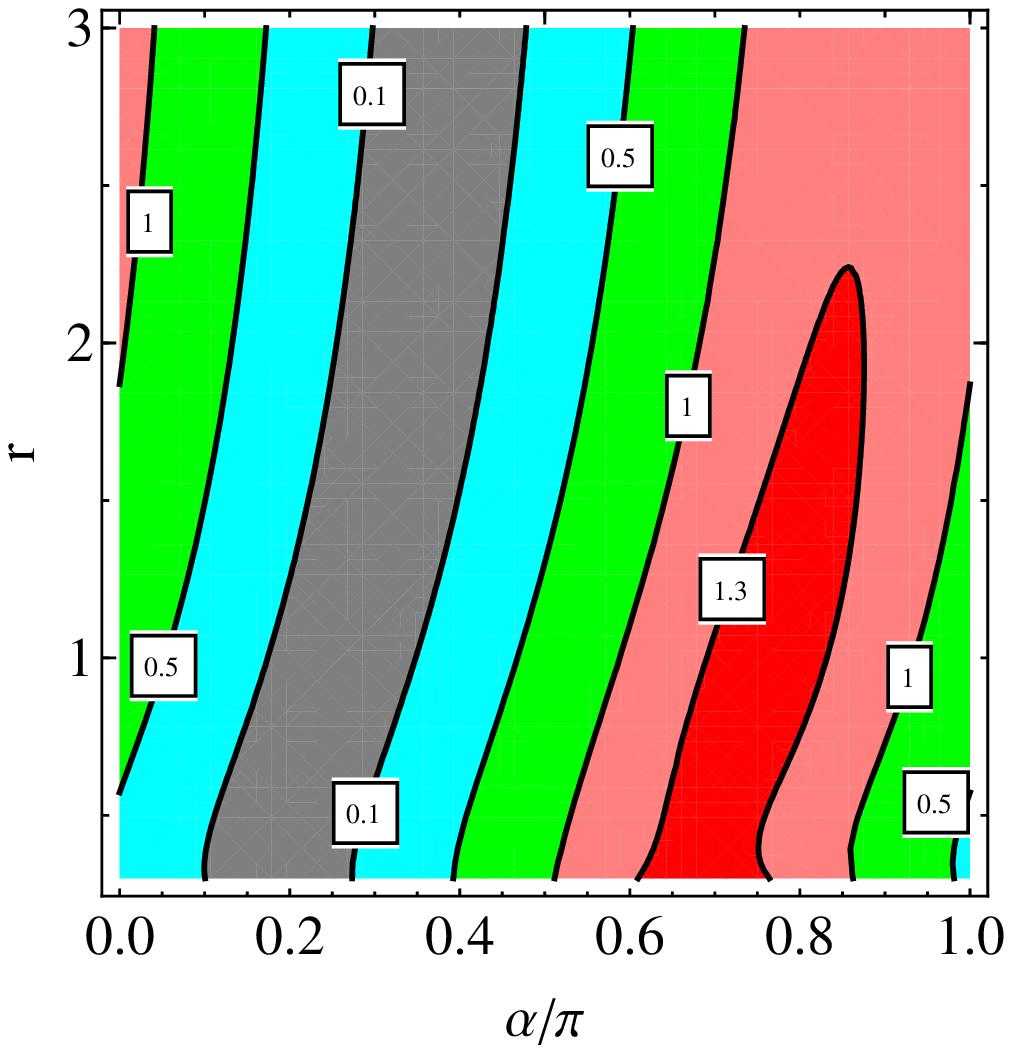}
\vspace*{-4mm}
\caption{Contour plots of the gluon-fusion ratio $\,\mR_{ggh}^{}$ in the parameter space
of $\,r\,$ versus $\,\alpha\,$,\, with light Higgs mass $\,M_h^{}=125\,\GeV$.
Different ranges of $\,\mR_{ggh}^{}$ are represented by the colored bands:
$\mR_{ggh}<0.1$ (gray), $0.1<\mR_{ggh}<0.5$ (light-blue),
$0.5<\mR_{ggh}<1$ (green), $1<\mR_{ggh}<1.6\, (1.3)$ (pink),
and $\mR_{ggh}>1.6\, (1.3)$ (red). We have input  $\,M_{W'}=400\,\GeV$
for the left panel and $M_{W'}=600\,\GeV$ for the right panel.
The heavy fermion mass is set as $\,M_F=2.5\,\TeV$ for both panels.
The maximal enhancement factors are $\,\mR_{ggh}\simeq 1.7$\, (left panel)
and $\,\mR_{ggh}\simeq 1.3$\, (right panel), respectively.}
\label{fig:ggh_sHf}
\end{center}
\end{figure}

The gluon fusion process gives the dominant Higgs production mechanism at the LHC.
It is induced by colored particles at the loop-level. For the SM case,
only the top-quark loop contribution dominates. But for the present model,
we have six extra vector-like heavy partners of the SM quarks with masses
$\,>1.8\,\TeV$, which will contribute to the gluon fusions as well.
The ratio of the gluon fusion cross sections between the 221 model and
the SM is readily given by the ratio between the partial decay widths
of $\,h\to gg$\, in Eq.\,(\ref{eq:hggwid}),
\beqn
\label{eq:xsec_ratio}
\mR_{ggh}^{} \,\equiv\,
\frac{\sigma[gg \to h]}{~\,\sigma[gg \to h]_{\rm SM}^{}~}
\,=\, \frac{\Gamma[h\to gg]}{~\,\Gamma[h\to gg]_{\rm SM}^{}~}\,.
\eeqn
We analyze this ratio over the parameter space of $\,(\alpha,\, r)$,\,
where $\,\al$\, is the Higgs mixing angle and the VEV ratio
$\,r = f_2^{} /f_1^{}\,$ is shown for the range of
$\,r=(0.5,\,2)=O(1)\,$.\,

\begin{figure}[t]
\begin{center}
\includegraphics[width=7.5cm,height=7cm]{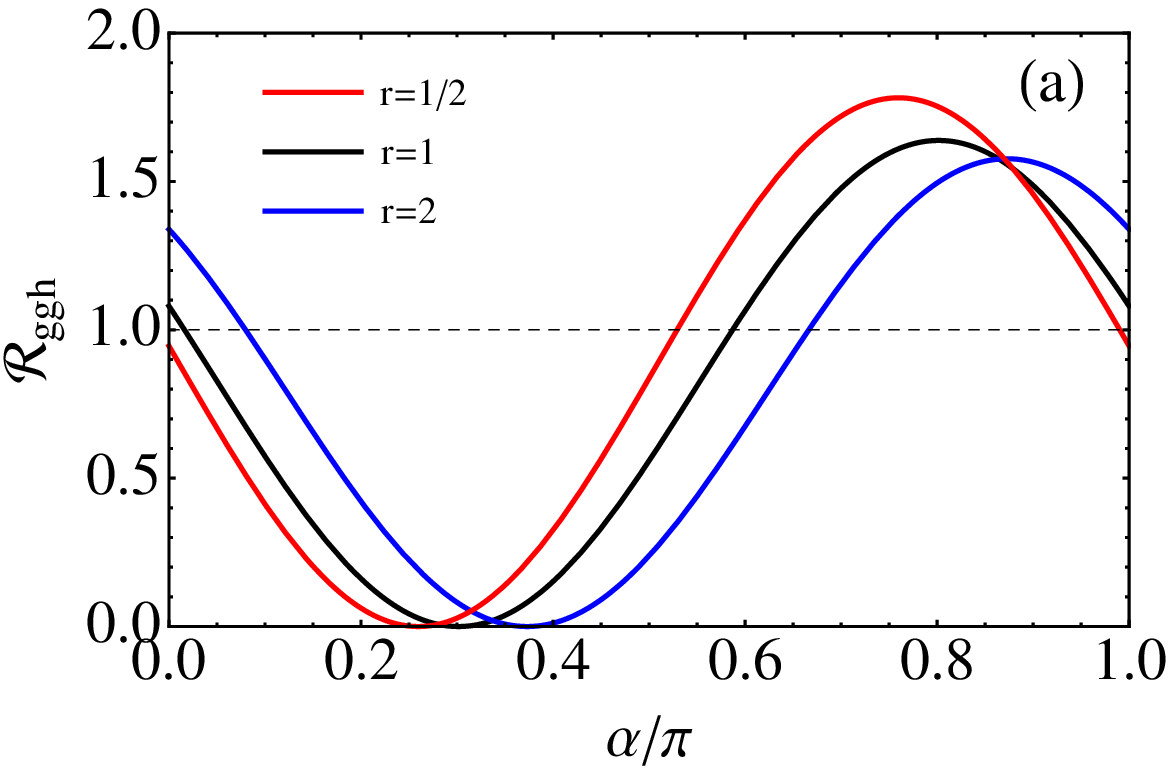}
\includegraphics[width=7.5cm,height=7cm]{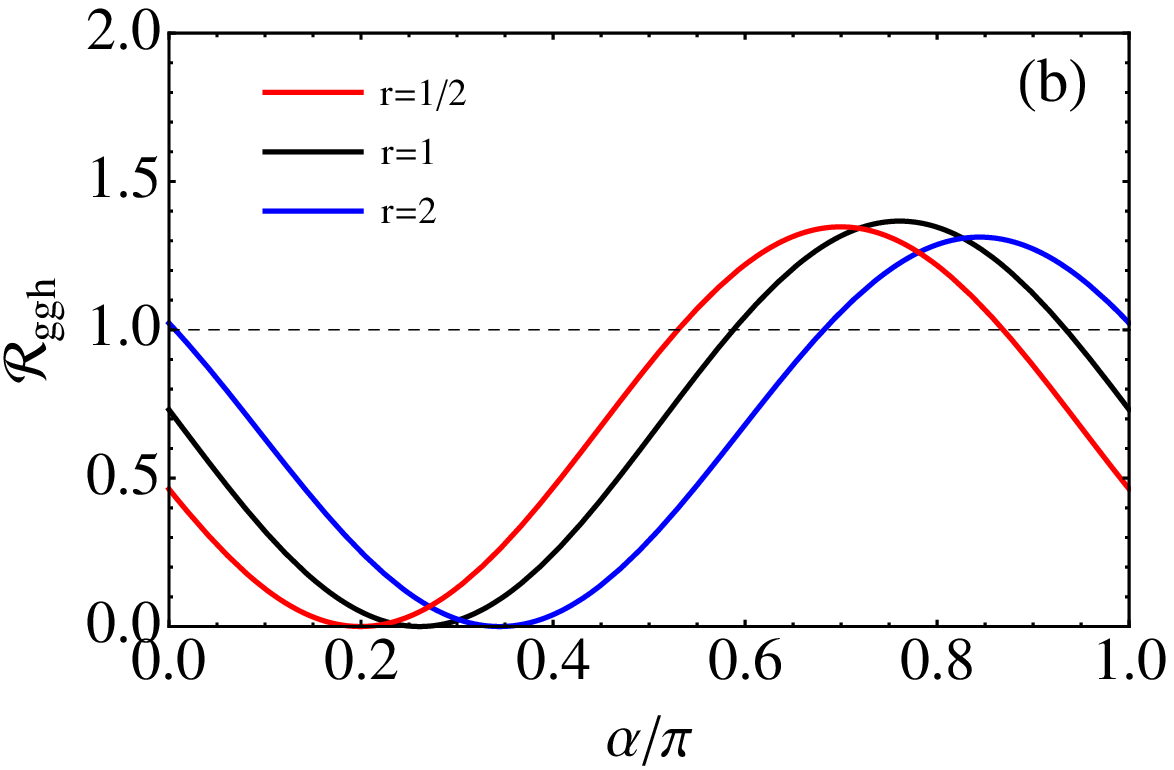}
\vspace*{-2mm}
\caption{Ratio $\mR_{ggh}^{}$ of $h^0$ production cross sections
via gluon fusion process as a function of Higgs mixing angle $\,\al$\,.\,
We identify the lighter Higgs mass $\,M_h^{} =125$\,GeV and
set the heavy fermion mass $M_F^{} = 2.5$\,TeV for both plots.
The panels (a) and (b) have input
$M_{W'}^{} = 400$\,GeV and $M_{W'}^{} = 600$\,GeV, respectively.
We have varied the VEV ratio, $\,r=(\hf,\,1,\,2)\,$,\, in each plot.
}
\label{fig:xsec}
\end{center}
\end{figure}

In Fig.\,\ref{fig:ggh_sHf},
we present the contour plots of the gluon-fusion ratio
$\,\mR_{ggh}^{}$\, over the parameter space of $\,(\alpha,\, r)$.\,
With the generic parameter inputs, we find that this ratio can be either moderately
enhanced or quite suppressed,
which periodically depends on the Higgs boson mixing angle $\,\alpha\in [0,\,\pi)$.\,
Here for most choices of the mixing angle $\alpha$,
the gluon fusion ratio $\mR_{ggh}^{}$ is not so sensitive to the VEV ratio $\,r\,$
since most $\mR_{ggh}^{}$ contours in Fig.\,\ref{fig:ggh_sHf}
are approximately vertical bands.
In Fig.\,\ref{fig:ggh_sHf}, we see that
the ratio $\mR_{ggh}$ reaches the peak values for relatively large
mixing angle $\,\alpha \simeq (0.6-0.9)\pi$\, and smaller VEV ratio
$\,r\lesssim 2\,$,\,  as marked by the red regions in both plots.
For these red regions of Fig.\,\ref{fig:ggh_sHf},
we find that $\,\mR_{ggh^{}}\simeq 1.6-1.8\,$ in the left panel,
and $\,\mR_{ggh}^{}\simeq 1.3\,$ in the right panel.
In the following analyses, we will take $\,r=\hf ,\, 2\,$
as two sample inputs, and compare them with the $r=1$ choice (with equal VEVs).

In Fig.\,\ref{fig:xsec}(a)-(b), we further display the gluon-fusion ratio
$\,\mR_{ggh}^{}\,$ as a function of the Higgs mixing angle $\,\alpha$\,,\,
for the parameter inputs $M_h=125\,\GeV$ and $M_F=2.5\,\TeV$.
The plots (a) and (b) have input
$M_{W'}^{} = 400$\,GeV and $M_{W'}^{} = 600$\,GeV, respectively.
In each panel, we have varied the VEV ratio, $\,r=(\hf,\,1,\,2)\,$.\,
We see that these production cross sections can be reasonably enhanced
relative to the SM. They are highly correlated with the top-quark Yukawa coupling
to the Higgs boson. The maximum values of the ratio $\mR_{ggh}^{}$ reach
around $\,\alpha = (0.7-0.9)\pi$\,,\, depending on the VEV ratio $r$.
From (\ref{eq:tth/H}), we see that
$\,G_{\bar{t} t h}\approx G_{\bar{t} t h}^{\rm SM}$\, for $\,r=1$.\,
This ratio $\,\mR_{ggh}^{}$\, may also vanish
at some other points, around $\,\alpha = (0.2-0.35)\pi$\,,\,
where the top-quark Yukawa coupling $\,G_{\bar{t}th}^{}$\, becomes nearly vanishing.
Comparing the two plots of Fig.\,\ref{fig:xsec}, we see
that $\,\mR_{ggh}^{}\,$ gets reduced as $M_{W'}^{}$ becomes larger.
This is a consequence from the suppression of heavy fermion Yukawa coupling,
$\,\xi_{hFF}^{} = \mO(m_W^2/M_{W'}^2)$\,.
In each plot, we also show the effects of different VEV ratios around $\,r=\mO(1)\,$,\,
which are consistent with the contours in Fig.\,\ref{fig:ggh_sHf}.

Next, we study the LHC signals of light Higgs boson
$\,h^0$\, via the most sensitive channels $(\gamma\gamma,\, WW^*,\, ZZ^*)$.
The signals is determined by the product of production cross section
and decay branching fraction.
It is convenient to analyze signal ratios
between our prediction and the SM expectation,
\beqn\label{eq:sigratio}
\mR_{XX}^{} ~\equiv~
\frac{\sigma[gg\to h]\times{\rm Br}[h\to XX]}
{~\sigma[gg\to h]_{\rm SM}^{}\times{\rm Br}[h\to XX]_{\rm SM}^{}~}\,.
\eeqn
In Fig.\,\ref{fig:h125_sig}, for the light Higgs boson $h^0$ with mass
$\,M_h^{}=125\,\GeV$,\,  we plot each ratio of
$(\mR_{\gamma\gamma}^{},\, \mR_{WW}^{},\, \mR_{ ZZ}^{})$
as a function of Higgs mixing angle $\,\al\,$.\,
The input parameters are set to be,
$\,M_{W'}^{} = 400\,\GeV\,(600\,\GeV)$, and $M = 2.5$\,TeV for the plot-(a) [plot-(b)].
In both plots, we have displayed the measured $(\gamma\gamma,\, WW,\, ZZ)$ rates
by the ATLAS and CMS experiments \cite{Atlas2012-7,CMS2012-7}, with central values plus
the $\pm 1\sigma$ error bars.  These are shown by the vertical gray lines, marked with \,$(\gamma\gamma_A^{},\,ZZ_A^{},\,WW_A^{})$\,
and \,$(\gamma\gamma_C^{}\,,ZZ_C^{}\,,WW_C^{})$.\,
Here the subscripts ``$_A$" and ``$_C$" denote the ATLAS and CMS data respectively,
which do not depend on $\,\al$\, in the horizontal direction.

\begin{figure}[]
\begin{center}
\includegraphics[width=7.5cm,height=7cm]{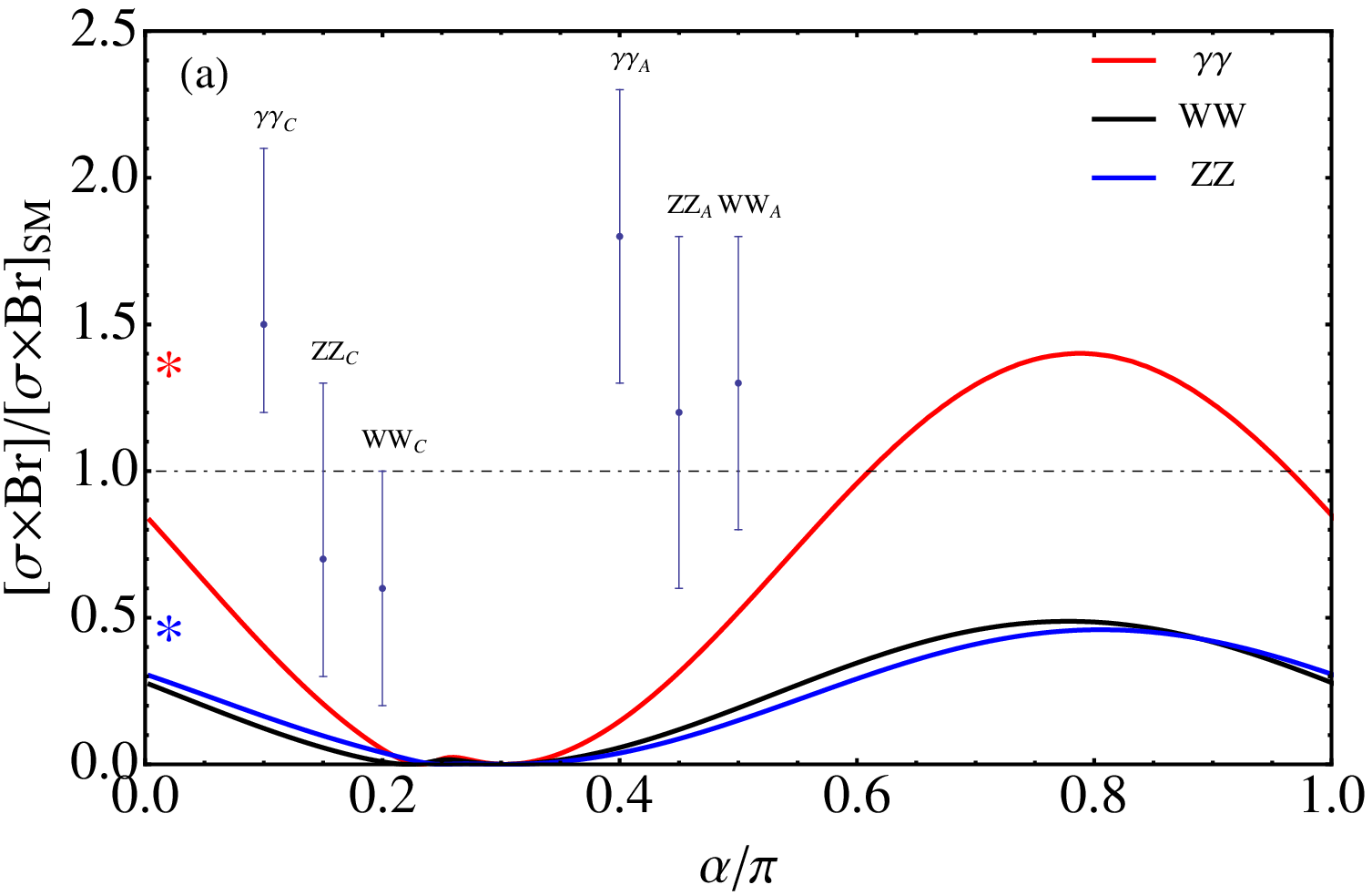}
\hspace*{2mm}
\includegraphics[width=7.5cm,height=7cm]{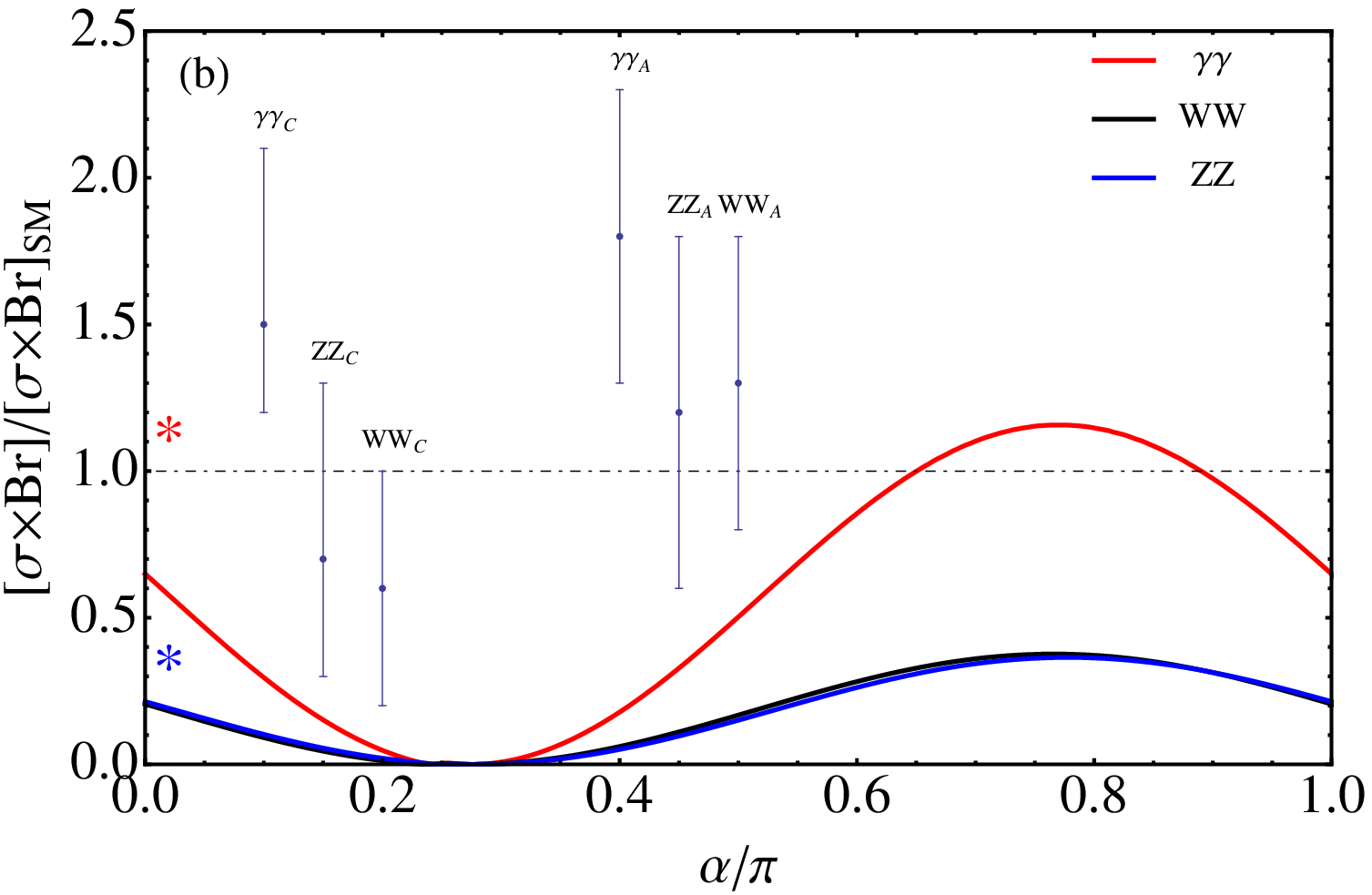}
\caption{Signal ratios \,$(\mR_{\gamma\gamma},\,  \mR_{WW},\, \mR_{ZZ})$\,
as functions of the Higgs mixing angle $\,\alpha$\,.\,
We set $\,M_h^{} =125$\,GeV and input $\,f_1^{}=f_2^{}$\,.\,
The heavy masses are taken to be, $\,M_F^{} = 2.5$\,TeV and
$\,M_{W'}^{} = 400\,$GeV [plot-(a)] or $\,M_{W'}^{} = 600\,$GeV [plot-(b)].
In each plot, at $\,\al =0\,$,\, we also show an interesting sample
with degenerate $h^0$ and $H^0$, where the predicted $\ga\ga$ rate is marked
by the red asterisk and the $ZZ^*$ ($WW^*$) rate is marked by the blue asterisk.
The ATLAS and CMS data for $\gamma\gamma$, $ZZ^*$, and $WW^*$ channels \cite{:2012gk,:2012gu}
are shown for each plot, where the subscripts ``$_A$'' and ``$_C$'' stand
for ATLAS and CMS, respectively. These data points do not depend on $\al$ and
their horizontal locations have no physical meaning,
except for the convenience of presentation.
}
\label{fig:h125_sig}
\end{center}
\end{figure}

The predicted $h^0\to\gamma\gamma$ signals can be larger than the SM Higgs boson with the same mass
$\,M_h=125\,$GeV,\, depending on the mixing angle $\,\al$\,.\,
For $M_{W'}=400$\,GeV, Fig.\,\ref{fig:h125_sig}(a) shows that
$\,\mR_{\dip}^{}>1\,$ holds
for $\,0.60\pi<\al <0.97\pi\,$.\,
For $M_{W'}=600$\,GeV, Fig.\,\ref{fig:h125_sig}(b) shows that
$\,\mR_{\dip}^{}>1\,$ holds
for $\,0.65\pi<\al <0.9\pi\,$.\,
We see that for the $\gamma\gamma$ channel, the maximal enhancement $\,\mR_{\dip}^{}=1.4\,$ for
$M_{W'}=400$\,GeV, and becomes $\,\mR_{\dip}^{}=1.2\,$ for $M_{W'}=600$\,GeV.
They are both consistent with the current LHC data as shown in Fig.\,\ref{fig:h125_sig}.
The upcoming data from the LHC\,(8\,TeV) and the next phase of LHC\,(14\,TeV)
will further discriminate this non-standard Higgs boson $h^0$ from the SM.

We also note that the $Z\gamma$ signals are generally suppressed relative
to the di-photon channel, which are reduced by the phase space suppression.
The $Z\gamma$ signals also become smaller for heavier $W'$.\,
This can be seen from inspecting the form factors in
(\ref{eq:3sitehZga_WpWp}) and (\ref{eq:3sitehZga_WWp}).

Signal rates in the $WW^*$ and $ZZ^*$ channels are quite close to each other,
and the minor differences originate from the $\mO(x^2)$ terms in Eq.\,(\ref{eq:VVh}).\,
Fig.\,\ref{fig:h125_sig} further shows that the $\mR_{VV}^{}$ ratios are
smaller than one over the full range of $\alpha$ and
for $\,\fx = \fy\,$.\footnote{The ratio $\,\mR_{VV}^{}\sim 1\,$ could be realized for
other situation with $\,f_1^{} \gg f_2^{}$\, or $\,f_1^{} \ll f_2^{}$.\,
But this would make $W'/Z'$ masses too heavy, well above 1\,TeV, and thus irrelevant
to the unitarity restoration. So this is not the parameter space we will consider
for our 221 model, as we mentioned in Sec.\,\ref{sec2}.}\,
This is because the $\,WWh\,$ and $\,ZZh\,$ couplings of (\ref{eq:VVh}) are always
suppressed relative to the SM couplings.
Fig.\,\ref{fig:h125_sig} shows that the maximal $WW^*$ and $ZZ^*$ signals
can reach about $\,\f{1}{3}-\f{1}{2}\,$ of the SM expectations.

\begin{figure}[t]
\begin{center}
\includegraphics[width=7.5cm,height=7cm]{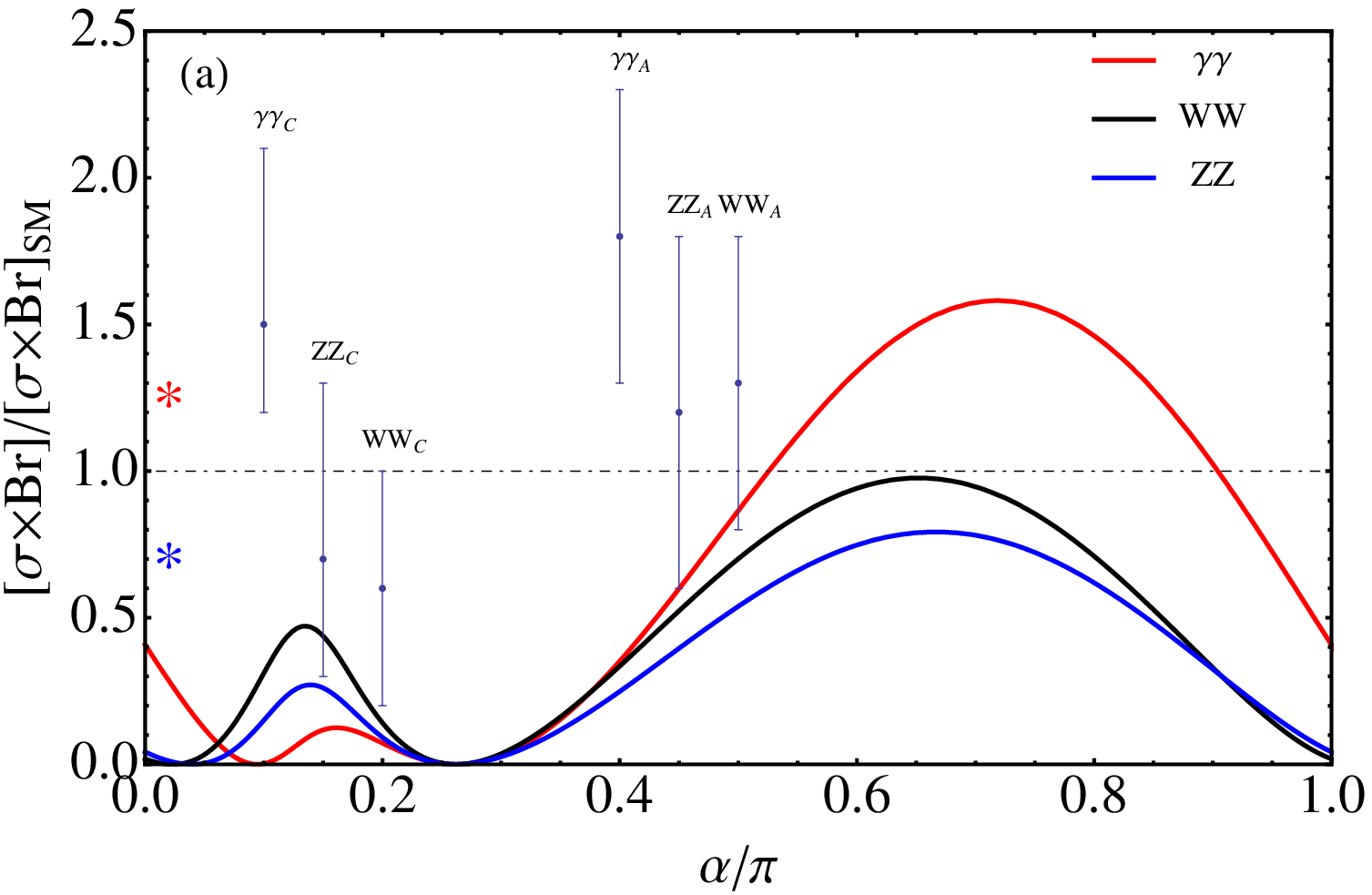}
\hspace*{2mm}
\includegraphics[width=7.5cm,height=7cm]{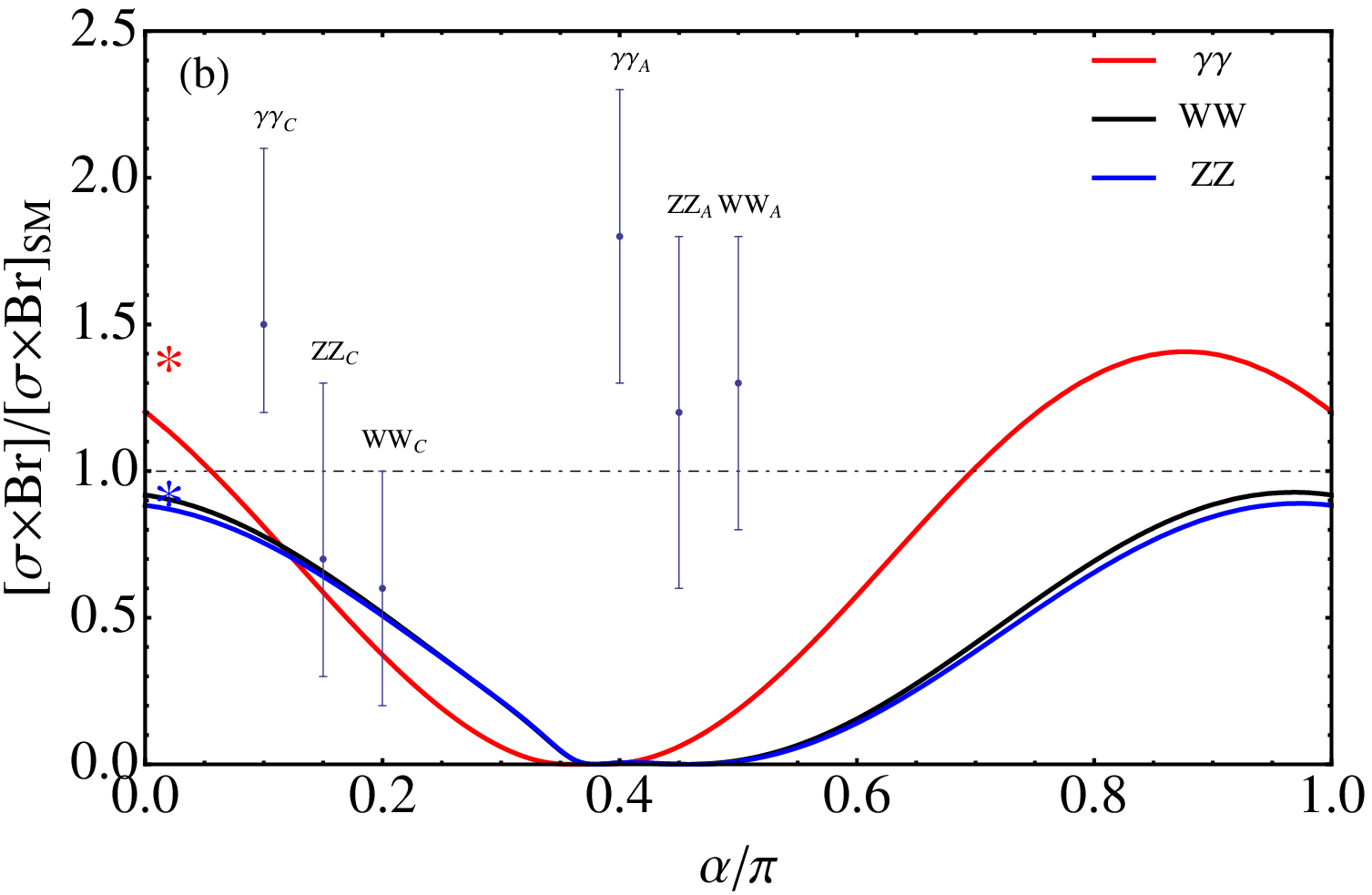}
\caption{Signal ratios \,$(\mR_{\gamma\gamma},\, \mR_{WW},\, \mR_{ZZ})$\,
as functions of the Higgs mixing angle $\,\alpha$\,.\,
The inputs are the same as Fig.\,\ref{fig:h125_sig}, except we set
$\,f_2/f_1=1/2$\, in plot-(a) and $\,f_2/f_1=2$\, in plot-(b).
For both panels we take $\,M_{W'}^{} = 400\,$GeV.
In each plot, at $\,\al =0\,$,\, we also show an interesting sample
with degenerate $h^0$ and $H^0$, where the predicted $\ga\ga$ rate is marked by the
red asterisk and the $ZZ^*$ ($WW^*$) rate is marked by the blue asterisk.
The ATLAS and CMS data for $\gamma\gamma$, $ZZ^*$, and $WW^*$ channels \cite{:2012gk,:2012gu}
are shown for each plot, where the subscripts ``$_A$'' and ``$_C$'' stand for ATLAS and CMS,
respectively. These data points do not depend on $\al$ and their horizontal locations
have no physical meaning, except for the convenience of presentation.}
\label{fig:h125-B}
\end{center}
\end{figure}

We further vary the input of VEV ratios according to (\ref{eq:Pranges}).
Different from Fig.\,\ref{fig:h125_sig} with $\,r=f_2^{}/f_1^{}=1\,$,
we redo the analysis in
Fig.\,\ref{fig:h125-B}(a) and \ref{fig:h125-B}(b) by inputting $\,r=\hf\,$
and $\,r=2\,$, respectively. The heavy masses are taken as,
$\,M_{W'}^{} = 400\,$GeV and $\,M_F^{} = 2.5$\,TeV, for both plots.
The signals via the $\,WW^*$\, and $\,ZZ^*$\, channels
exhibit stronger dependence on the VEV ratio $\,r\,$ than the di-photon channel.
Especially, in contrast to Fig.\,\ref{fig:h125_sig}(a), we see from
Fig.\,\ref{fig:h125-B}(a)-(b) that for $\,r=\hf\,$ (\,$r=2$\,),
the maximal ratio of $\,WW^*$\, rate over
the SM has raised to about $1.0$ $(0.9)$\,,\, and that of
$\,ZZ^*$\, mode is shifted up to $0.8~(0.9)$\,,\, while the maximal $\,\ga\ga$\,
signals become about a factor $1.6$ ($1.4$) of the SM expectation.
These predictions are in perfect agreement with the current LHC data
shown in the same figure.

Based on our analysis in the previous section,
we present an interesting case under
$\,\lambda_1^{}f_1^2=\lambda_2^{}f_2^2$\, and $\,\lambda_{12}^{}=0$,\,
where two Higgs bosons $(h^0,\,H^0)$ become degenerate in mass, so we have
$\,M_h^{}=M_H^{}=125\,\GeV$.
This corresponds to the Higgs mixing angle $\,\alpha=0$.\,
In practice, we can allow the masses of
$(h^0,\,H^0)$ to be nearly degenerate, consistent with the present
mass-difference and uncertainties of the $125-126$\,GeV resonances found
in the ATLAS and CMS detectors \cite{Atlas2012-7,CMS2012-7}.
According to the formulae of (\ref{eq:VVh})-(\ref{eq:tth/H}),
we note that for $\,\alpha=0$\,,\,
the Higgs couplings with gauge bosons and fermions take the simple relations,
$\,\xi_{hVV}^{}/\xi_{HVV}^{} \simeq r^3\,$,~
$\,\xi_{hff}^{}/\xi_{Hff}^{} \simeq r\,$,\, and
$\,\xi_{hFF}^{}/\xi_{HFF}^{} \simeq -[r/(1+r^2)^2](x^4M_F^2/m_f^2)\,$.\,
Numerically, we show the degeneracy signals for the $\ga\ga$ and $ZZ^*/WW^*$ channels
in Figs.\,\ref{fig:h125_sig}-\ref{fig:h125-B}, where
the red asterisks represent the predictions for the $\ga\ga$ mode and the blue asterisks
denote the results for the $ZZ^*$ ($WW^*$) mode.
To summarize, we present the predicted signal rates for the degeneracy samples,
with $\,r=1\,$,
\beqs
\label{eq:deg-r-1}
\beqn
&& \hspace*{-10mm}
\mR_{\gamma\gamma}^{\rm deg}=1.37\,,~~~\mR_{ZZ}^{\rm deg}=0.46\,,~~~\mR_{WW}^{\rm deg}=0.47\,,
~~~~~~(\text{for}~M_{W'}^{}=400\,\GeV,~ r=1);
\\[2mm]
&& \hspace*{-10mm}
\mR_{\gamma\gamma}^{\rm deg}=1.15\,,~~~\mR_{ZZ}^{\rm deg}=0.36\,,~~~\mR_{WW}^{\rm deg}=0.37\,,
~~~~~~(\text{for}~M_{W'}^{}=600\,\GeV,~ r=1).
\eeqn
\eeqs
and with $\,r=\hf,\,2\,$,
\beqs
\label{eq:deg-r2-05-2}
\beqn
&& \hspace*{-10mm}
\mR_{\gamma\gamma}^{\rm deg}=1.27\,,~~~\mR_{ZZ}^{\rm deg}=0.58\,,~~~\mR_{WW}^{\rm deg}=0.72\,,
~~~~~~(\text{for}~M_{W'}^{}=400\,\GeV,~ r=\hf );
\\[2mm]
&& \hspace*{-10mm}
\mR_{\gamma\gamma}^{\rm deg}=1.39\,,~~~\mR_{ZZ}^{\rm deg}=0.89\,,~~~\mR_{WW}^{\rm deg}=0.93\,,
~~~~~~(\text{for}~M_{W'}^{}=400\,\GeV,~ r=2).
\eeqn
\eeqs
It is clear that they agree well with the current LHC observations,
and can be further discriminated from the SM Higgs boson.
In passing, during the finalization of this paper, we notice a different study appeared\,\cite{Gunion:2012gc}, which considered
the two nearly degenerate neutral Higgs bosons in the context of
GUT-scale NMSSM scenarios and found enhanced $\ga\ga$ signal rate as well as broadened mass peaks.
Our collider phenomena with the nearly degenerate $(h^0,\,H^0)$
differs from \cite{Gunion:2012gc} substantially, due to the new gauge bosons
and vector-like heavy fermions, as well as the absence of charged Higgs boson and
superpartners in our model.

\vspace*{3mm}
\subsection{\hspace*{-2mm}Signatures of h$^0$ via Associated Production and Vector Boson Fusion}
\label{sec:AP_VBF}

We now turn to discussing the other two production channels,
the associated production process $\,q_1^{}\bar{q}_2^{} \to\! V h\,$
and the vector boson fusion (VBF) process
$\,q_1^{} q_2^{} \to h q_3^{}q_4^{}\,$.\,
In both productions, the $hVV$ couplings
and the decay branching fractions of $\,h^0$\,
differ from the SM Higgs boson.
Let us consider the Higgs decays into the SM fermions $\,h^0\to f\bar{f}\,$,\,
with the final states such as $\,f\bar{f}=b\bar{b},\,\tau\bar{\tau}$\,.\,
Using the couplings from (\ref{eq:VVh}), (\ref{eq:tth/H}) and (\ref{eq:Vff}),
we can derive the signal ratios of the current model over the SM as follows,
\beqs
\label{eq:APandVBF}
\beqn
\label{eq:AP}
&& \hspace*{-10mm}
\f{\sigma[q_1^{}\bar{q}_2^{} \to\! V h] \times {\rm Br}[h \to\! f\bar{f}]}
  {\sigma[q_1^{}\bar{q}_2^{} \to\! Vh]_{\rm SM}\times {\rm Br}[h\to\! f\bar{f}]_{\rm SM}^{}}
~\simeq~  \xi_{VVh}^2 \xi_{hff}^2 \f{\,\Gamma_h^{\rm SM}\,}{\,\Gamma_h^{}\,} \,,
\\[2mm]
&& \hspace*{-10mm}
\f{\sigma[q_1^{} q_2^{} \to h q_3^{}q_4^{}] \times {\rm Br}[h\to\! f\bar{f}]}
  {\sigma[q_1^{} q_2^{} \to h q_3^{}q_4^{}]_{\rm SM}^{}
   \times{\rm Br}[h \to\! f\bar{f}]_{\rm SM}^{}}
~\simeq~  \xi_{VVh}^2 \xi_{hff}^2 \f{\,\Gamma_h^{\rm SM}\,}{\,\Gamma_h^{}\,}
\,,~~
\label{eq:VBF}
\eeqn
\eeqs
where the relevant $V$-$q$-$\bar{q}'$ couplings equal the SM couplings
to good accuracy (cf.\ Sec.\,\ref{sec2.4.2}) and the small $\mO(x^4)$ corrections
can be safely dropped here. Hence, with (\ref{eq:VVh}) and (\ref{eq:tth/H}), we
derive coupling ratio on the RHS of (\ref{eq:APandVBF}),
%
\beqa
\label{eq:AP-VBF-2}
\xi_{hVV}^2\xi_{hff}^2
~\simeq~ \f{~(r^3\ca\!-\sa)^2(r\ca\!-\sa)^2\,}{(1+r^2)^4} \,,
\eeqa
%
which only explicitly depends on the VEV ratio $\,r = \fy/\fx\,$
and Higgs mixing angle $\,\al$\,.\,
For the current analysis, we will vary $\,r$\, within $\mO(1)$\,.\,
Meanwhile it is useful to note the two special limits, $\,r\to 0\,$ and $\,r\to\infty\,$.\,
For $\,r\to 0\,$ (\,$\fx\gg\fy$\,), we have
$\,\xi_{hVV}^2\xi_{hff}^2\to \sin^4\!\al\,$;\, while for
$\,r\to \infty\,$ ($\,\fy\gg\fx\,$), we arrive at
$\,\xi_{hVV}^2\xi_{hff}^2\to \cos^4\!\al\,$.\,

\begin{figure}[t]
\begin{center}
\includegraphics[width=7.7cm,height=7.2cm]{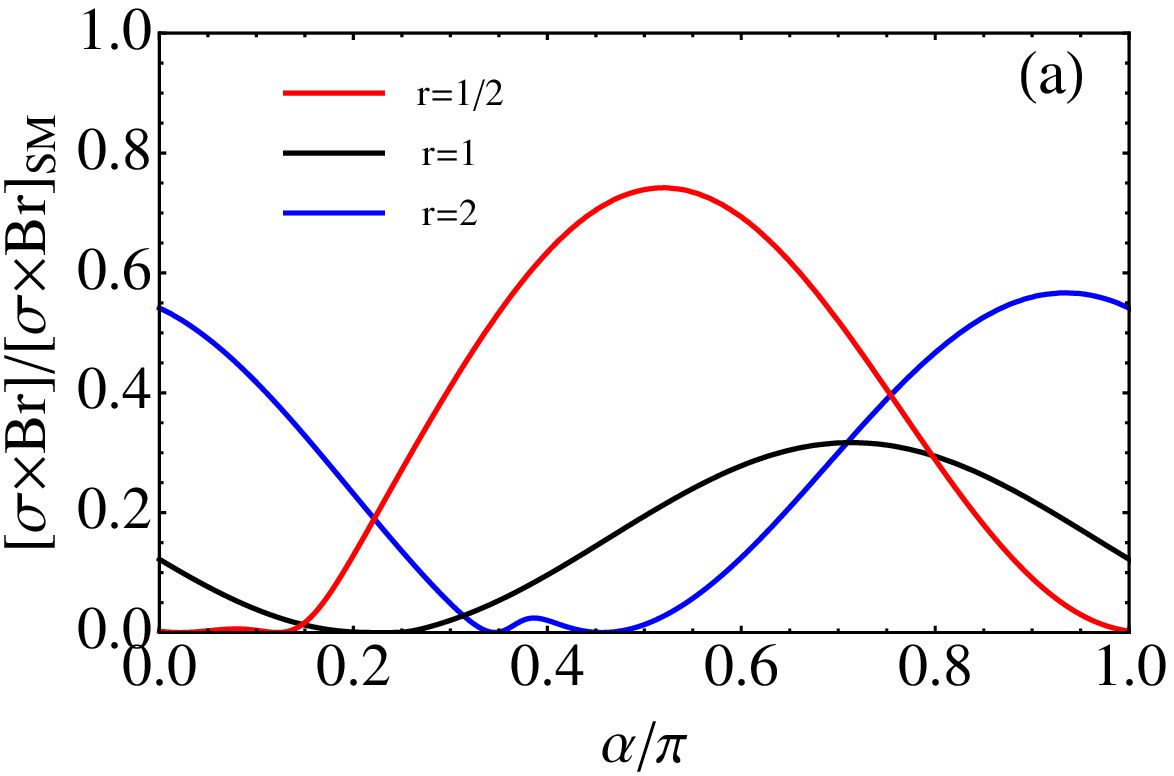}
\includegraphics[width=7.7cm,height=7.2cm]{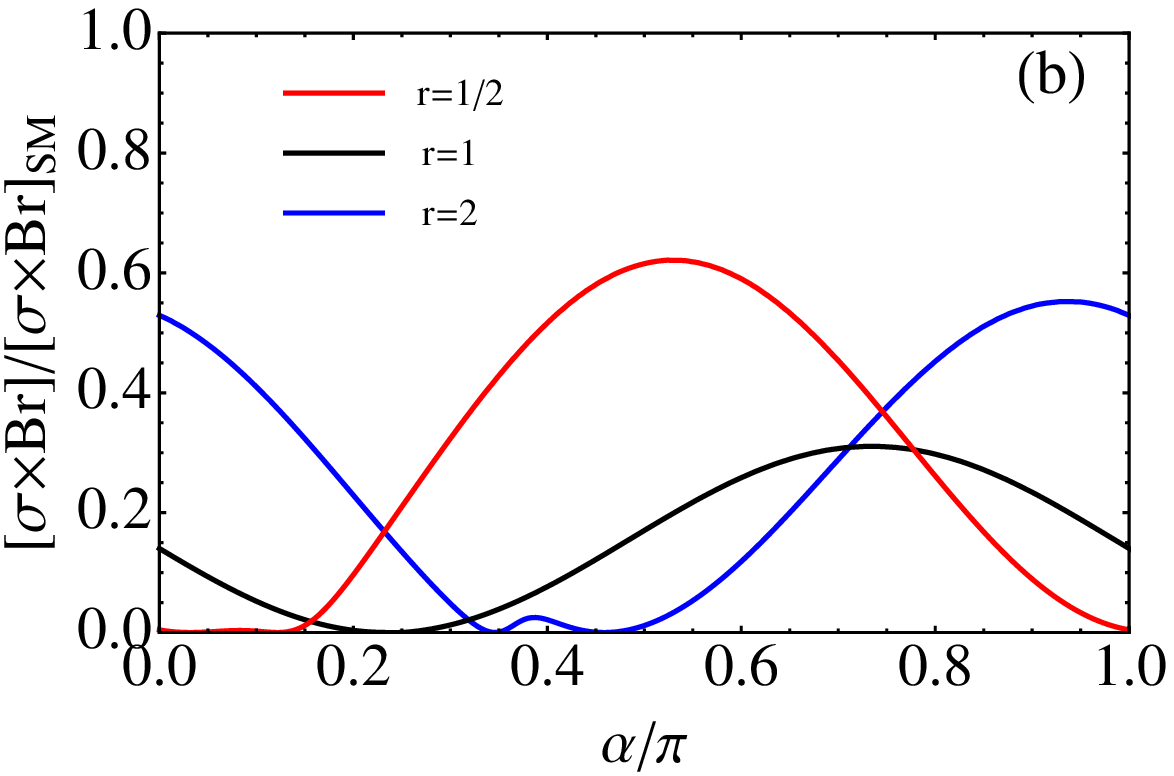}
\vspace*{-4mm}
\caption{Signals of $\,\sigma(q\bar{q}'\to V h) \times {\rm Br}(h \to b \bar{b})$\,
or $\,\sigma(q\bar{q}'\to h\,q_3^{}q_4^{}) \times {\rm Br}(h \to \tau\bar{\tau})$\,
in the present model over the corresponding SM expectations,
as a function of the Higgs mixing angle $\,\alpha\,$.\,
We identify the lighter Higgs mass $\,M_h^{} =125$\,GeV and
set the heavy fermion mass $M_F^{} = 2.5$\,TeV for both plots.
The panels (a) and (b) have input
$M_{W'}^{} = 400$\,GeV and $M_{W'}^{} = 600$\,GeV, respectively.
We have varied the VEV ratio, $\,r=(\hf,\,1,\,2)\,$,\, in each plot.
}
\label{fig:xsecRatio_AP_400}
\end{center}
\end{figure}

In Fig.\,\ref{fig:xsecRatio_AP_400}, we plot the ratios
(\ref{eq:APandVBF})-(\ref{eq:AP-VBF-2})
for the present model as a function of the Higgs mixing angle $\,\al\,$.\,
We identify the lighter Higgs boson mass  $\,M_h^{} =125\,\GeV$ and
choose the heavy fermion mass $M = 2.5$\,TeV.
In the plots (a) and (b) we have input $M_{W'}^{} = 400$\,GeV and
$M_{W'}^{} = 600$\,GeV, respectively.
In each plot, we have varied the VEV relations,
$\,f_2 = f_1/2$,\, $\,f_2 = f_1$,\, and $\,f_2 = 2f_1$,\,
corresponding to $\,r=(\hf,\,1,\,2)$,\, respectively.
Inspecting each panel, we note
that for smaller $\,r\,$ such as $\,r=\hf$\,,\, the curve is similar
to the asymptotic behavior $\,\sin^4\!\al\,$ (under $\,r\to 0\,$),
while for larger $\,r\,$ such as $\,r=2$\,,\, the curve becomes closer
to the asymptotic behavior $\,\cos^4\!\al\,$ (under $\,r\to\infty\,$).
As expected, our signals are always smaller than the SM expectation
due to the general suppressions in the tree-level $hVV$ and $hff$ couplings
of (\ref{eq:VVh})-(\ref{eq:tth/H}).
From Fig.\,\ref{fig:xsecRatio_AP_400},
we see that for the VEV ratio $\,r\,$ varying within $\,\mO(1)\,$,\,
the $\,h^0$\, signal rates can reach up to about
$\,\f{1}{3}-\f{3}{4}\,$ of the SM expectations.

Finally, we note that for the case of two (nearly) degenerate Higgs states
$h^0$ and $H^0$ with masses around $125-126$\,GeV, the signals at $\,\alpha = 0\,$
will get enhanced.
We summarize the corresponding maximal ratios of signal rates $\,\mR\,$ as follows,
\beqs
\label{eq:deg-R-MW1}
\beqn
\label{eq:deg-R-MW1=400}
\hspace*{-5mm}
\mR^{\rm deg} ~=~ (0.55,\,{0.32},\,0.74)\,,  &~~~~~~&
\text{for}\,~ r=\(2,\,1,\,\hf \) ~\,\text{and}~\, M_{W'}^{}=400\,\GeV;
\\[2mm]
\label{eq:deg-R-MW1=600}
\hspace*{-5mm}
\mR^{\rm deg} ~=~ (0.54,\,{0.31},\,0.62)\,,  &~~~~~~&
\text{for}\,~ r=\(2,\,1,\,\hf\) ~\,\text{and}~\, M_{W'}^{}=600\,\GeV.
\eeqn
\eeqs
These should be compared to the non-degenerate predictions of
Fig.\,\ref{fig:xsecRatio_AP_400} for the choice of $\,\alpha =0\,$.\,


\vspace*{3mm}
\section{\hspace*{-2mm}LHC Signatures of the Heavier Higgs Boson H$^{\bf 0}$}
\label{sec4}

 In this section, we proceed to study the LHC signals of the heavier CP-even Higgs boson $H^0$.
 For the SM Higgs boson in the large mass-ranges, the ATLAS experiment
 has excluded the SM Higgs boson mass window of
 (130.7,\,523)\,GeV at 95\%\,C.L.\,\cite{Atlas2012-7}\cite{AtlasHH},
 while the CMS has put exclusion limit within
 the mass-range of (128,\,600)\,GeV at 95\%\,C.L.\,\cite{CMS2012-7}\cite{CMSHH}.
 The dominant decay channels for a heavy SM
 or SM-like Higgs boson are known to be $\,h^0_{\rm SM}\to(WW,\,ZZ)$\,.\,
 For the present analysis, we will study our signal predictions of the heavier Higgs
 boson $H^0$ through the most sensitive channels of
 $\,H\to ZZ\to 4\ell$\, and $\,H\to WW\to 2\ell 2\nu$\,,\,
 in comparison with the ATLAS and CMS searches.
 We find parameter regions where $H^0$ is relatively light
 and well below 600\,GeV, but is hidden and escapes the current LHC searches in
 the $\,H\to WW,\,ZZ$ channels so far. We demonstrate that
 the current LHC\,(8\,TeV) is starting to probe the $H^0$ Higgs boson
 over some parameter ranges, and the next runs at LHC\,(14\,TeV)
 with higher integrated luminosities will have good potential
 to discover or exclude $H^0$ over significant parameter space,
 providing a decisive test of this model against the conventional SM.

\vspace*{2mm}
\subsection{\hspace*{-2mm}Decay Branching Fractions and Productions of H$^0$}
\label{sec4.1}

The gauge and Yukawa couplings of $H^0$ Higgs boson are derived in section\,\ref{sec2.4}.
For the present study, we first analyze the decay branching fractions of $H^0$
over the mass-range of $\,M_H^{}=130-600$\,GeV.
In Fig.\,\ref{fig:HBR600} and Fig.\,\ref{fig:HBR400},
we compute the $H^0$ decay branching fractions for the VEV ratio
$\,r=1\,$ and $\,r=\hf\,$,\, respectively, where we also set the
sample heavy fermion mass $\,M_F^{}=2.5\,$TeV.
In Figs.\,\ref{fig:HBR600}-\ref{fig:HBR400},
we present the results with $\,M_{W'}^{}=600\,\GeV$ for plot-(a)(b),
while for plot-(c)(d), we set $\,M_{W'}^{}=400\,\GeV$.
In addition, we input the sample value of Higgs mixing angle $\,\al = 0.6\pi\,$ for
the plot-(a)(c) of Figs.\,\ref{fig:HBR600}-\ref{fig:HBR400},
and $\,\al = 0.8\pi\,$ for their plot-(b)(d).

Over most of the parameter region, Figs.\,\ref{fig:HBR600}-\ref{fig:HBR400} show
that the $WW$ and $ZZ$ final states are the dominant decay channels of $H^0$.\,
Accordingly, the detection of $H^0$ should be most sensitive via
$\,gg\to H^0\to ZZ\to 4\ell$\,
and $\,gg\to H^0\to WW\to 2\ell 2\nu$\,.\,
We note that for a lighter mass of the new gauge boson,
such as $\,M_{W'}=400\,$GeV in plot-(c)(d)
of Figs.\,\ref{fig:HBR600}-\ref{fig:HBR400},
the decay channel $\,H\to VV'\,$ is open for $\,M_H^{}> M_{V'}^{}+m_{V}^{}\,$.\,
As shown in (\ref{eq:gWWpH}) of Sec.\,\ref{sec2.4}, we see that the
$\,HVV'\,$ couplings receive enhancement from the heavy mass $M_{V'}^{}$,
and thus can make the $\,H^0\to VV'\,$ decay modes significant
for the large $M_H$ mass ranges.

\begin{figure}[t]
\begin{center}
\includegraphics[width=7.6cm,height=7.2cm]{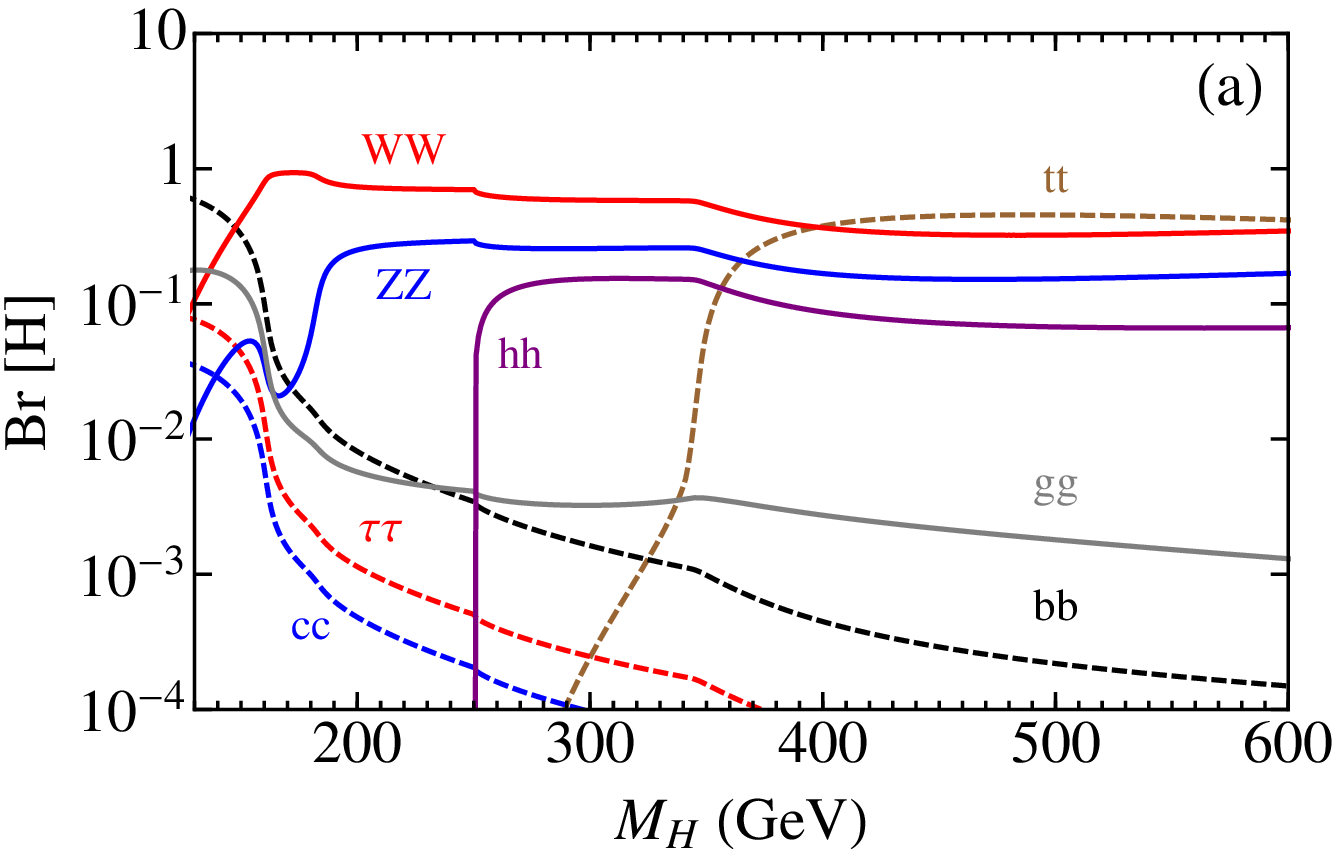}
\includegraphics[width=7.6cm,height=7.2cm]{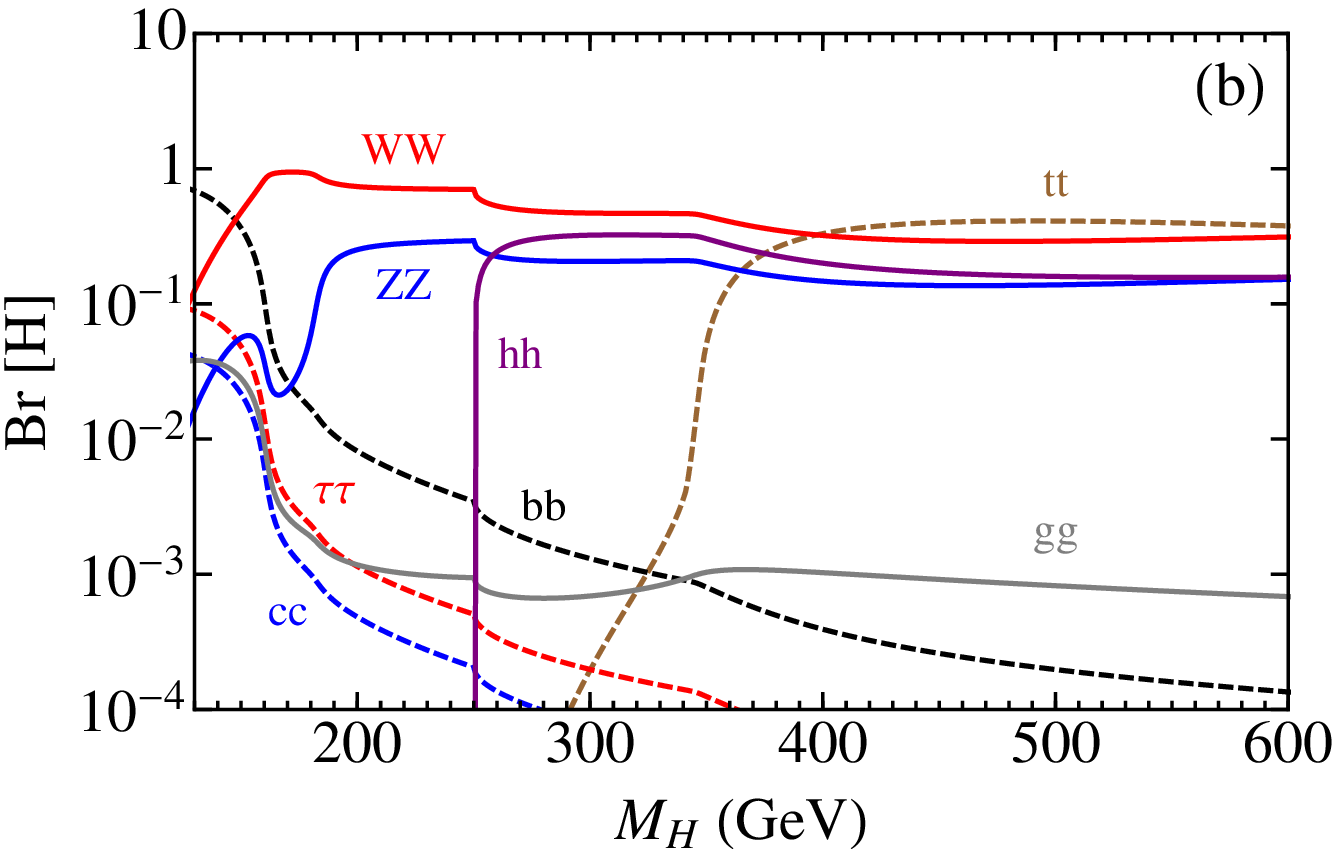}
\includegraphics[width=7.6cm,height=7.2cm]{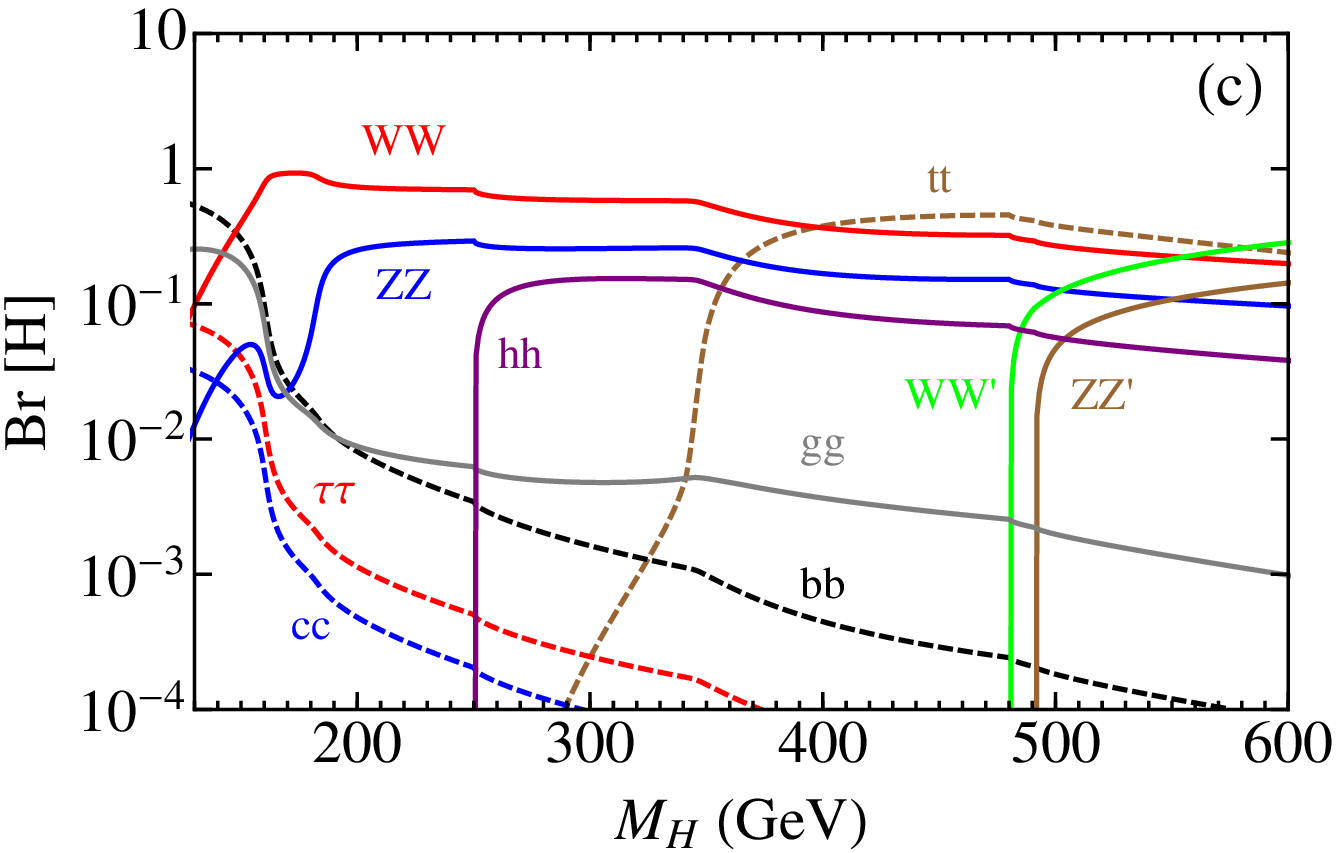}
\includegraphics[width=7.6cm,height=7.2cm]{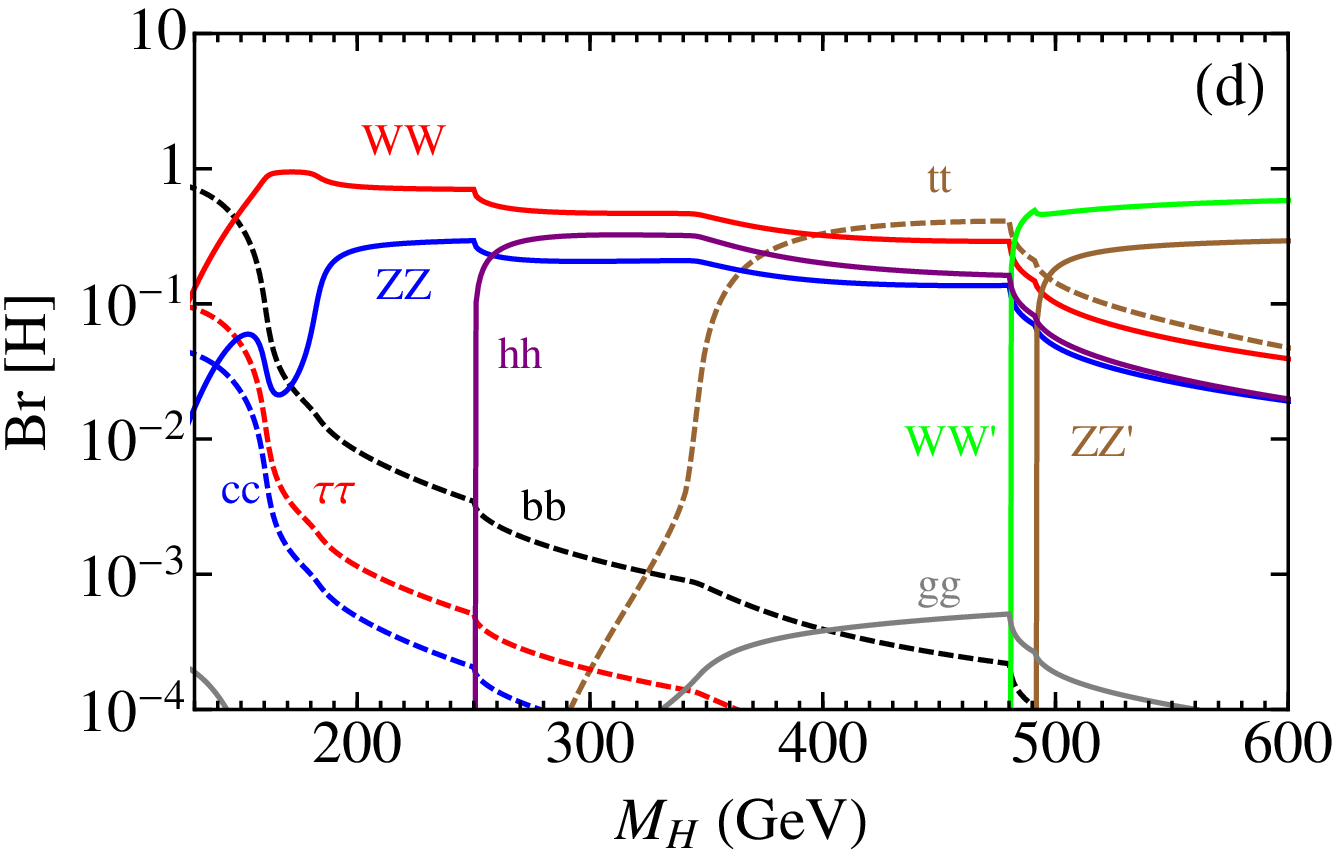}
\vspace*{-2mm}
\caption{Decay branching fraction of the heavier Higgs boson $H^0$ of our model.
We have inputs $r=1$ and $\,M_F^{} = 2.5\,$TeV.
The Higgs mixing angle is taken to be
$\,\alpha=0.6\pi$\, for plot-(a)(c) and $\,\alpha=0.8\pi$\, for plot-(b)(d).
We set the $W'$ mass $\,M_{W'}^{} = 600\,$GeV for plot-(a)(b) and
$\,M_{W'}^{} = 400\,$GeV for plot-(c)(d).}
\label{fig:HBR600}
\end{center}
\end{figure}

\begin{figure}[h]
\begin{center}
\includegraphics[width=7.6cm,height=7.2cm]{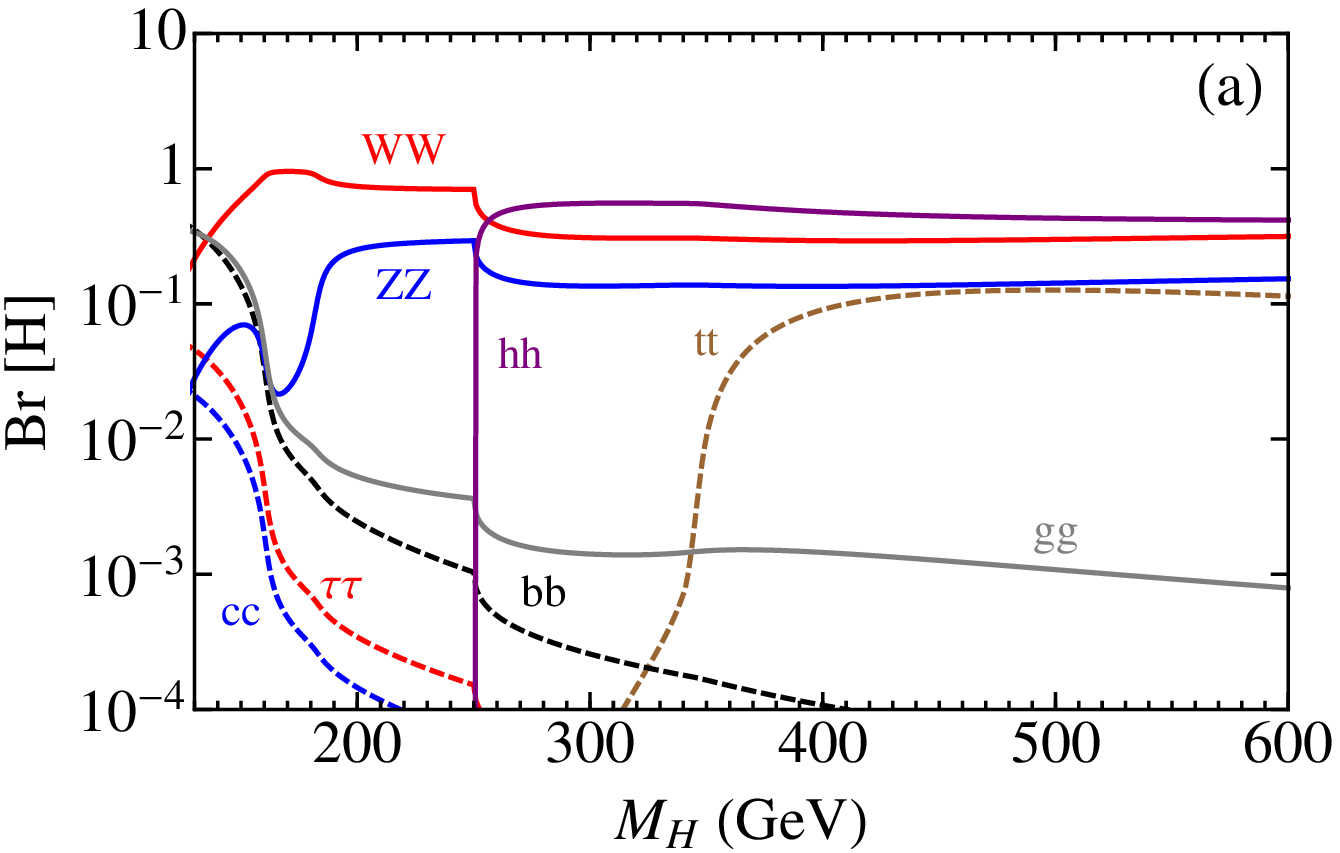}
\includegraphics[width=7.6cm,height=7.2cm]{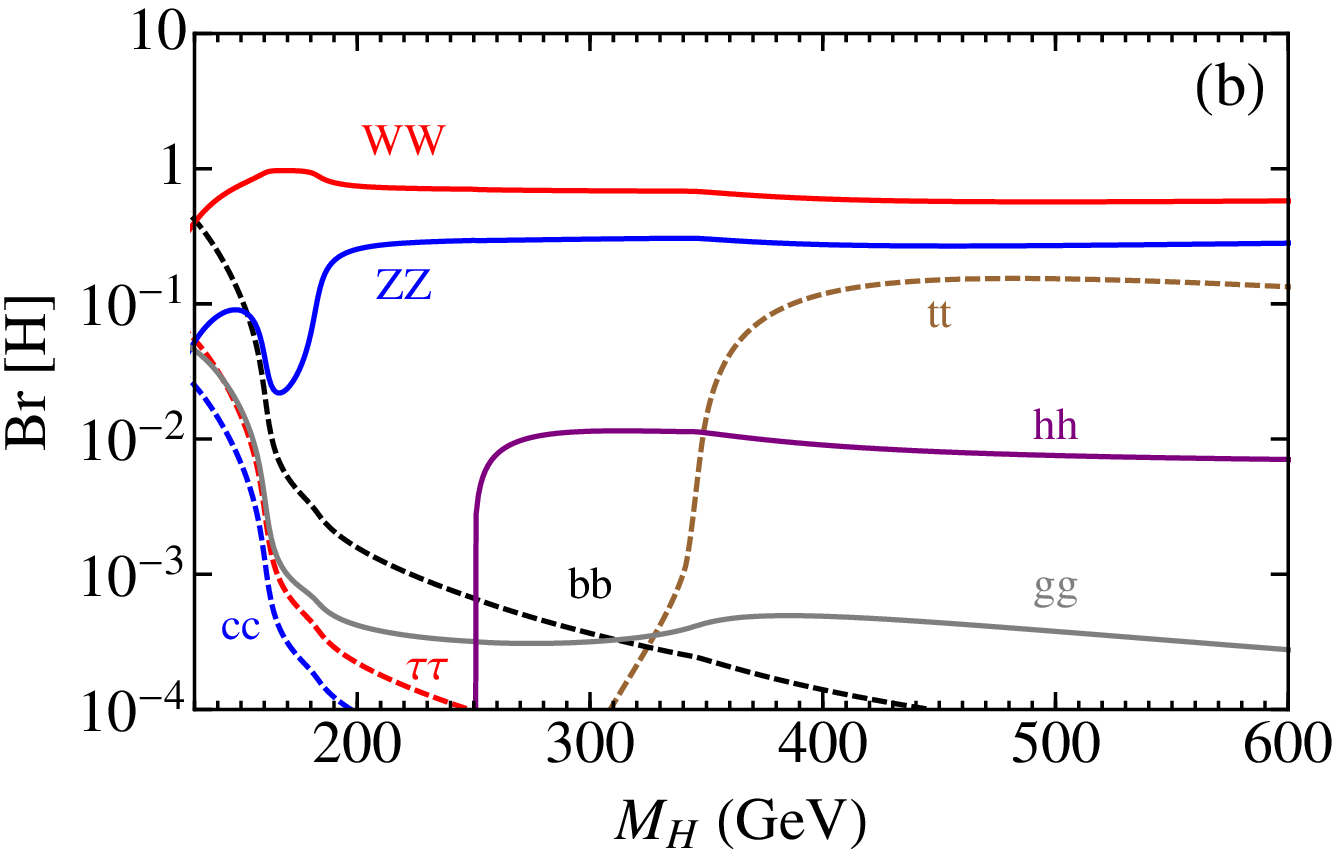}
\includegraphics[width=7.6cm,height=7.2cm]{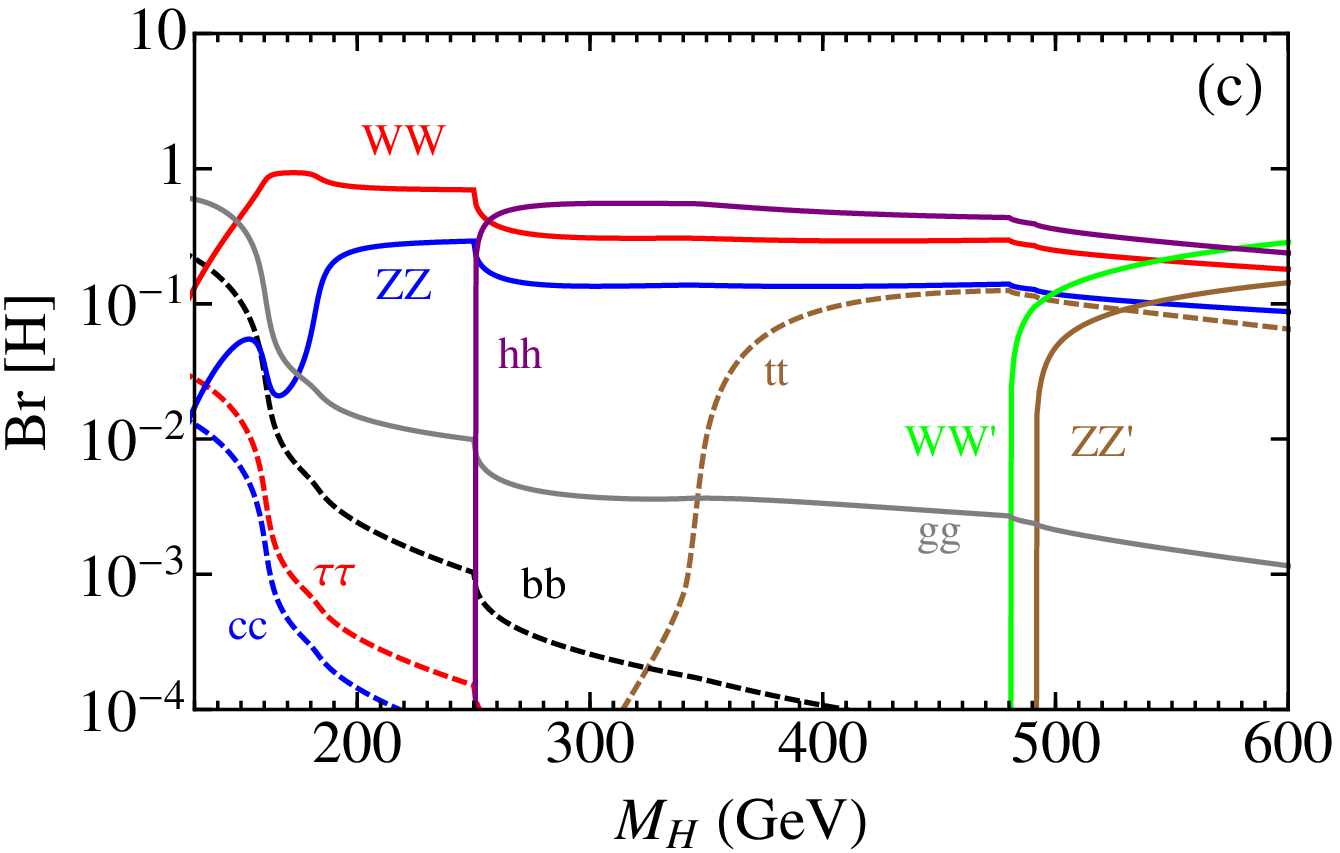}
\includegraphics[width=7.6cm,height=7.2cm]{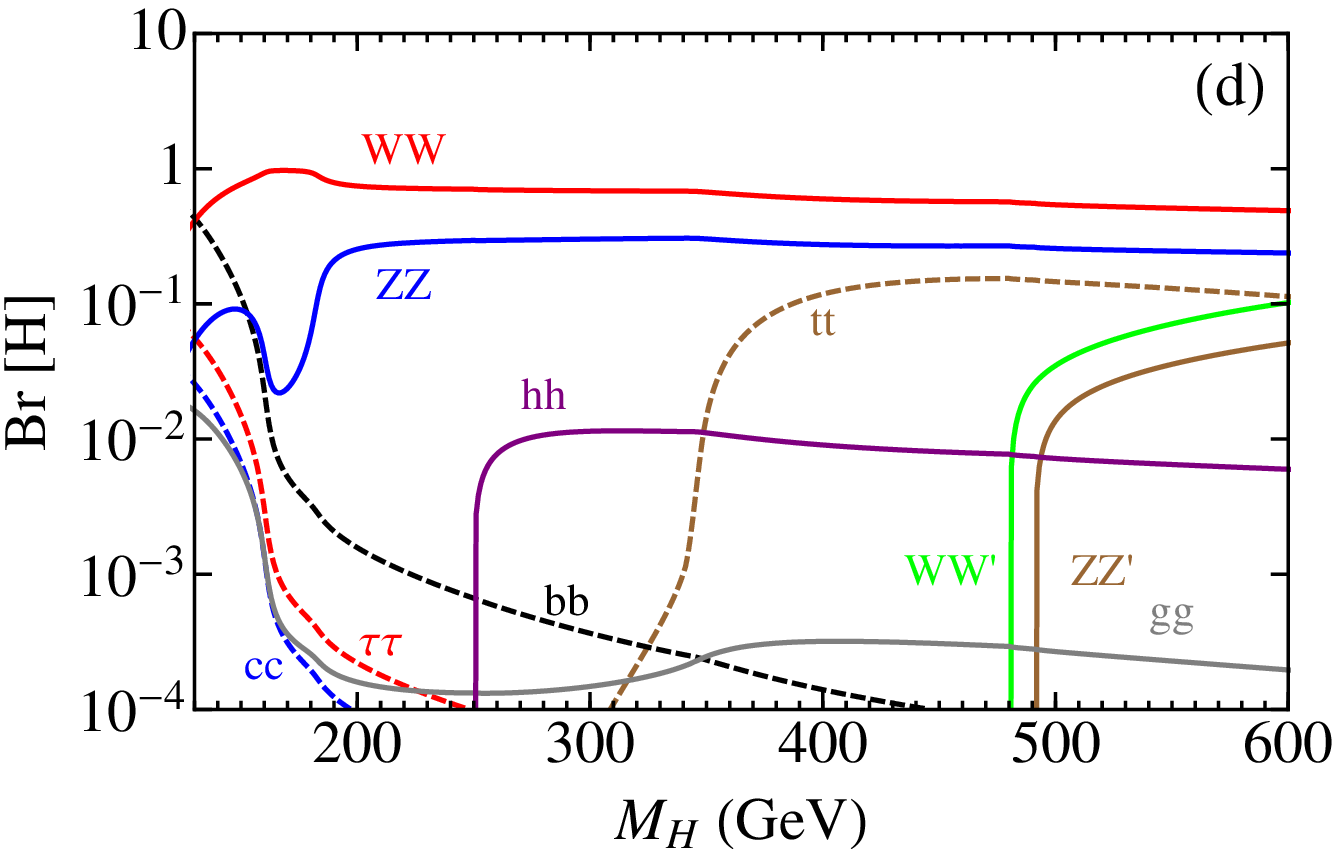}
\vspace*{-2mm}
\caption{Decay branching fraction of the heavier Higgs boson $H^0$ of our model.
We have input $r=\hf\,$ and $\,M_F^{} = 2.5\,$TeV.
The Higgs mixing angle is taken to be
$\,\alpha=0.6\pi$\, for plot-(a)(c) and $\,\alpha=0.8\pi$\, for plot-(b)(d).
We set the $W'$ mass $\,M_{W'}^{} = 600\,$GeV for plot-(a)(b) and
$\,M_{W'}^{} = 400\,$GeV for plot-(c)(d).}
\label{fig:HBR400}
\end{center}
\end{figure}

In addition, we can derive the cubic scalar vertex $H$-$h$-$h$ from the Higgs potential
(\ref{eq:V}) with the coupling,
\beqn
G_{Hhh}^{} \,&=&\,
-\hf\sin 2\alpha \(M_h^2+\hf M_H^2\)
\(\frac{\cos\alpha}{f_1^{}}+\frac{\sin\alpha}{f_2^{}} \)\,.
\eeqn
Thus, when the heavy Higgs boson mass becomes $\,M_H^{}\geqq 2 M_h^{}$\,,\,
the decay channel $\,H^0\to h^0 h^0$\, will be opened,
with the partial decay width,
\beqn
\Gamma\!\left[H^0\to h^0 h^0\right] \,&=&\,
\frac{G_{Hhh}^2}{\,8\pi M_H^{}\,}\sqrt{1-\frac{4M_h^2}{M_H^2}\,}\,.
\eeqn
By identifying the lighter Higgs boson mass  $\,M_h^{}=125\,\GeV$,\,
this decay mode is open for the mass-range $\,M_H^{}\geqq 250\,\GeV$.\,
We find that the branching fraction of $\,H^0\to h^0 h^0$\, may be comparable to the
major channels $\,H^0\to WW,\,ZZ,\,tt$\, in the relevant mass-range of $H^0$,
as shown in Fig.\,\ref{fig:HBR600}(b)(d) for $(r,\,\alpha)=(1,\,0.8\pi)$
and Fig.\,\ref{fig:HBR400}(a)(c) for $(r,\,\alpha)=(\hf,\,0.6\pi)$.

\begin{figure}[h]
\begin{center}
\includegraphics[width=7.7cm,height=7.2cm]{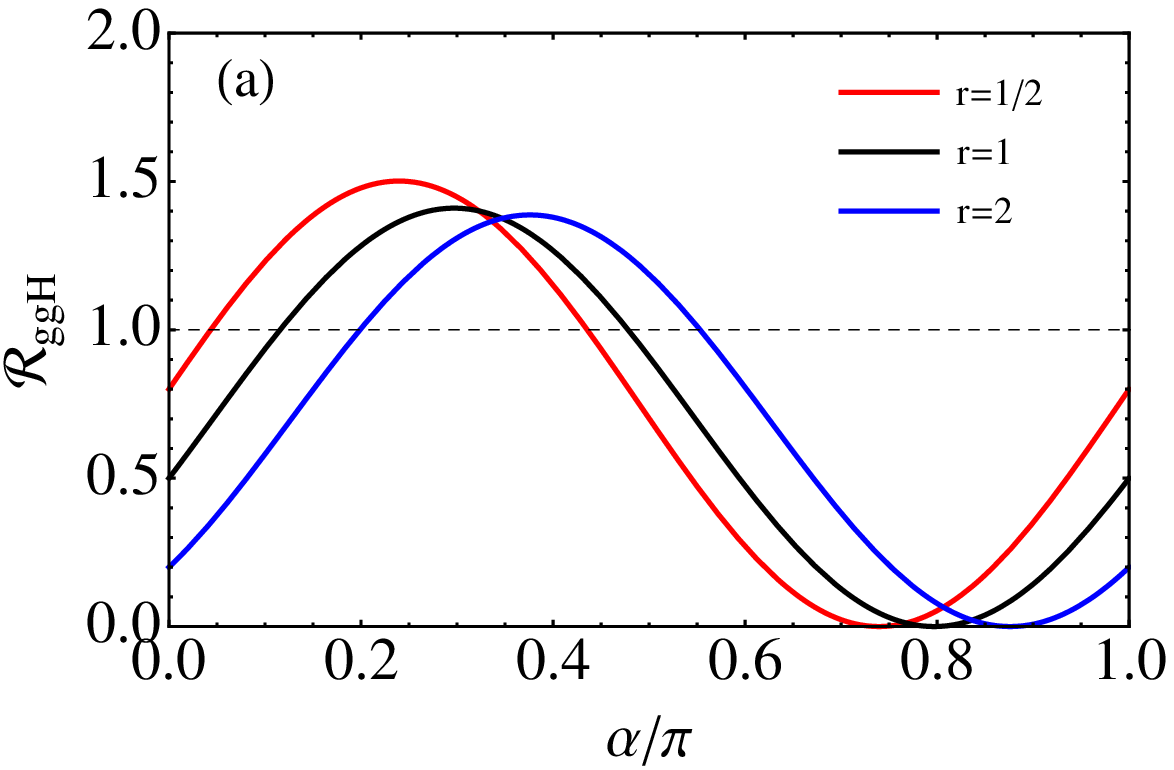}
\includegraphics[width=7.7cm,height=7.2cm]{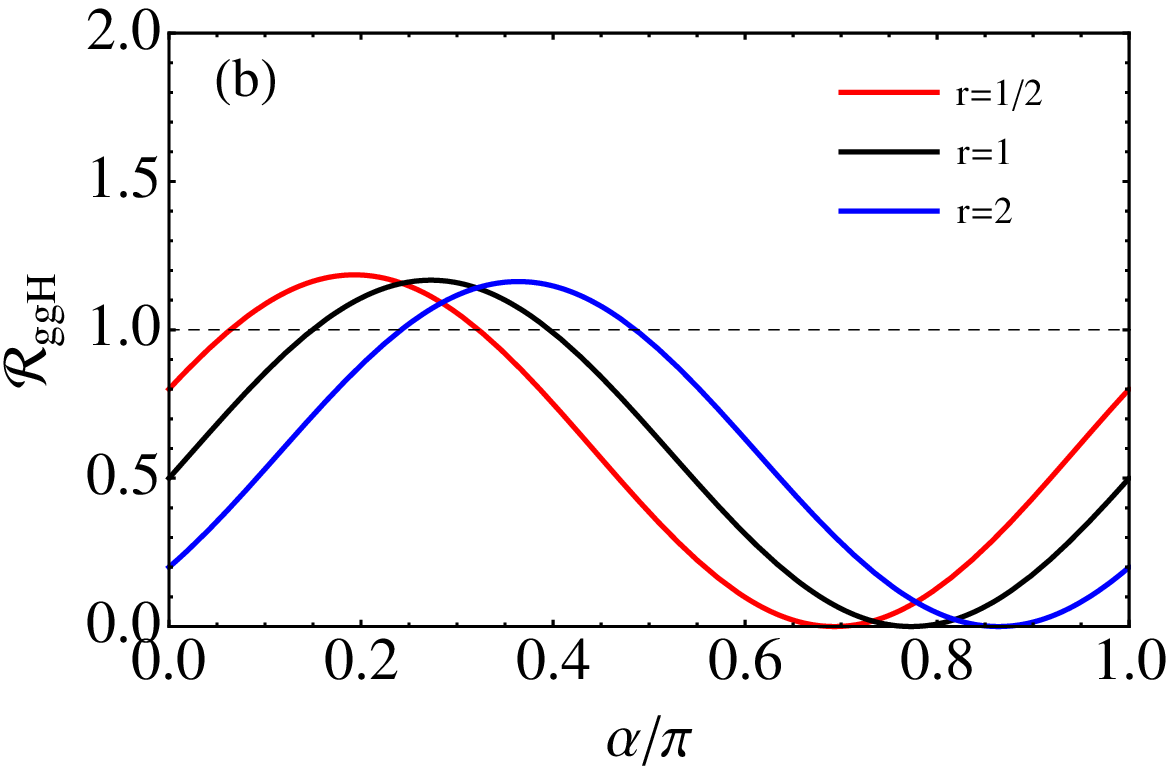}
\includegraphics[width=7.7cm,height=7.2cm]{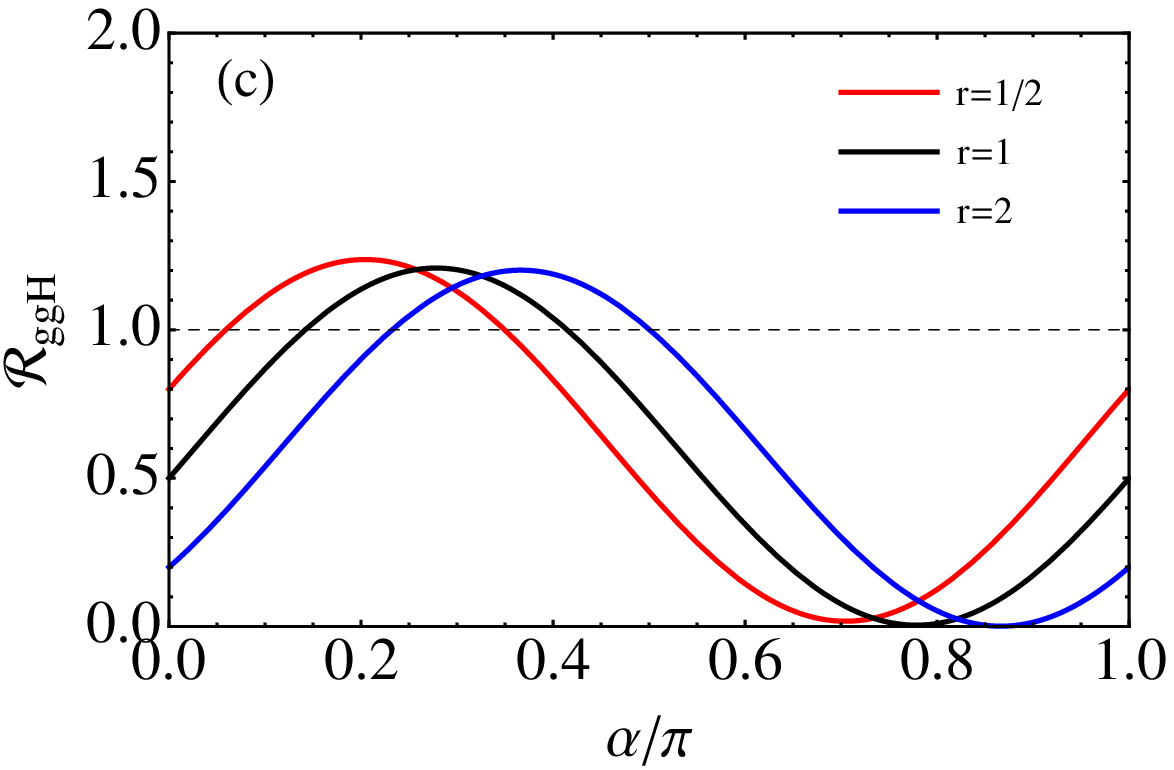}
\includegraphics[width=7.7cm,height=7.2cm]{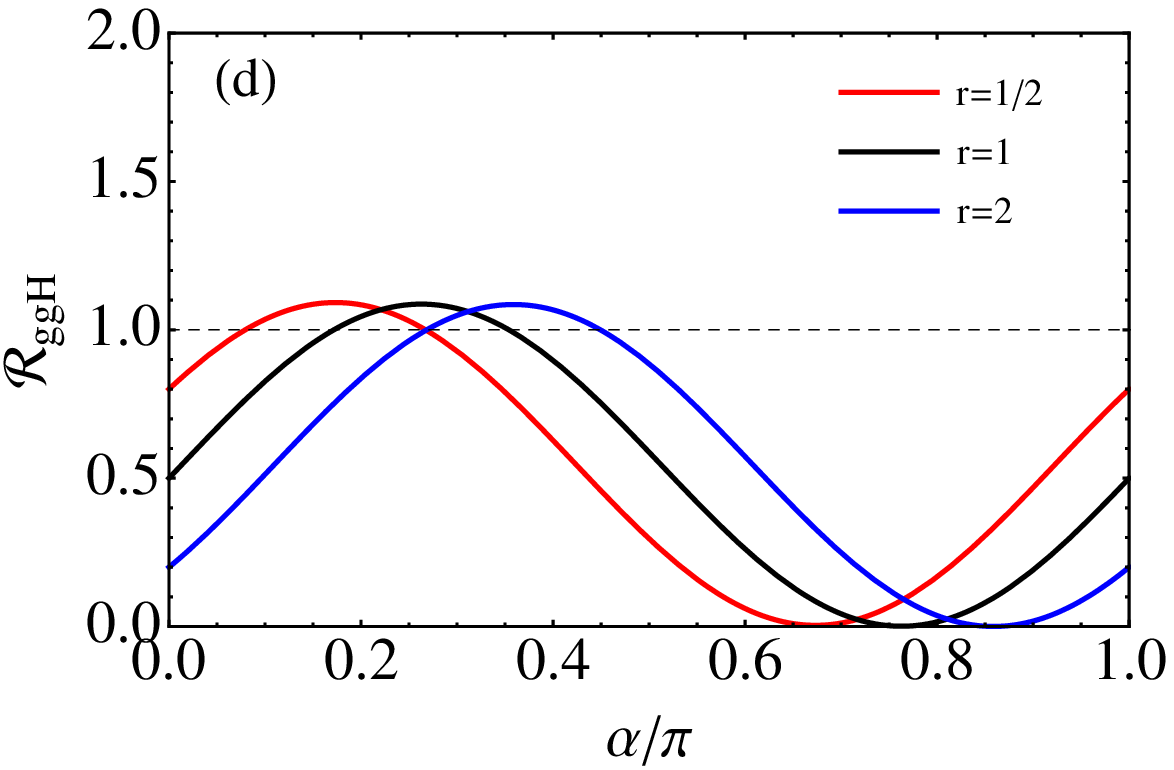}
\vspace*{-4mm}
\caption{Ratio $\mR_{ggH}$ of $H^0$ production cross sections via gluon fusion process
as a function of the Higgs mixing angle $\,\alpha$\,.\, We set the heavy Higgs mass,
$M_H^{}=250\,\GeV$ for plot-(a)(b), and $M_H^{}=500\,\GeV$ for plot-(c)(d).
The plots (a,c) and (b,d) have $\,M_{W'}=400\,$GeV and $\,M_{W'}=600\,$GeV,
respectively. All plots have input the heavy fermion mass $M_F^{}=2.5\,\TeV$.
}
\label{fig:ggH}
\end{center}
\end{figure}

In parallel to Eq.\,(\ref{eq:xsec_ratio}), we can define the ratio $\mR_{ggH}$
of the gluon-fusion production cross sections for
the heavy Higgs boson $H$ over a hypothetical SM Higgs boson with the same mass,
\beqn
\label{eq:xsec-ratio-H}
\mR_{ggH}^{} \,\equiv\,
\frac{\sigma[gg \to H]}{~\,\sigma[gg \to h]_{\rm SM}^{}~}
\,=\, \frac{\Gamma[H\to gg]}{~\,\Gamma[h\to gg]_{\rm SM}^{}~}\,.
\eeqn
In Fig.\,\ref{fig:ggH}, we also present this ratio $\mR_{ggH}$
as a function of the Higgs mixing angle $\,\alpha$\,.\,
We input the heavy Higgs mass,
$M_H^{}=250\,\GeV$ for plot-(a)(b), and $M_H^{}=500\,\GeV$ for plot-(c)(d).
The plots (a,c) and (b,d) have taken $\,M_{W'}=400\,$GeV and $\,M_{W'}=600\,$GeV,
respectively. We also set the sample heavy fermion mass $M_F^{}=2.5\,\TeV$ for all plots.
From Fig.\,\ref{fig:ggH}, we see that for the sample inputs of $\,r=(1,\, \hf ,\,2)$\,,\,
the signal ratio $\mR_{ggH}$ reaches its peak around $\,\alpha \simeq (0.15-0.35)\pi$,\,
and falls into its minimum at $\,\alpha\simeq (0.7-0.9)\pi$\,.\,
This should be compared to the production of the light Higgs boson $h^0$
as shown earlier in Fig.\,\ref{fig:xsec}.
The maximal and minimal values of the ratios $\mR_{ggH}$ (Fig.\,\ref{fig:ggH})
and $\mR_{ggh}$ (Fig.\,\ref{fig:xsec}) have different locations because the
corresponding $H$-$t$-$\bar{t}$ and $h$-$t$-$\bar{t}$ Yukawa couplings
in (\ref{eq:tth/H}) depend on the mixing angle $\alpha$ differently.

\vspace*{3mm}
\subsection{\hspace*{-2mm}LHC Potential of Detecting H$^0$}
\label{sec4.2}

Next, we analyze the potential for probing the heavier Higgs boson $H^0$ at LHC.\,
This is important for discriminating the present model from the conventional
SM containing only one CP-even state $\,h^0_{\rm SM}\,$.\,

We study the LHC processes,
$\,gg\to H^0\to ZZ\to 4\ell\,$ and $\,gg\to H^0\to WW\to 2\ell 2\nu\,$.\,
Thus, we consider the following signal rates of $H^0$ over
that of a hypothetical SM Higgs boson with the same mass,
\beqs
\beqa
\mR_{ZZ}^{} &=&
\f{\,\sigma(gg\!\to\!H)\!\times\!\textrm{Br}(H\!\to\!ZZ\!\to\!4\ell)\,}
  {~[\sigma(gg\!\to\!h)\!\times\!\textrm{Br}(h\!\to\!ZZ\!\to\!4\ell)]_{\rm SM}^{}~}\,,
\\[3mm]
\mR_{WW}^{} &=&
\f{\,\sigma(gg\!\to\!H)\!\times\!\textrm{Br}(H\!\to\!WW\!\to\!2\ell 2\nu)\,}
  {~[\sigma(gg\!\to\!h)\!\times\!\textrm{Br}(h\!\to\!WW\!\to\!2\ell 2\nu)]_{\rm SM}^{}~}\,.
~~~~~~
\eeqa
\eeqs

We present our results in Fig.\,\ref{fig:Hsignalr1} with $\,r=1$\,
and Fig.\,\ref{fig:Hsignalr05} with $\,r=1/2$,\, respectively.
In these two figures, we have also imposed the current 95\%\,C.L.\ exclusions
of ATLAS\,\cite{AtlasHH} and CMS\,\cite{CMSHH} on the above signal ratios,
shown as brown and black dashed curves.
For the Higgs mixing angle in the range
of $\,\al \lesssim 0.6\pi\,$,\, the LHC data have already put some nontrivial
constraints on the $H^0$ mass below about 340\,GeV.
Especially, the parameter region around $\,\al = 0.25\pi\,$
is already largely excluded by the current data.
This is expected
since Fig.\,\ref{fig:ggH} shows that the production cross section of $\,gg\to H\,$
reaches its peak values around $\,\alpha \simeq (0.15-0.35)\pi\,$,\,
and its minimal values around $\,\alpha \simeq (0.7-0.9)\pi\,$.\,
It is useful to note that for the
the parameter region with a small mixing angle $\,\alpha \simeq (0.15-0.35)\pi\,$
is also disfavored by requiring the consistency of our light Higgs boson $h^0$
with the current LHC signals, as shown in Figs.\,\ref{fig:h125_sig}-\ref{fig:h125-B}.

\begin{figure}[t]
\begin{center}
\includegraphics[width=7.6cm,height=7.2cm]{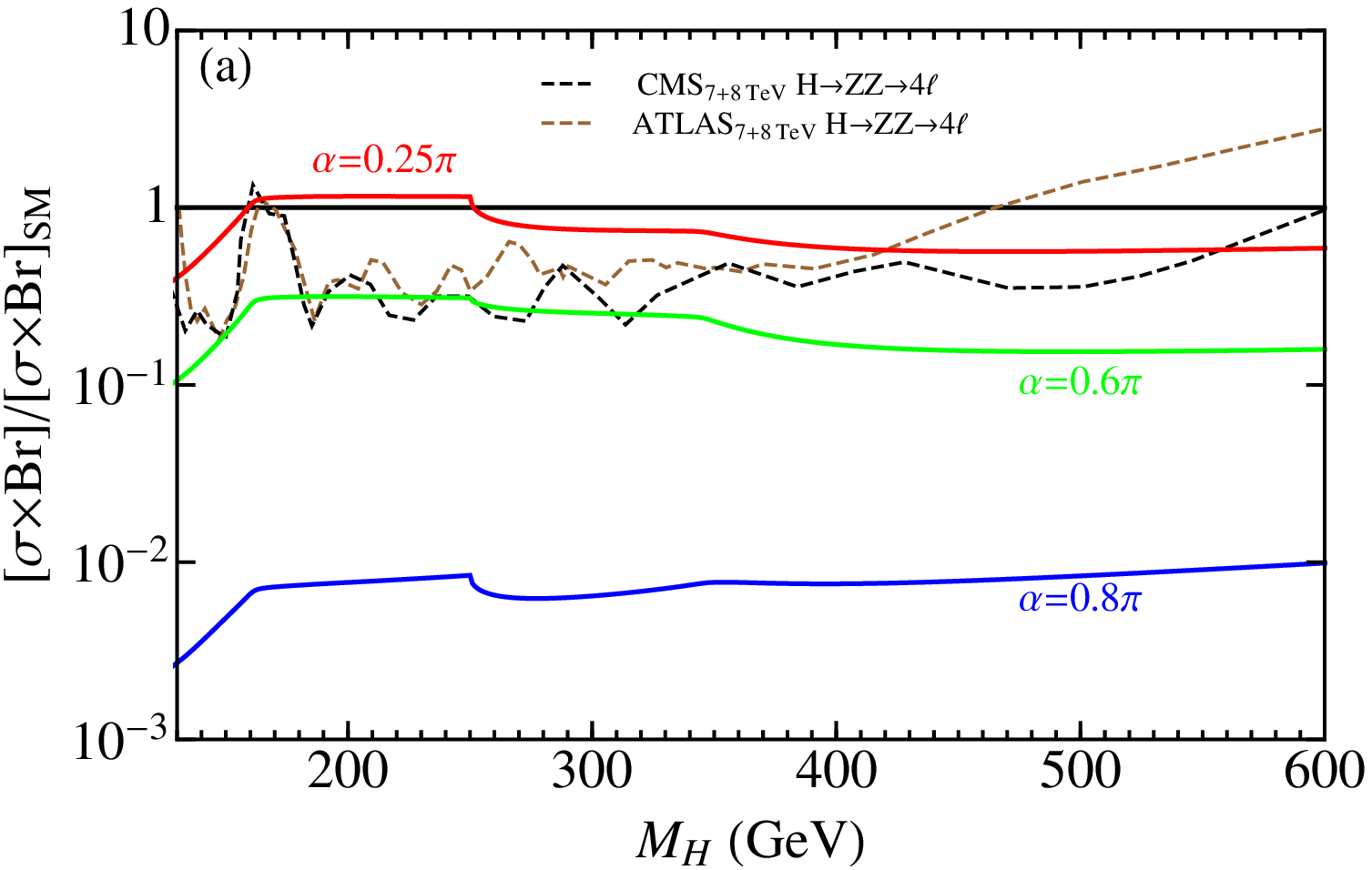}
\includegraphics[width=7.6cm,height=7.2cm]{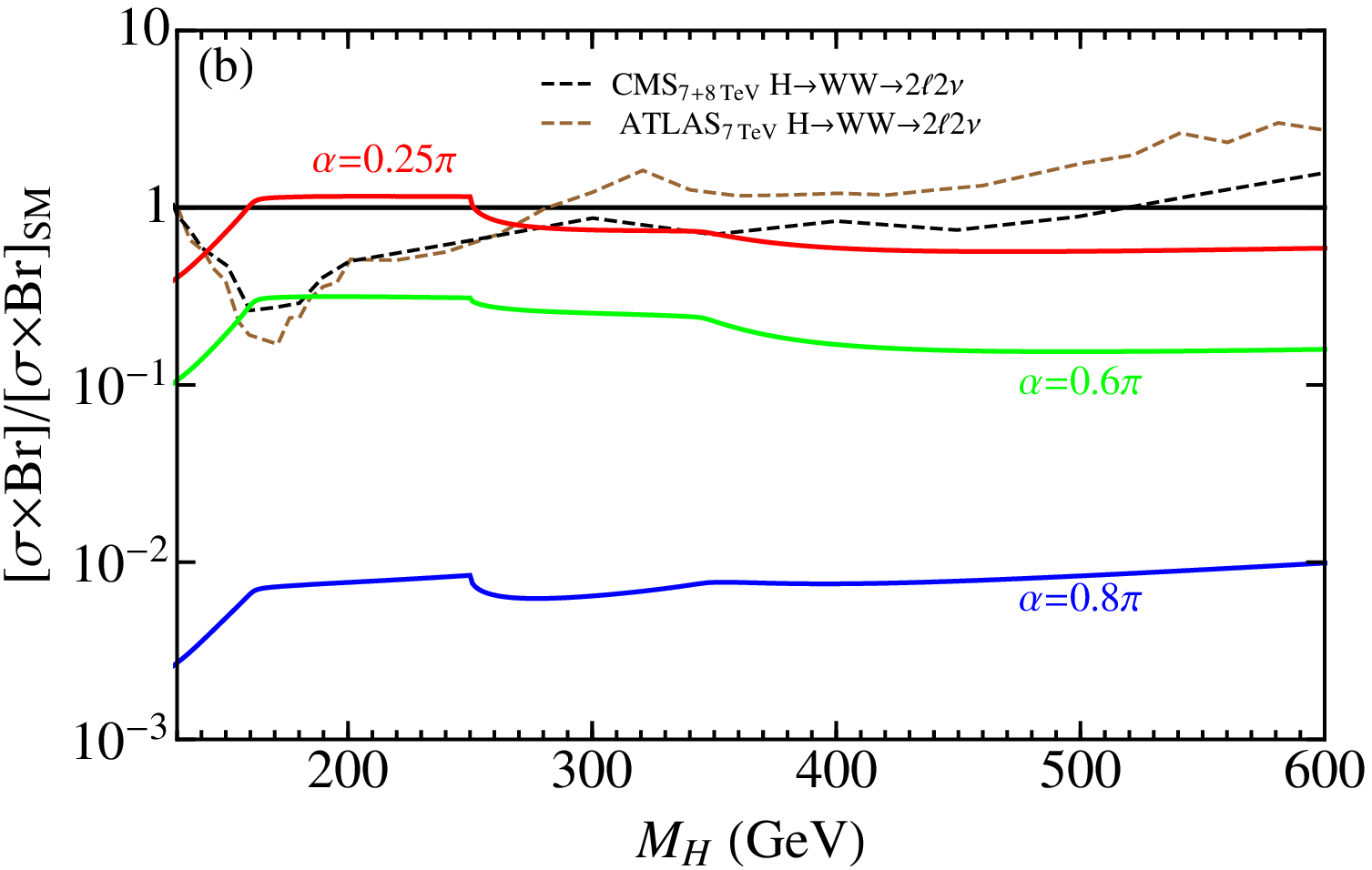}
\includegraphics[width=7.6cm,height=7.2cm]{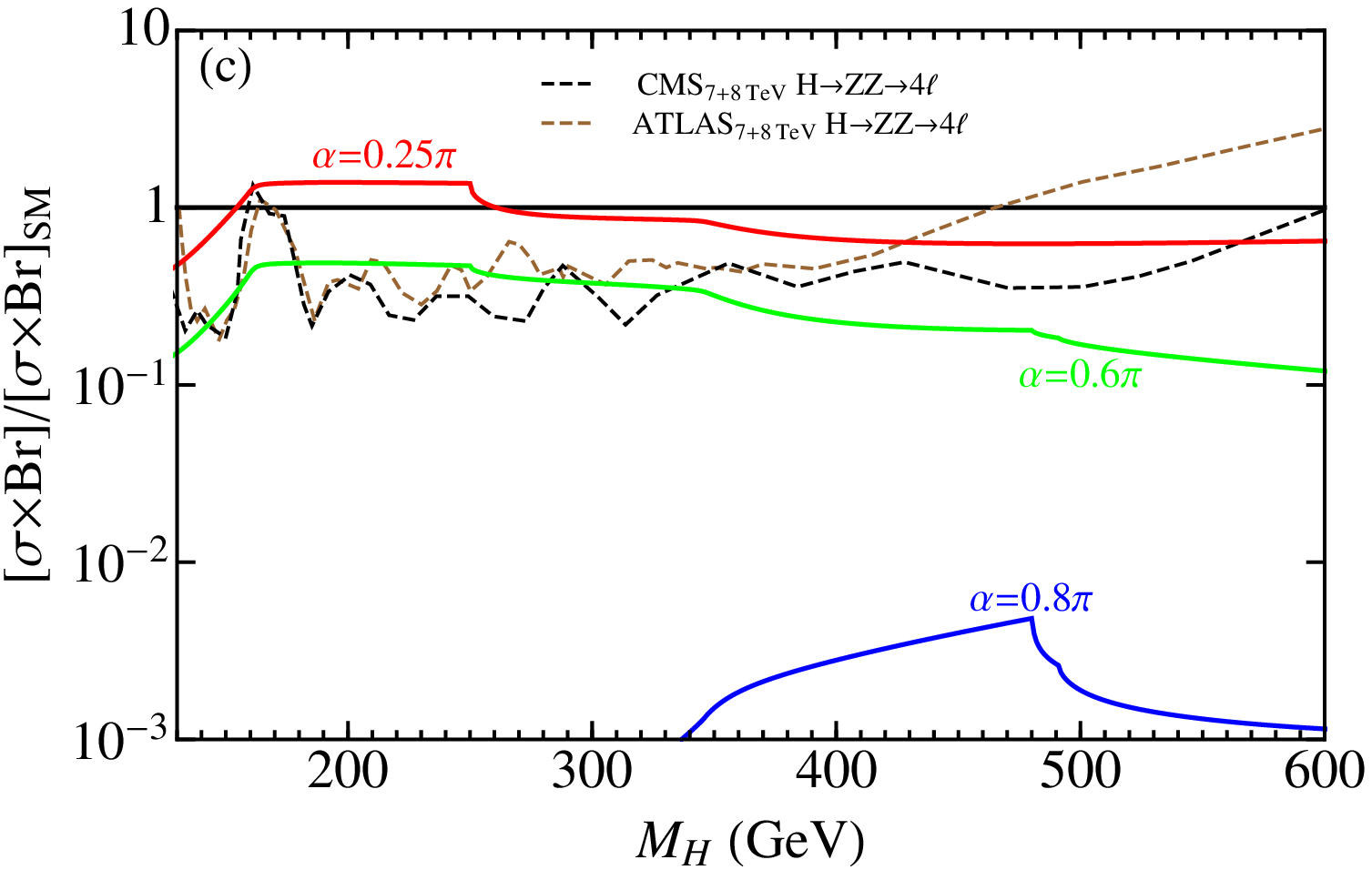}
\includegraphics[width=7.6cm,height=7.2cm]{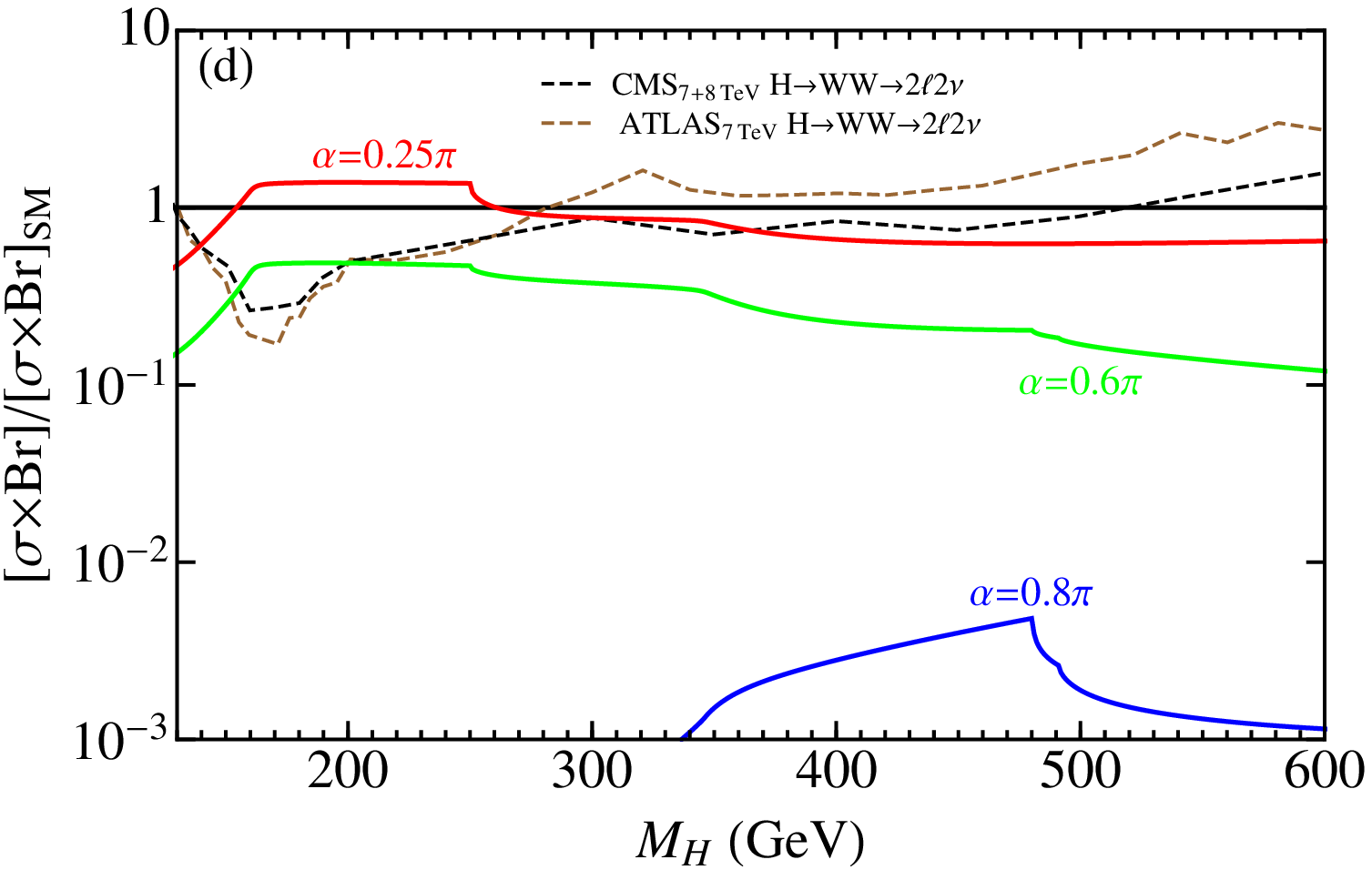}
\vspace*{-2mm}
\caption{Signal ratios $\,\mR_{ZZ}^{}\,$ and $\,\mR_{WW}^{}\,$
for the heavy Higgs boson searches at the LHC.
Plots (a) and (c) present the ratio $\,\mR_{ZZ}^{}\,$,
while plots (b) and (d) display the ratio $\,\mR_{WW}^{}\,$.\,
We have input the $W'$ mass $M_{W'}=600\,\GeV$ for plots (a)-(b), and
$M_{W'}=400\,\GeV$ for plots (c)-(d).
The other inputs are taken to be $\,r=f_2^{}/f_1^{}=1\,$,\, and $\,M_F^{}=2.5\,$TeV.\,
The blue, red, and green curves in each plot correspond to three different Higgs mixing angles,
$\,\alpha=(0.8\, \pi,\,0.6\,\pi,\, 0.25\,\pi)$,\, respectively.
The 95\%\,C.L.\ exclusion curves of ATLAS (brown dashed) and CMS (black dashed)
are imposed for comparison.}
\label{fig:Hsignalr1}
\end{center}
\end{figure}
\begin{figure}[]
\vspace*{8mm}
\begin{center}
\includegraphics[width=7.6cm,height=7.2cm]{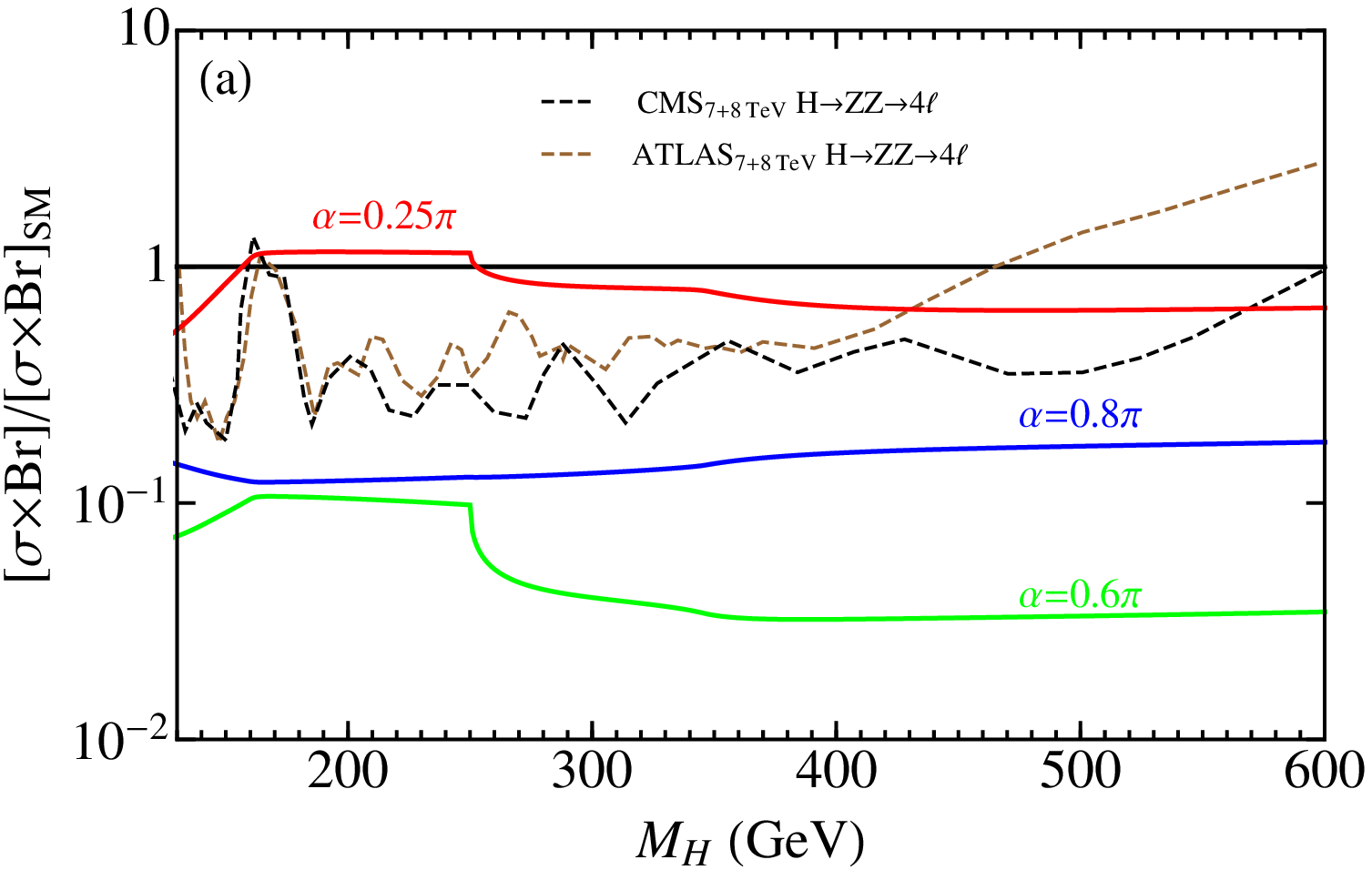}
\includegraphics[width=7.6cm,height=7.2cm]{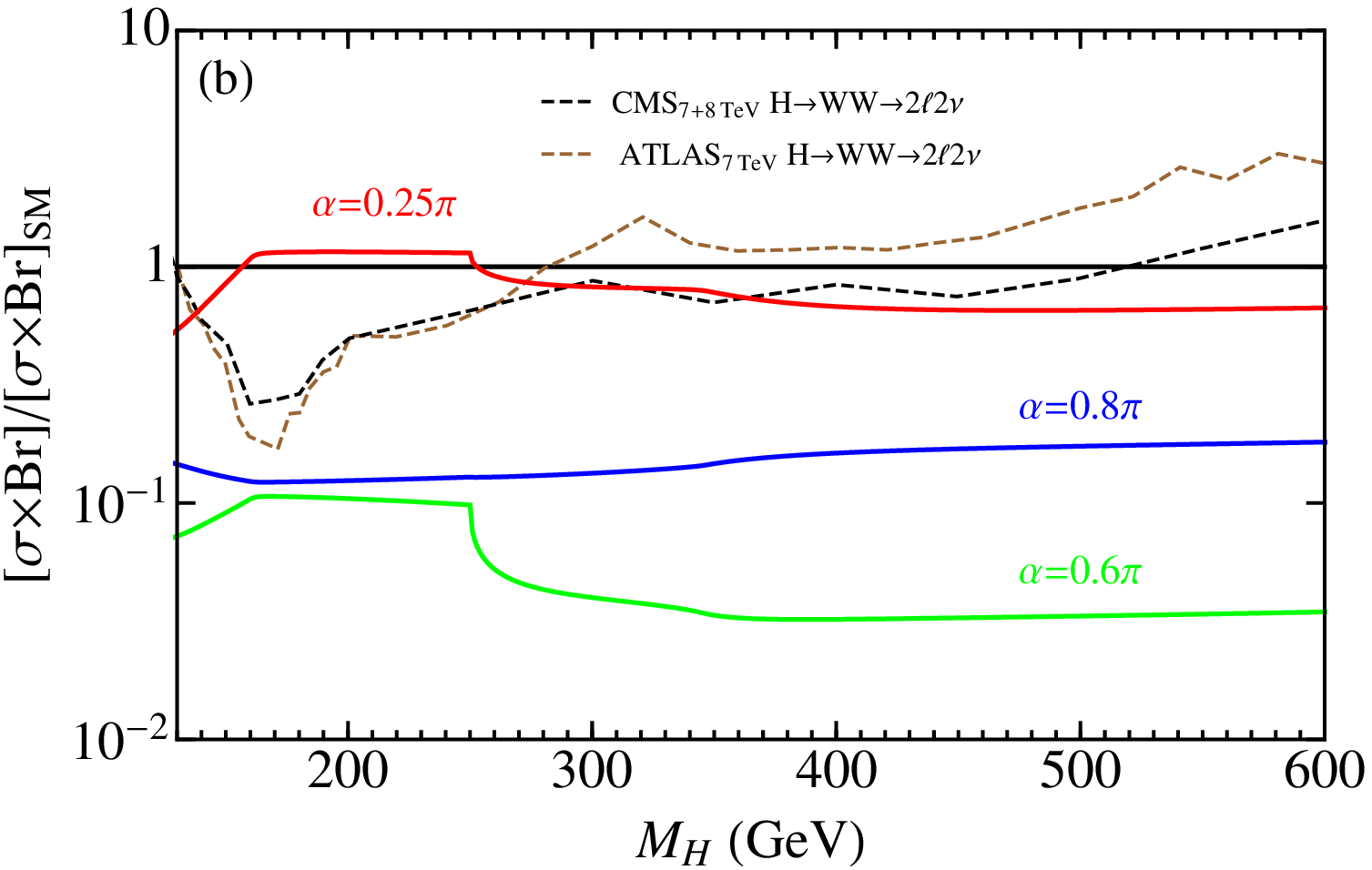}
\includegraphics[width=7.6cm,height=7.2cm]{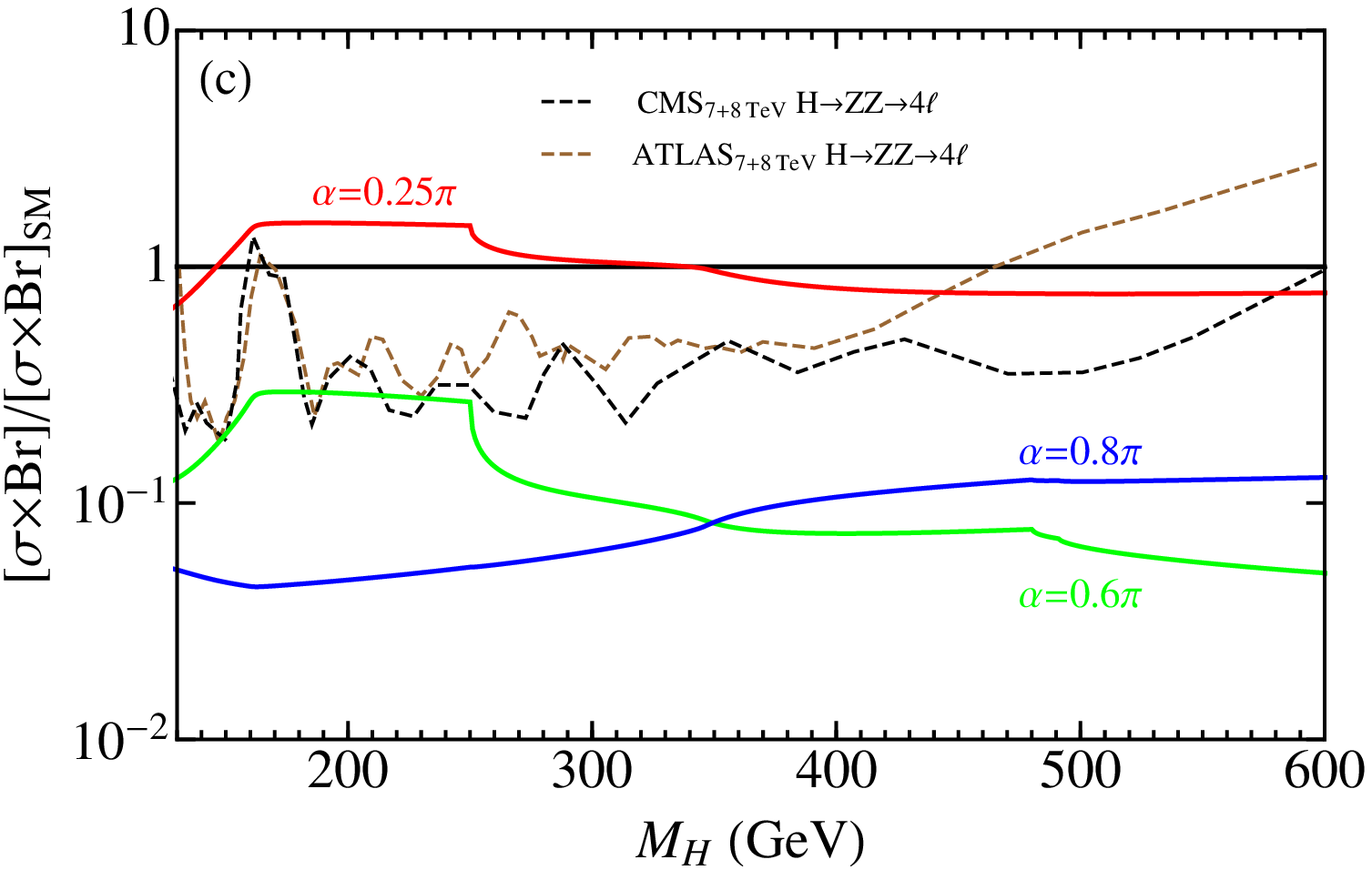}
\includegraphics[width=7.6cm,height=7.2cm]{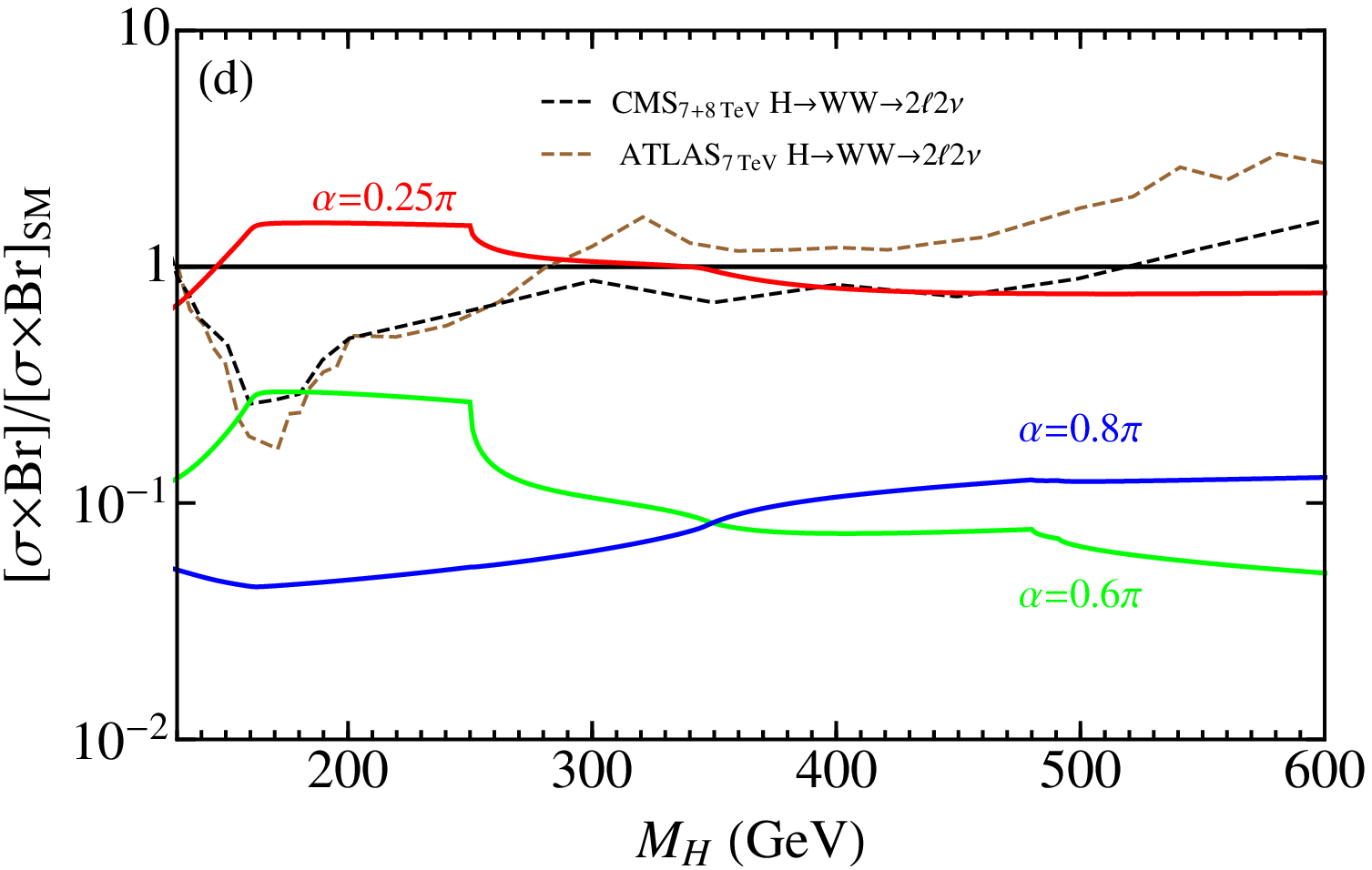}
\vspace*{-2mm}
\caption{Signal ratios $\,\mR_{ZZ}^{}\,$ and $\,\mR_{WW}^{}\,$
for the heavy Higgs boson searches at the LHC.
Plots (a) and (c) present the ratio $\,\mR_{ZZ}^{}\,$,
while plots (b) and (d) display the ratio $\,\mR_{WW}^{}\,$.\,
Different from Fig.\,\ref{fig:Hsignalr1}, we have input $\,r=f_2^{}/f_1^{}=1/2\,$
for all plots. The $W'$ mass is taken to be $M_{W'}=600\,\GeV$ for plots (a)-(b), and
$M_{W'}=400\,\GeV$ for plots (c)-(d). The heavy fermion mass is $\,M_F^{}=2.5\,$TeV.\,
The blue, red, and green curves in each plot correspond to three different Higgs mixing angles,
$\,\alpha=(0.8\, \pi,\,0.6\,\pi,\, 0.25\,\pi)$,\, respectively.
The 95\%\,C.L.\ exclusion curves of ATLAS (brown dashed) and CMS (black dashed)
are imposed for comparison.}
\label{fig:Hsignalr05}
\end{center}
\end{figure}

From Figs.\,\ref{fig:Hsignalr1}-\ref{fig:Hsignalr05}, we see
that both signal ratios $\,\mR_{ZZ}^{}$\, and $\,\mR_{WW}^{}$\,
decrease when the Higgs mixing angle increases from
$\,\alpha =0.25\,\pi$\, to $\,\alpha =0.8\,\pi$.\,
In Fig.\,\ref{fig:Hsignalr1} with $\,r=1$,\,
for the $h^0$ signals become maximal around $\,\alpha=0.8\,\pi$\, (Fig.\,\ref{fig:h125_sig}),
the corresponding signals of the heavier $H^0$ Higgs boson via the $(WW,\,ZZ)$
channels are much suppressed relative to the SM expectation, by a factor
of $\,\mR_{WW/ZZ} = \mO(10^{-2})$\, for $\,M_{W'}^{}=600\,$GeV, and
$\,\mR_{WW/ZZ} =\mO(10^{-2}-10^{-3})$\, for $\,M_{W'}^{}=400\,$GeV.
Thus, in this case the $H^0$ becomes hidden
and escapes the current LHC detections. On the other hand, the situation
gets changed for a mildly reduced Higgs mixing angle such as $\,\al = 0.6\pi\,$,\,
for which we derive the corresponding signal rates for the light Higgs boson $h^0$
from Fig.\,\ref{fig:h125_sig},
\beqs
\beqn
\mR_{\gamma\gamma}^{}=0.96\,, ~~~~ \mR_{ZZ}^{}=0.30\,, ~~~~ \mR_{WW}^{}=0.35\,,
&~~~~~& (\textrm{for}~M_{W'}=400\,\GeV),
\\[2mm]
\mR_{\gamma\gamma}^{}=0.82\,, ~~~~ \mR_{ZZ}^{}=0.25\,, ~~~~ \mR_{WW}^{}=0.27\,,
&~~~~~& (\textrm{for}~M_{W'}=600\,\GeV).
\eeqn
\eeqs
In these samples, the $\ga\ga$ rates are slightly lower than
the SM expectation, and the $ZZ^*$ and $WW^*$ signals are about a factor $1/3$ of the SM.
These are still consistent with the current ATLAS and CMS observations\,\cite{Atlas2012-7,CMS2012-7}
shown in Fig.\,\ref{fig:h125_sig} [Sec.\,\ref{sec3.2}].
In the meantime, the current experimental data from both ATLAS and CMS are
still insufficient to discover or exclude such a non-SM Higgs boson $H^0$
with $\,\alpha \gtrsim 0.6\,\pi$\,.\,
An increase of the integrated luminosity up to $\,\mL_{\rm tot}^{} = 40 - 60\,{\rm fb}^{-1}$\,
at the LHC\,(8\,TeV) for both detectors
should provide more effective probe of $\,H^0$\, in this case.
The next phase of the LHC\,(14\,TeV) with an integrated luminosity of
$\,100\,{\rm fb}^{-1}$\, will do a much better job for
detecting such a heavy Higgs boson $\,H^0$\,.

Furthermore,  Fig.\,\ref{fig:Hsignalr05} shows that
the suppressions on the $H^0\to WW,ZZ$ signals become moderate with $\,r=\hf\,$.\,
We find that the heavy Higgs boson signal predictions with $\,\alpha =0.8\,\pi$ are
much larger than the corresponding Fig.\,\ref{fig:Hsignalr1} ($\,r=1\,$),\,
and become larger than (or comparable to) the case of $\,\alpha=0.6\,\pi$ in
Fig.\,\ref{fig:Hsignalr05} ($\,r=\hf\,$).\,
Fig.\,\ref{fig:Hsignalr05} also shows that the signal rates with
$\,\alpha =(0.6-0.8)\,\pi$ are within a factor of $\mO(10)$ from
the current LHC exclusion limits via the
$\,H^0\to ZZ\to 4\ell$\, and $\,H^0\to WW\to 2\ell 2\nu$\, channels.
Hence, we anticipate exciting search results of $H^0$ from the LHC\,(8\,TeV)
after analyzing the full data sample of 2012. Much more sensitive probes of $H^0$
will be done in the next phase of the LHC running at the 14\,TeV collision energy.

Besides the major channels of $\,gg\to H^0\to WW,ZZ$,\, we also note that
the scalar channel $\,gg\to H^0\to h^0h^0\to (b\bar{b})(b\bar{b})\,$ is also
useful when $\,M_H^{} > 2M_h^{}\simeq 250\,$GeV.\,
Especially, the new CMS analysis shows that a $b$-tagging efficiency
of $\,70\%-85\%$\, can be realized \cite{b-tag-new}.  Thus, tagging the $4b$ final
state only reduces our signal rate by a factor of \,$(24-52)\%\simeq \fr{1}{4}-\fr{1}{2}$.\,
Such $4b$ final state is distinctive because we can require
each pair of energetic $b\bar{b}$ jets to reconstruct the $125$\,GeV resonance
of $h^0$ and the $4b$ jets to further reconstruct the heavy $H^0$ resonance.
This should effectively suppress the SM backgrounds.

In addition, from (\ref{eq:gWWpH}) we note that the $HVV'$ coupling
is enhanced by the factor $\,M_{V'}/v\,$.\,
Thus, for $H^0$ in the mass-range $\,M_H^{} > M_{V'}^{}+M_{V}\,$,\,
the decay channels $\,H^0\to WW',\,ZZ'\,$ become dominant,
as shown in Figs.\,\ref{fig:HBR600}-\ref{fig:HBR400}.
For the mass-range $\,M_H^{} > M_{V'}^{}+M_{V}^{}\,$,\,
it is useful to search for the decay channels
$\,H\to W'W,Z'Z\to Z^0W^+W^- \to 3\ell+jj+ /{\hspace*{-2.5mm}}E_T^{}\,$,\,
with distinctive tri-lepton signals plus dijets and missing $E_T^{}$,\,
where we have $\,Z\to \ell^+\ell^-\,$ and $\,WW\to (\ell\,\nu)(jj)\,$
with $\,\ell = e,\mu\,$.\, Finally, we expect that more sensitive detections of
$\,H^0\,$ should come from next runs of the LHC at the 14\,TeV collision energy
and with much higher integrated luminosities around $100-300\,\ifb$.
The future Linear Colliders (either ILC or CLIC) should help to make
further precision probes of the $H^0$ signals.

Finally, we note that due to the ideal fermion delocalization\,\cite{iDLF},
the $W'$ couplings to the SM fermions vanish and $Z'$ couplings
with the light fermions are suppressed by $\,m_W^{}/M_{W'}^{}\,$.\,
Thus, the major decay modes of $(W',\,Z')$ gauge bosons are,
$\,W'\to WZ\,$ and $\,Z'\to WW\,$, or $\,W'\to Wh\,$ and $\,Z'\to Zh\,$\,.
The partial decay widths of $\,V'\to V_1V_2\,$,\,
with $V_1V_2=WZ\,(WW)$ for $V'=W'\,(Z')$ are derived as follows,
%
\beqa
\Gamma[V' \!\to\! V_1V_2] &=&
\frac{\,G_{V'V_1V_2}^2 M_{V'}^3\,}{192 \pi m_1^2}\!
\left[\frac{M_{V'}^2+10m_{12}^2}{m_{2}^2}
      +\frac{m_{12}^4+8m_1^2 m_2^2}{M_{V'}^2 m_2^2} \right]\!
\left[\!\(\!1 \!-\frac{m_+^2}{M_{V'}^2}\)\!
      \(\!1 \!-\frac{m_-^2}{M_{V'}^2}\)\!\right]^{\f{3}{2}} \!,~~~~~
\nn\\
&& \hspace*{-10mm}
\label{eq:WptoWZ_wid}
\eeqa
with the masses $\,(m_1^{},\,m_2^{}) = (m_{V_1}^{},\,m_{V_2}^{})$,\,
$\,m_\pm^{}=m_1^{}\pm m_2^{}\,$,\, and
$\,m_{12}^2=m_1^2+ m_2^2\,$.\,
For the other partial decay width of $\,V'\to Vh\,$
with $V=W\,(Z)$ for $V'=W'\,(Z')$, we deduce the following,
\beqa
\Gamma[V' \!\to\! Vh] &=&  \frac{\,\xi_{hVV'}^2 m_V^2 M_{V'}^{}\,}{12\pi v^2}
\!\left[ 2+\frac{\,(M_{V'}^2\!+\!m_V^2\!-\!M_h^2)^2\,}{4m_V^2 M_{V'}^2}  \right]\!
\left[1+\frac{\,(M_h^2\!-\!m_V^2)^2\,}{M_{V'}^4}-2\frac{\,m_V^2\!+\! M_h^2\,}{M_{V'}^2}
\right]^{\hf}\! .~~~~~
\nn\\
&& \hspace*{-10mm}
\label{eq:WptoWh_wid}
\eeqa
%
Here the gauge couplings $\,G_{V'V_1V_2}=G_{W'WZ},G_{Z'WW}$\,
and the Higgs-gauge couplings
$\,\xi_{hVV'}^{}=\xi_{hWW'}^{}$, $\xi^{}_{hZZ'}$\, are given by
(\ref{eq:VVV}) and (\ref{eq:hVV-hFF}), respectively.

\begin{figure}[]
\vspace*{8mm}
\begin{center}
\includegraphics[width=7.4cm,height=7cm]{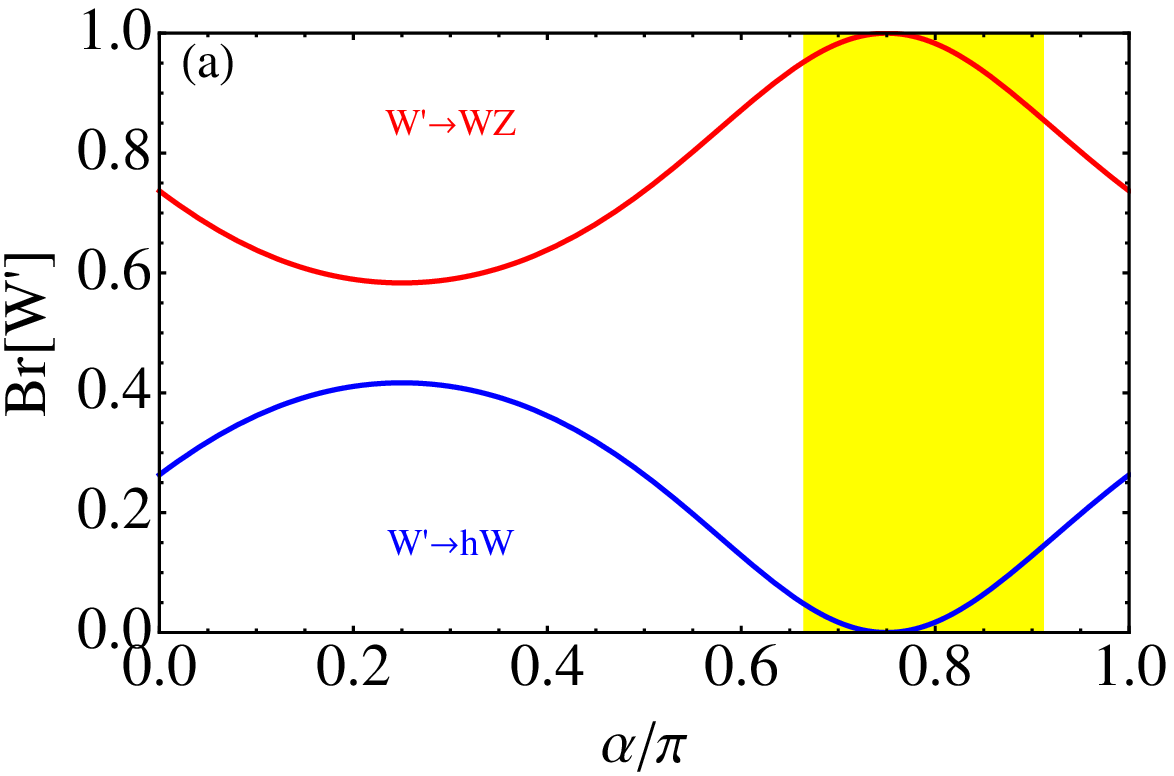}
\includegraphics[width=7.4cm,height=7cm]{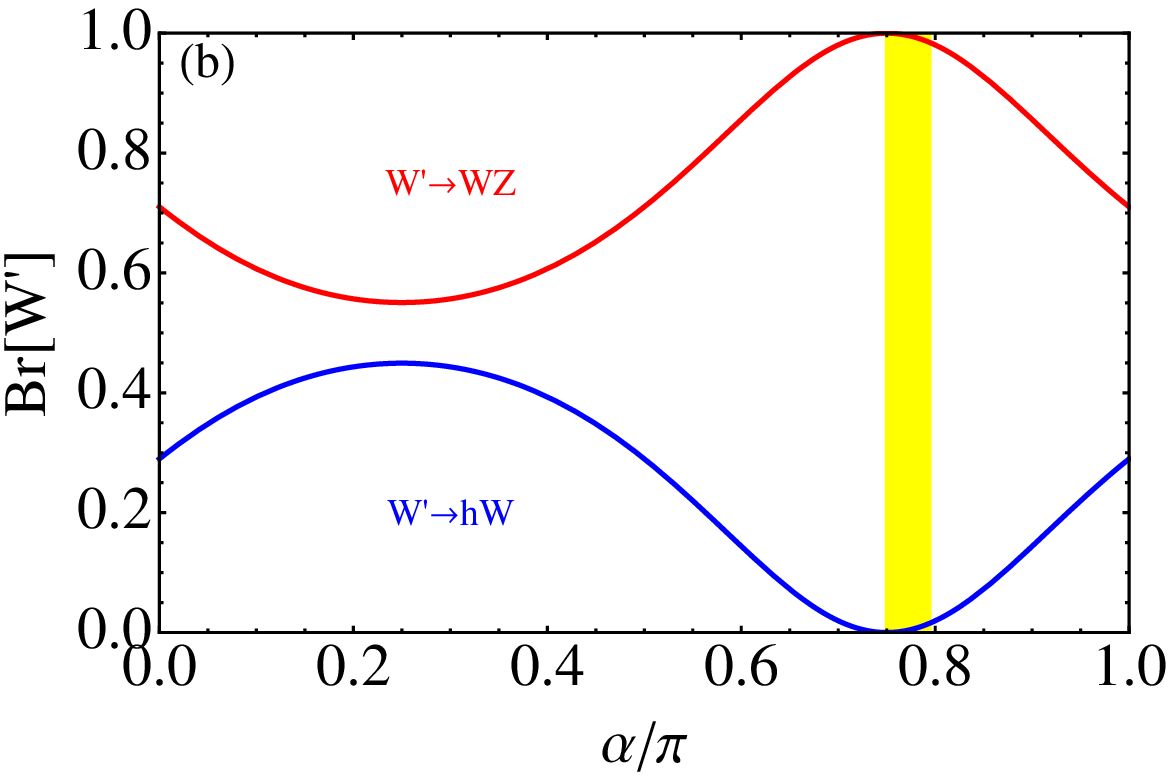}
\includegraphics[width=7.4cm,height=7cm]{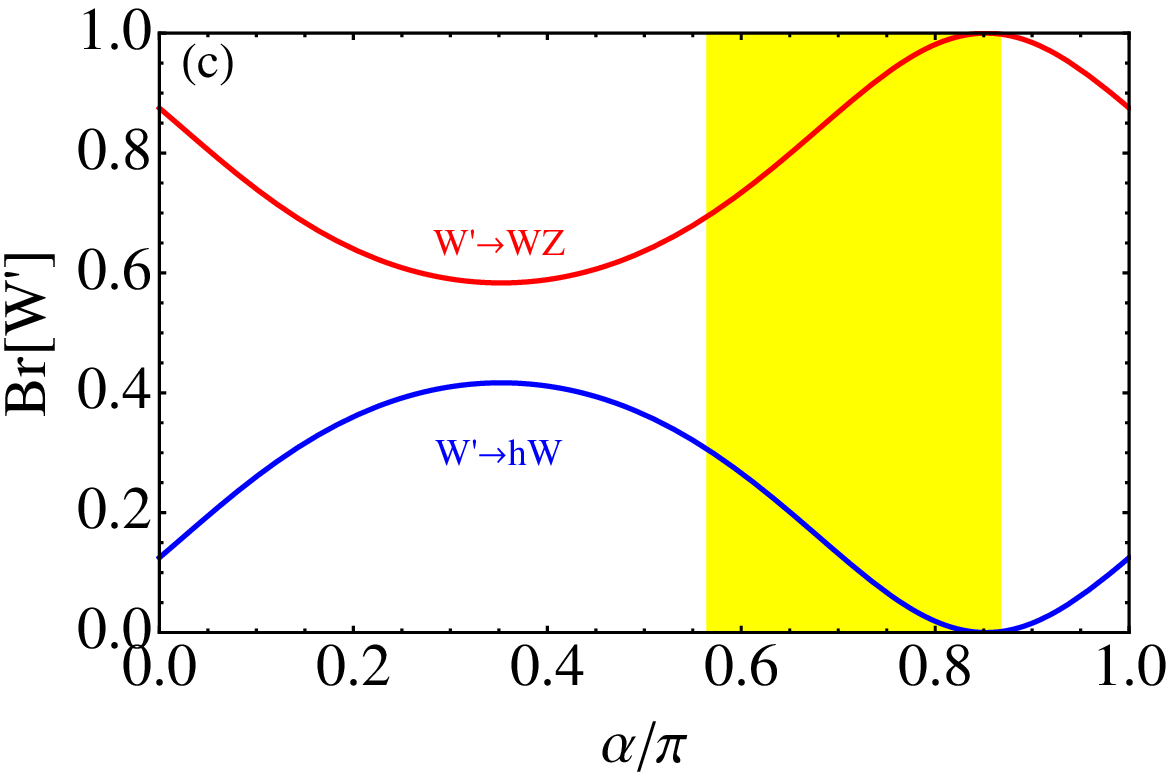}
\includegraphics[width=7.4cm,height=7cm]{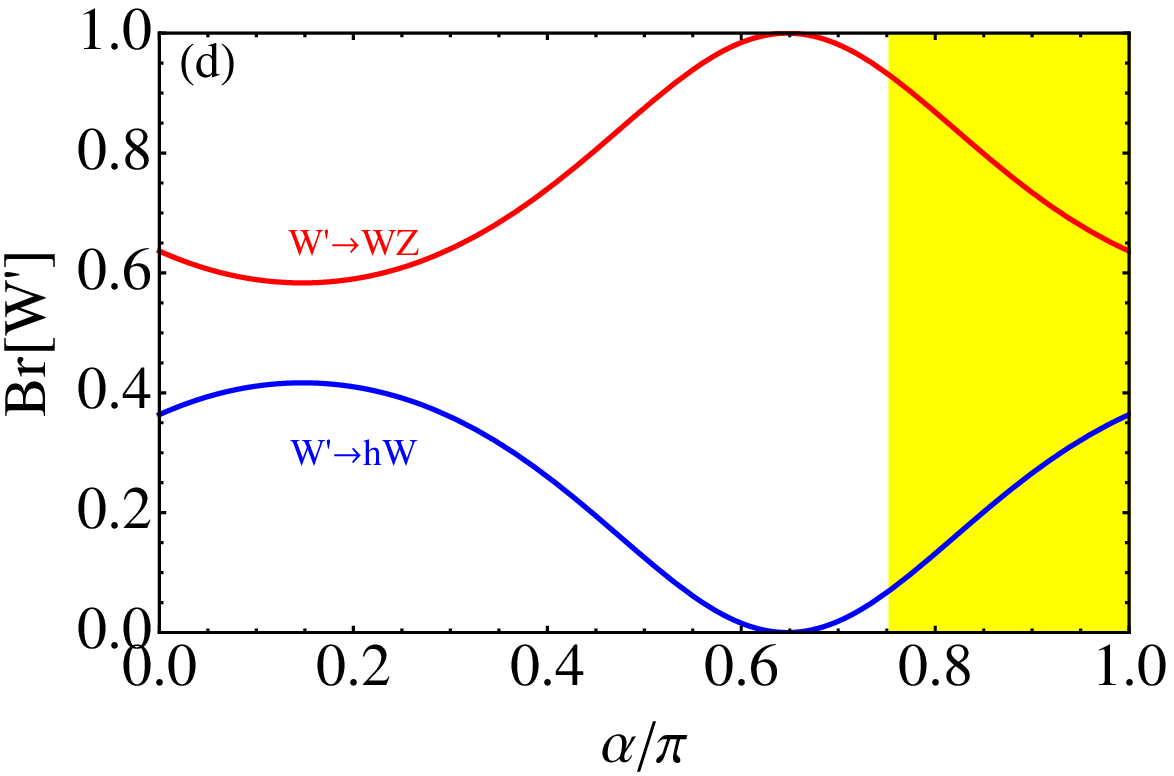}
\vspace*{-2mm}
\caption{Decay branching fractions of $W'$ versus Higgs mixing angle $\alpha$ for
four sample inputs:
$\,(r\,,M_{W'})=(1\,,400\,\GeV)\,$ [plot-(a)],
$\,(r\,,M_{W'})=(1\,,600\,\GeV)$ [plot-(b)],
$\,(r\,,M_{W'})=(\hf\,,400\,\GeV)$ [plot-(c)],
and $(r\,,M_{W'})=(2\,,400\,\GeV)$ [plot-(d)].
The light Higgs boson mass $\,M_h=125\,\GeV\,$.
The yellow strips correspond to the allowed $\alpha$ ranges
by the consistency of our predicted $\,h^0\to\ga\ga$\, signals
with the ATLAS/CMS data within $\pm 1\sigma$\,. }
\label{fig:BRWp}
\end{center}
\end{figure}

Taking into account these two major channels,
we present the decay branching fractions of $W'$ in Fig.\,\ref{fig:BRWp}
by varying Higgs mixing angle $\,\alpha\in [0,\,\pi)$.
It turns out that $\,{\rm Br}[W'\to WZ]>{\rm Br}[W'\to Wh]$\,
always holds over the full range of parameter $\,\alpha$\,.\,
In particular, we marked the ranges of the mixing angle $\alpha$ in yellow strips
where the corresponding $\,h^0\to\ga\ga$\, signal predictions
(as shown in Fig.\,\ref{fig:h125_sig}-\ref{fig:h125-B})
are consistent with the ATLAS and CMS measurements within $\pm 1\sigma$.
For such preferred parameter region of $\,\alpha$\,,\,
we see that the decay branching fraction
$\,{\rm Br}[W'\to Wh]$\, is significantly suppressed relative to
$\,{\rm Br}[W'\to WZ]$.\,
The situation for $\,{\rm Br}[Z'\to Zh]$\, versus $\,{\rm Br}[Z'\to WW]$\,
is similar.


\vspace*{3mm}
\section{\hspace*{-2mm}Conclusions}
\label{sec5}

During the finalization of this paper, it is our pleasure
to learn the LHC announcement on July\,4, 2012
\cite{Atlas2012-7,CMS2012-7,:2012gk,:2012gu}
of finding a Higgs-like new particle around $125-126$\,GeV.
This has created great excitements to search for new physics of
the Higgs sector and the origin of the electroweak symmetry breaking.
Some other interesting theoretical interpretations have appeared
very recently \cite{others}.

In this work, we have systematically studied the Higgs sector of a simple
$\,\gSU(2) \otimes \gSU(2)\otimes \gU(1)\,$ gauge model (the 221 model),
which has ideal fermion delocalization\,\cite{iDLF} and thus allows relatively light
new gauge bosons $\,(W',\,Z')\,$ below 1\,TeV scale to participate in the unitarization
of longitudinal $\,V_L^{}V_L^{}\,$ scattering ($V=W,Z$) together with the Higgs bosons
$\,(h,\,H)$.\, This may be viewed as an effective UV completion of the Higgsless
3-site nonlinear moose model\,\cite{3site}.
In Sec.\,\ref{sec2}, we analyzed the structure of the model and derived
all relevant Higgs, gauge and fermion couplings.
Then, we derived the general sum rules (\ref{eq:SR})-(\ref{eq:SR-2})
for the exact cancellation of asymptotic $E^2$ amplitudes of the longitudinal
$\,V_L^{}V_L^{}\,$ scattering at high energies.
From this we revealed that
{\it such $E^2$ cancellations are achieved by the joint role of
exchanging both spin-1 new gauge bosons and spin-0 Higgs bosons.}\,
This was explicitly demonstrated in Fig.\,\ref{fig:wwzzamp}-\ref{fig:wzwzamp}.
We further derived the unitarity bound on the mass of the heavier Higgs state $H^0$
(Fig.\,\ref{fig:MH}) when the lighter Higgs boson $h^0$ weighs about $125-126$\,GeV.

In Sec.\,\ref{sec3}-\ref{sec4}, we presented systematical analyses of
the LHC phenomenology of the 221 model, focusing on the Higgs signatures
in connection with the $125-126$\,GeV resonance $\,h^0\,$
and the probe of additional new Higgs state $\,H^0\,$ beyond the SM.
We identified the lighter Higgs state $h^0$ with the LHC observed mass 125\,GeV
and set the Higgs VEV ratio $\,r=\fy /\fx = \mO(1)\,$ as input.
The Higgs sector only contains two free parameters, namely,
the heavier Higgs boson mass $\,M_H^{}$\, and the Higgs mixing angle $\,\al\,$.\,
Hence, the parameter space of this model is highly predictive.

In Sec.\,\ref{sec3}, we analyzed the production of $\,h^0$\, via gluon fusions
and its decays via the most sensitive channels of $\,\ga\ga\,$,\, $ZZ^*$\,
and $\,WW^*\,$,\, as shown in Figs.\,\ref{fig:Br3site}-\ref{fig:xsec}.
The $\,h^0\,$ signal rates over the SM expectation
are depicted in Figs.\,\ref{fig:h125_sig}-\ref{fig:h125-B} via
$\,\ga\ga\,$,\, $ZZ^*$,\, and $\,WW^*\,$  channels.
We found that due to new contributions of the heavy vector-like quarks
in the current model the $\,gg\to h^0$\, production rates can get enhanced
for proper ranges of the Higgs mixing angle $\,\al$\, (cf.\ Fig.\,\ref{fig:xsec}).
The maximal enhancements of the $\ga\ga$ rates are about a factor of
$\,1.4-1.6$\, of the SM expectations, depending on the Higgs mixing angle $\,\al\,$
and the $W'$ mass. The signal rates in the $ZZ^*$ and $\,WW^*$ channels
are generally suppressed and can reach up to about $\,1/2\,$ of the SM values
for $\,r=1\,$,\, and about a factor $\,0.8-1$\, of the SM for $\,r=\hf,2\,$.\,
These are compared with the current measurements of ATLAS\,\cite{Atlas2012-7}
and CMS\,\cite{CMS2012-7}, and are found to reach good agreements.
We further studied an interesting case with (nearly) degenerate $h^0$ and $H^0$
Higgs bosons around $125-126$\,GeV. As shown in Figs.\,\ref{fig:h125_sig}-\ref{fig:h125-B}
and Eqs.\,(\ref{eq:deg-r-1})-(\ref{eq:deg-r2-05-2}),
we found that the predicted maximal $\,\ga\ga\,$ rates are higher than the SM by about
$\,15\%-37\%\,$ for $\,r=1$\, and $\,27\%-39\%\,$ for $\,r=\hf,2\,$,\,
while the $\,ZZ^*$ and $\,WW^*\,$ signals can raise to about
$\,\f{1}{3}-\f{1}{2}\,$ of the SM expectations for $\,r=1$\,
and $\,58\%-93\%\,$ of the SM for $\,r=\hf,2\,$.\,
They also agree with the latest LHC findings\,\cite{Atlas2012-7,CMS2012-7}.
In addition, we studied the signal rates of $\,h^0$\, via associate production
and vector boson fusion channels, as shown in Fig.\,\ref{fig:xsecRatio_AP_400}.
It turns out that the predicted signals are generally suppressed by the tree-level
$\,hVV\,$ and $\,hff\,$ couplings. For the VEV ratio $\,f_2^{}/f_1^{}\,$
varying within $\,\mO(1)\,$,\, we found that the $\,h^0$\, signal rates
can reach up to about $\,\f{1}{3}-\f{3}{4}\,$ of the SM results.
Hence, a combined analysis of $\,h^0\,$ signals
in all three types of processes (including gluon fusions, $Vh$ associate productions
and vector boson fusions) will help to discriminate our $\,h^0\,$
Higgs boson from the SM.

In Sec.\,\ref{sec4}, we further analyzed the decays and productions of
the heavier Higgs boson $\,H^0$\, at the LHC. The detection of such a second
new heavier $\,H^0$\, state is important for discriminating the present model
from the SM. The two major channels are
$\,gg\to H^0\to ZZ\to 4\ell\,$ and
$\,gg\to H^0\to WW\to 2\ell2\nu\,$.\,
The signal rates of $\,H^0\,$ over the SM expectations are shown in
Figs.\,\ref{fig:Hsignalr1}-\ref{fig:Hsignalr05}.
The current search limits of ATLAS and CMS already start to probe
$\,H^0$\, via the $\,ZZ\,$ and $\,WW$\, channels for the Higgs mixing angle
$\,\al \lesssim 0.6\pi\,$ and mass-range below about 340\,GeV.
For the mixing angle in the $\,\al \sim (0.6-0.8)\pi\,$
range, the $H^0$ signal rates via $ZZ$ and $WW$ channels become lower
and will require higher integrated luminosities at the LHC\,(14\,TeV).
But, in the mass-range $\,M_H^{} > 250\,$GeV, the decay mode $\,H^0\to h^0h^0\,$
may be sizable for $\,\al \sim 0.8\pi\,$ [Fig.\,\ref{fig:HBR600}(b)(d) with $\,r=1\,$]
or for $\,\al \sim 0.6\pi\,$ [Fig.\,\ref{fig:HBR400}(a)(c) with $\,r=\hf\,$].
Thus, the scalar channel $\,gg\to H^0\to h^0h^0\to (b\bar{b})(b\bar{b})\,$
will be useful since it gives rise to the distinctive $4b$ final state with each pair
of $b\bar{b}$ dijets reconstructing the 125\,GeV resonance of $h^0$.
Furthermore, we note that in the mass-range
$\,M_{H}^{}> M_{W'}^{}+m_{W}^{}\,$,\,
the $H^0$ Higgs state opens up new decay channels of
$\,H^0\to W'W,Z'Z\to Z^0W^+W^- \to 3\ell+jj+ /{\hspace*{-2.5mm}}E_T^{}\,$,\,
with distinctive tri-lepton signals plus dijets and missing $E_T^{}$,\,
where $\,Z\to \ell^+\ell^-\,$ and $\,WW\to (\ell\,\nu)(jj)\,$
with $\,\ell = e,\mu\,$.\,
The $\,H^0\to W'W/Z'Z$\, modes may become the dominant decays
for $\,M_{H}^{}> M_{W'}^{}+m_{W}^{}\,$,\,
as shown in Fig.\,\ref{fig:HBR600}(c,d) and Fig.\,\ref{fig:HBR400}(c,d).
This is useful to give further probes of $\,H^0$\, at the LHC.

\addcontentsline{toc}{section}{Acknowledgments\,}
\vspace*{2mm}
\section*{Acknowledgments}
\vspace*{-2mm}

We thank R.\ Sekhar Chivukula, Neil Christensen, and Masaharu Tanabashi
for useful discussions. This research was supported by the NSF of China
(under grants 11275101, 10625522, 10635030, 11135003) and
the National Basic Research Program of China (under grant 2010CB 833000).
HJH thanks CERN Theory Division for hospitality during the finalization of this paper.


\vspace*{8mm}
\appendix

\noindent
{\bf\large Appendices}

\section{\hspace*{-1mm}Decays of the Lighter Higgs Boson}
\label{appA}

In this Appendix\,A, we analyze all decay modes for the lighter Higgs boson
$\,h^0\,$ in the present model, which will differ from the SM Higgs boson due to
its non-standard couplings with gauge bosons and fermions shown in Sec.\,\ref{sec2}.
There are three types of decay modes for $\,h^0\,$,\, namely,
(i) the weak gauge bosons $\,WW/ZZ$\,,\,
(ii) the SM fermions, and
(iii) the loop-induced radiative decay modes of $\,(gg,\, \gamma\gamma,\, Z\gamma)\,$.

The partial decay widths for $\,h\to VV$\, ($V=W,Z$) final states differ from
that of the SM Higgs boson by a common coupling factor according to Eq.\,(\ref{eq:VVh}),
\beqn
\label{eq:hVVwid}
&&
\frac{\Gamma[h \to VV]~}{~~\Gamma[h \to VV]_{\rm SM}~}
~\simeq~ \xi_{hVV}^2\,\leqq 1\,.
\eeqn
This shows that the partial decay widths of $\,h^0\,$ to the weak gauge bosons
tend to be smaller than the SM values.
For $h^0$ decays into the SM fermions, the partial decay widths differ
from the SM values by the corresponding Yukawa coupling factor
as in Eq.\,(\ref{eq:tth/H}),
\beqn
\label{eq:hffwid}
\frac{~\Gamma[h\to f \bar f]~~~}{~~\Gamma[h\to f \bar f]_{\rm SM}~}
~&\simeq&~ \xi_{hff}^2 \,\leqq 1\,,
\eeqn
where $\,f\,$ stands for a given SM fermion.

For the type (iii) loop-induced decay channels, there are two sources of
the differences from the SM Higgs boson.
One is the non-standard $h^0$ couplings to the SM particles,
such as the $hVV$ couplings in (\ref{eq:VVh}) and $hff$ couplings in (\ref{eq:tth/H}).
The other source is the existence of extra new states, which enter the loop contributions.
For the $\,h\to gg$\, decays shown in Fig.\,\ref{fig:h2gg}, besides the top-quark,
six new heavy vector-like partners of the SM quarks will also contribute to the loop.
The decay width of $\,h\to gg$\, differs the corresponding SM value by the ratio,
\beqn
\label{eq:hggwid}
&&
\frac{~\Gamma[h\to gg]\,~~~}{~~\Gamma[h\to gg]_{\rm SM}^{}~}
~=~
\frac{~\displaystyle\left| \sum_{f=t, Q} \xi_{hff}^{} A_{1/2}^H(\tau_f^{}) \right|^2
}{\displaystyle \left| \sum_{f= t} A_{1/2}^H(\tau_f) \right|^2}\,,
\eeqn
with $\,\tau_f^{}\equiv m_h^2/(4m_f^2)$\,.\,
The fermion-loop form factor is given by
\beqs
\beqn
&&
A_{1/2}^H(\tau)\,=\,2[\tau+(\tau-1)f(\tau)]\tau^{-2}\,,
\\[2mm]
&&
f(\tau)= \begin{cases}
~ \arcsin^2\!\sqrt{\tau} \,, \qquad & \tau\leqq 1\,,
\\[2.5mm]
~ -\frac{1}{4}\!
\left[\dis\ln\frac{1+\sqrt{1-\tau^{-1}}}{1-\sqrt{1-\tau^{-1}}}-i\pi\right]^2\!.~~
\qquad & \tau>1 \,.
\end{cases}
\eeqn
\eeqs
In the heavy fermion mass limit  $\,M_h^2\ll 4m_f^2$\,,\,
the asymptotic behavior is, $\,A_{1/2}^H(\tau)\to 4/3$\,.\,
So, the loop contributions from the extra heavy quarks are comparable to that
of the top-quark loop, and thus will enhance the Higgs production via
the gluon fusion process. On the other hand, the coupling factors
of the Higgs boson with the extra heavy fermions (\ref{eq:fFFh}) become,
$\,\xi_{hQQ}^{} = \mO(m_W^2/M_{W'}^2)\ll 1$\,.\,
Thus, the contributions from the heavy fermions only give moderate corrections
to the Higgs productions, without drastic enhancement.
This feature differs from other extensions, such as the fourth family SM fermions.

\begin{figure}[t]
\begin{center}
\includegraphics[width=6cm,height=3cm]{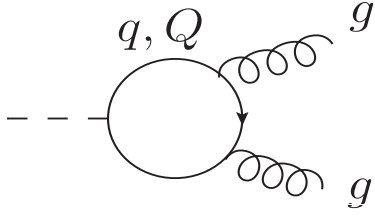}
\caption{Loop-induced radiative decays of the light CP-even
Higgs boson $h^0$ into gluon pairs, which include loop contributions
from both the SM quarks $q$ and new vector-like quarks $Q$ in the present model.}
\label{fig:h2gg}
\end{center}
\end{figure}

The modification to the $\,\Gamma[h\to \gamma\gamma]$\, from the SM case
is more complicated, since both fermion-loops and gauge-boson-loops enter this decay channel,
as shown in Fig.\,\ref{fig:h2gaga}. Schematically, the non-standard di-photon
partial width of $h^0$ is expressed as follows at the leading order,
\beqs
\beqn
\frac{~\Gamma[h \to \gamma \gamma]~~~}
     {~~\Gamma[h \to \gamma \gamma]_{\rm SM}~}
~&=&~
\frac{~\displaystyle\left|\sum_{f=t,Q} N_{c,f}^{}Q_f^2
      \xi_{hff}^{} A_{1/2}^H(\tau_f^{}) \,+\!\!
      \sum_{V=W\,, W'}\xi_{hVV}^{} A_1^H(\tau_V^{})\right|^2}
     {\displaystyle\left|\sum_{f=t} N_{c, f}^{}Q_f^2 A_{1/2}^H(\tau_f^{})
      + A_1^H(\tau_W^{}) \right|^2}\,,
\hspace*{10mm}
\label{eq:hgagawid}
\\
A_1^H(\tau) &=& -[2\tau^2+3\tau+3(2\tau-1)f(\tau)]\tau^{-2}\,,
\eeqn
\eeqs
where $N_{c, f} =3\,(1)$ for colored (uncolored) states.
$Q_f^{}$ denotes the electric charge for each fermion.
$\,A_1^H(\tau)$\, is the form factor of vector boson loop,
with $\,\tau_V^{}\equiv M_h^2/(4M_V^2)$\,.\,
For the $\,h\to\gamma\gamma$\, decay channel, there are both
$W$-loops and $W'$-loops contributing to the partial decay width.
The heavy vector boson limit of $\,M_h^2\ll 4M_V^2$\, gives
the asymptotic behavior $\,A_1^H(\tau)\to -7$\,.\,
Therefore, the contributions from the heavy gauge boson $W'$-loops
can be compatible with the usual $W$-loop contribution.
In most parameter space of this model, the pre-factor
$\,\xi_{hW'W'}^{}\,$ is not really suppressed, so
the $W'$-loop contributions should be retained.
Altogether, the $\,\Gamma[h \to \gamma \gamma]\,$ in this model
can be comparable to that of the SM Higgs,
even though $\,G_{hWW}^{}\leqq G_{hWW}^{\rm SM}$\, always holds.
This feature differs from the case of $\,\Gamma[h\to WW/ZZ]$\,,\,
which is always lower than the SM values.

\begin{figure}[t]
\begin{center}
\includegraphics[width=4cm,height=2.5cm]{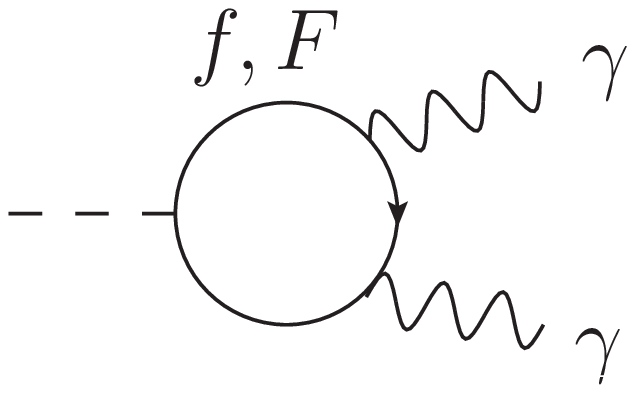}
\includegraphics[width=4cm,height=2.5cm]{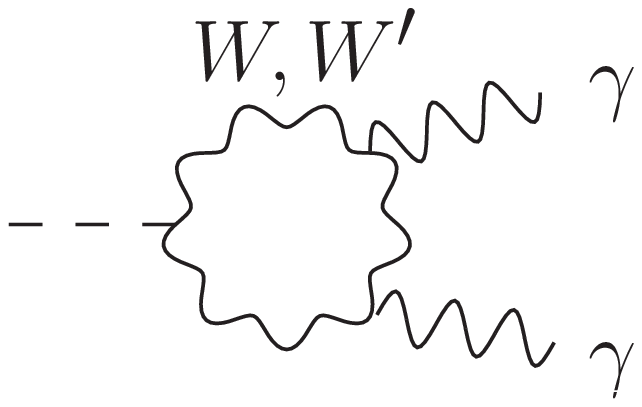}
\includegraphics[width=4cm,height=2.5cm]{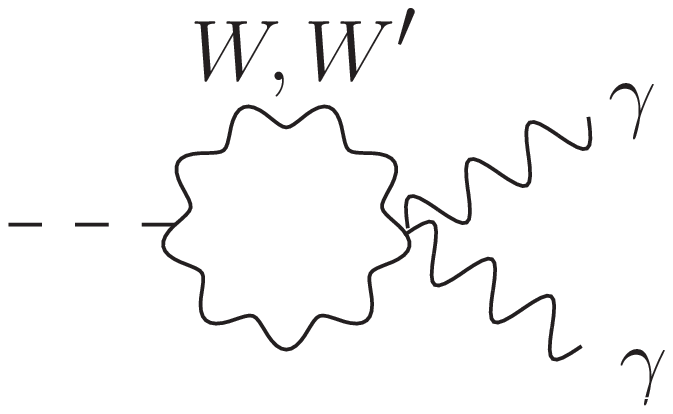}
\caption{Loop-induced radiative decays of the light CP-even Higgs $h^0$
into di-photons in the present model.}
\label{fig:h2gaga}
\end{center}
\end{figure}

The decay width $\,\Gamma[h \to Z \gamma]$\, has similar feature to
$\,\Gamma[h \to \gamma \gamma]$,\, and can be comparable to the SM values.
But, in this radiative decay process,
there are diagrams including both SM and heavy fields in the loop
due to the existence of $\,G_{WW'h}^{}\,$,\, $G_{WW'Z}^{}$,\,
and so on. We summarize all Feynman diagrams contributing to this decay channel
in Fig.\,\ref{fig:h2zga}. The explicit formula for
$\,\Gamma[h \to Z \gamma]_{\rm SM}^{}\,$
contains contributions from the SM fermion-loops and the $W$-loops,
\beqs
\beqn
\label{eq:SMhzga}
&&\Gamma[h\to Z\gamma]_{\rm SM}
\,=\,\frac{\alpha}{(4\pi)^3}\frac{M_h^3}{8\pi}\Big(1-\frac{m_Z^2}{M_h^2} \Big)^3\Big|\mA_{ff}^{}+\mA_{WW}^{}\Big|^2\,,
\label{eq:SMhZga}
\\[2mm]
&&\mA_{ff} ~\equiv~
\sum_f \frac{eN_{c, f}^{}}{s_W c_W v}Q_f \hat{v}_f \mA_{1/2}^H(\tau_f, \lambda_f)\,,
\label{eq:SMhZga_ff}
\\[2mm]
&&\mA_{WW}~\equiv~ \frac{e}{s_W v}\mA_1^H(\tau_W, \lambda_W)\,,\label{eq:SMhZga_WW}
\eeqn
\eeqs
with $\,\hat{v}_f=2T_{3f}-4Q_f s_W^2$\, and
$Q_f$ denoting the electric charges.
Here, the form factors are defined as,
\beqs
\beqn
&&\mA_{1/2}^H(\tau, \lambda)\equiv I_1(\tau, \lambda)-I_2(\tau, \lambda)\,,\\
&&\mA_1^H(\tau, \lambda)\equiv c_W\Big\{4\Big(3-\frac{s_W^2}{c_W^2}\Big)I_2(\tau, \lambda)+\Big[(1+2\tau)\frac{s_W^2}{c_W^2}-(5+2\tau)\Big]I_1(\tau,\lambda) \Big\}\,.
\eeqn
\eeqs
where we have defined $\,\tau_i^{}\,\equiv\, M_h^2/(4 m_i^2)$\, and
$\,\lambda_i^{} \,\equiv\, m_Z^2/(4 m_i^2)$\, for $\,i=f, V\,$.\,
The relevant functions in the above form factors are defined as follows,
\beqs
\beqn
&& I_1^{}(\tau, \lambda)=\frac{1}{2(\lambda-\tau)}+
\frac{1}{2(\lambda-\tau)^2}\Big[f(\tau)-f(\lambda)\Big]+\frac{\lambda}{(\lambda-\tau)^2}
\Big[g(\tau)-g(\lambda) \Big], ~~~~~~~~
\label{eq:I1}
\\[3mm]
&& I_2^{}(\tau, \lambda)=\frac{1}{2(\tau-\lambda)}\Big[f(\tau)-f(\lambda)\Big]\,,
\label{eq:I2}
\\[3mm]
&& g(\tau)=\begin{cases}
~ \sqrt{\tau^{-1}-1}\arcsin\!\sqrt{\tau}, \qquad & \tau \leqq 1\,,
\\[2.5mm]
 ~ -\frac{\sqrt{1-\tau^{-1}}}{2}\!
 \Big[\ln\frac{1+\sqrt{1-\tau^{-1}}}{1-\sqrt{1-\tau^{-1}}}-i\pi\Big] ,~~
\qquad & \tau>1 \,.
\end{cases}
\eeqn
\eeqs

\begin{figure}[]
\begin{center}
\includegraphics[width=4cm,height=2.5cm]{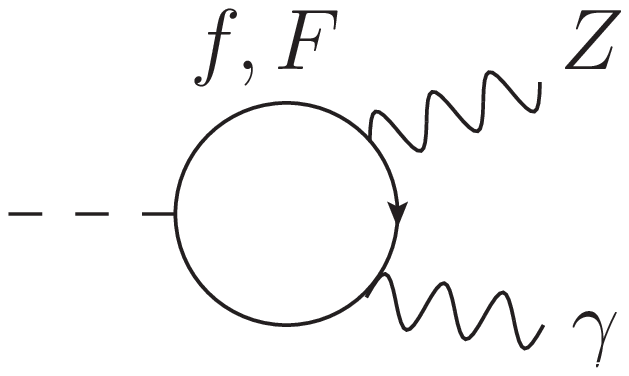}
\includegraphics[width=4cm,height=2.5cm]{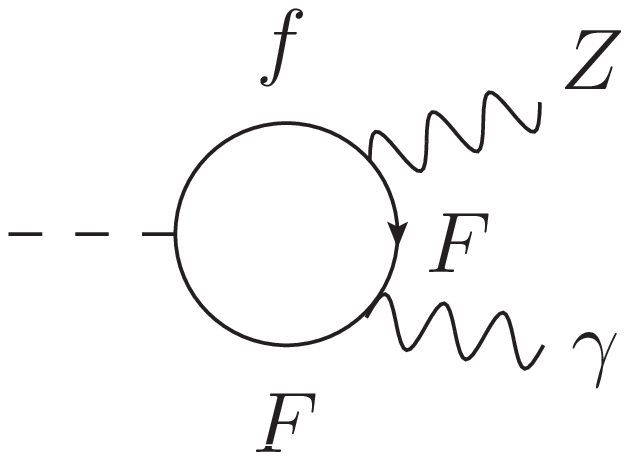}
\includegraphics[width=4cm,height=2.5cm]{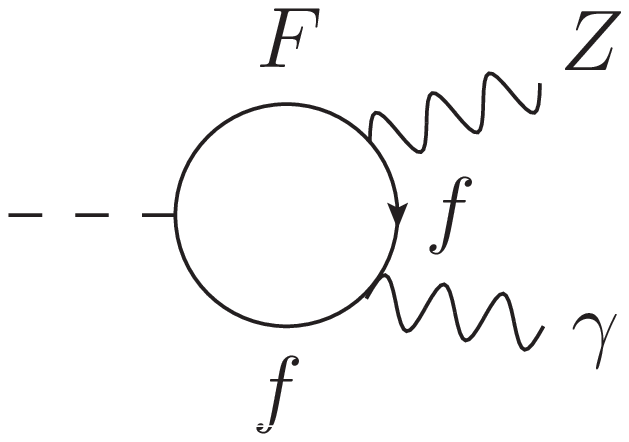}\\
\includegraphics[width=4cm,height=2.5cm]{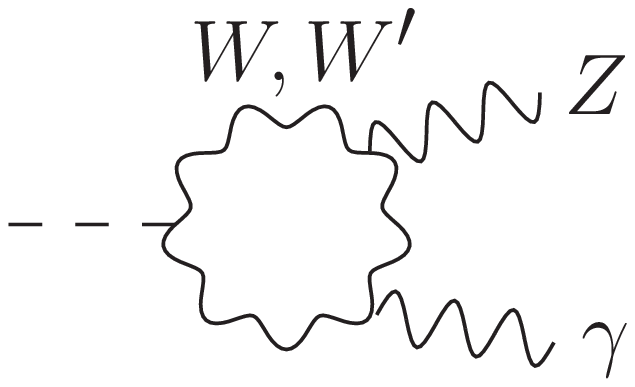}
\includegraphics[width=4cm,height=2.5cm]{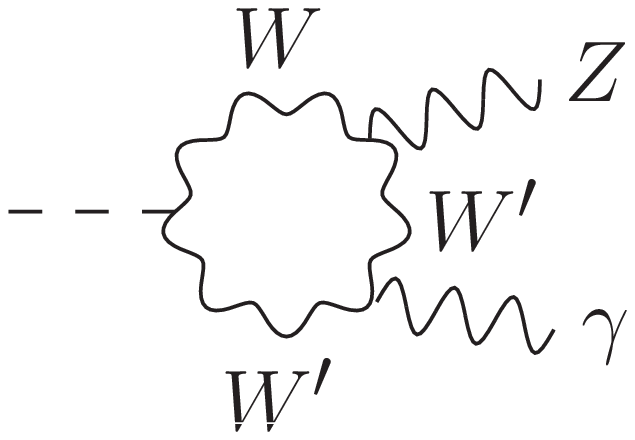}
\includegraphics[width=4cm,height=2.5cm]{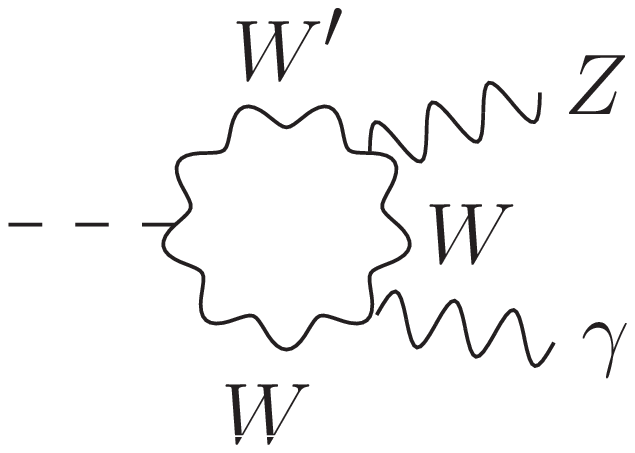}\\
\includegraphics[width=4cm,height=2.5cm]{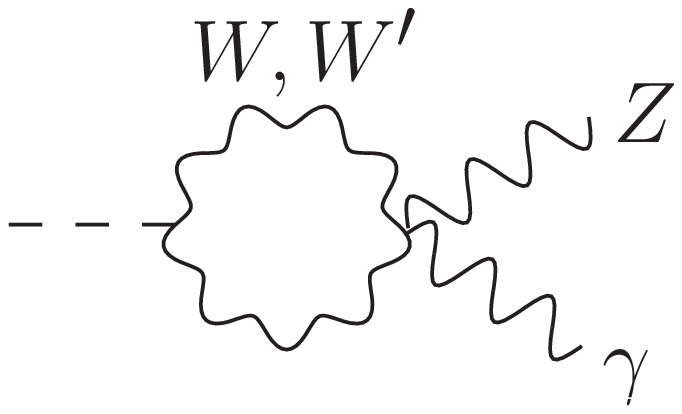}
\includegraphics[width=4cm,height=2.5cm]{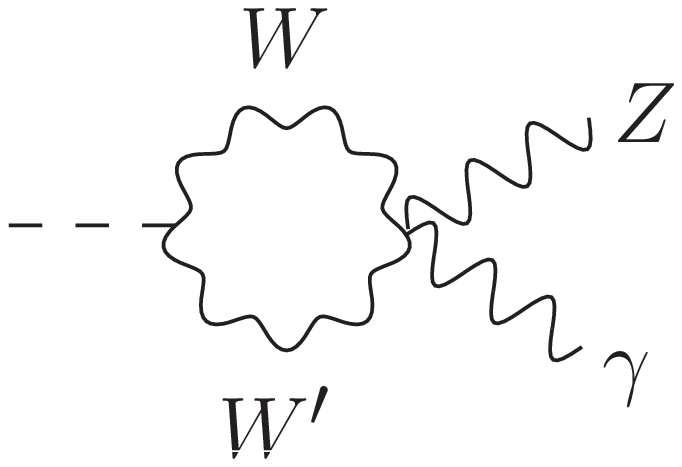}
\includegraphics[width=4cm,height=2.5cm]{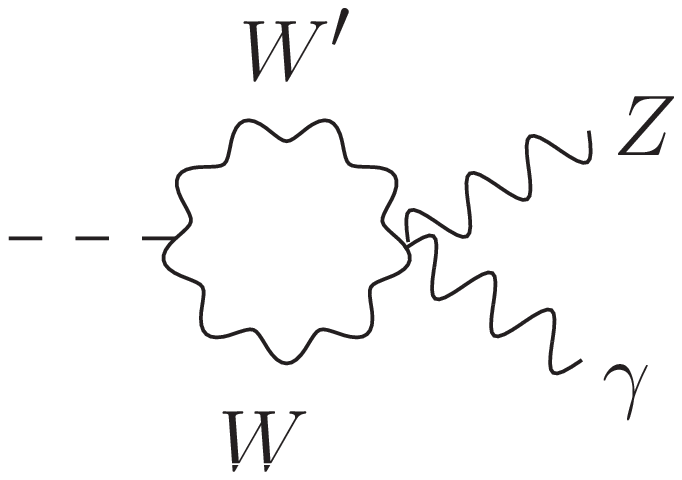}
\caption{Loop-induced radiative decays of the light CP-even Higgs boson $h^0$
into the $\,Z\gamma\,$ final state in the present model.}
\label{fig:h2zga}
\end{center}
\end{figure}

For the $\,h^0\to Z\gamma$\, decay channel in the present model,
the (\ref{eq:SMhZga}) should be corrected by including all loop contributions
given in Fig.\,\ref{fig:h2zga}. Analytically, we can express this partial decay width,
\beqn
\label{eq:3sitehzga}
\Gamma[h\!\to\! Z\gamma] &\,=\,&
\frac{\alpha}{(4\pi)^3}\frac{M_h^3}{8\pi}\Big(1\!-\!\frac{m_Z^2}{M_h^2} \Big)^3\Big|\mA_{ff}^{}+\mA_{FF}^{}+\mA_{fF}^{}
+\!\!\sum_{V=W, W'}\!\!\!\mA_{VV}+\mA_{WW'}\Big|^2,
\hspace*{9mm}
\label{eq:3sitehZga}
\eeqn
where all the form factors read as follows,
\beqs
\beqn
\mA_{ff}^{}
&=&\xi_f Q_f\hat{v}_f N_{c,f}\frac{e}{s_W c_W v}\mA_{1/2}^H(\tau_f,\lambda_f)\,,
\label{eq:3sitehZga_ff}
\\[2mm]
\mA_{FF}^{}
&=&\xi_F Q_F\hat{v}_F N_{c,F}\frac{e}{s_W c_W v}\mA_{1/2}^H(\tau_F, \lambda_F)\,,
\label{eq:3sitehZga_FF}
\\[2mm]
\mA_{fF}^{}
&\simeq& \frac{4 m_f}{M_F^2}\left[g_{h\overline{F}_R f_L}
g_{\overline{F}_L f_L Z}+(L\leftrightarrow R)\right]\(3+4\ln\frac{m_f}{M_F}\)
\nonumber\\
&&
-\frac{4}{M_F}\left[g_{h\overline{f}_R F_L}g_{\overline{f}_L F_L Z}+(L\leftrightarrow R)
\right] ,
\label{eq:3sitehZga_fF}
\\[2mm]
\mA_{WW}^{}
&=&\frac{e}{s_W v}\xi_{hVV}\mA_1^H(\tau_W,\lambda_W)\,,
\label{eq:3sitehZga_WW}
\\[2mm]
\mA_{W'W'}^{}
&=& \frac{e}{s_W v}\xi_{hV'V'}\mA_1^H(\tau_{W'},\lambda_{W'})\,,
\label{eq:3sitehZga_WpWp}
\\[2mm]
\mA_{WW'}^{}
&\simeq& \frac{e}{s_W v}\xi_{hVV'}\frac{r}{\,1\!+\!r^2\,}
\left[\frac{7}{2}+\frac{1}{18 M_{W'}^2}(5M_h^2
\right.
\nonumber
\\[2mm]
&& \left.\left.
-45m_W^2-47m_Z^2+324m_W^2\ln\frac{M_{W'}}{m_W}\) \right] .~~~~~~
\label{eq:3sitehZga_WWp}
\eeqn
\eeqs
In the above, we have taken both heavy fermions mass limit
$\,M_F^{}\gg (M_h^{},\,m_f^{},\,m_Z^{})$,\,
and heavy $W'$ mass limit
$\,M_{W'}^{} \gg (M_h^{},\, m_{W}^{},\,m_Z^{})$.


\vspace*{3mm}
\section{\hspace*{-1mm}Decays of the Heavier Higgs Boson}
\label{appB}

In this Appendix\,B,
we further analyze the dominant decay modes for the heavy Higgs boson
$\,H^0\,$ in the present model. For the LHC analysis in Sec.\,\ref{sec4},
we mainly consider $H^0$ for the mass-range of $\,M_H^{} = 130-600$\,GeV.
The major decay modes of $H^0$ include,
$\,H^0\to WW,\, ZZ,\, WW',\, ZZ',\,, t \bar t\,, b\bar b,\, \tau\bar{\tau}$\,.

The partial decay widths for $\,H^0\to WW,ZZ$\, channels
differ from the SM values by a common coupling factor according to Eq.\,(\ref{eq:VVh}),
\beqn
\label{eq:HVVwid}
&&
\frac{~\Gamma[H \to VV]~~~}{~~\Gamma[H \to VV]_{\rm SM}~}
~\simeq~ \xi_{HVV}^2\, \leqq 1\,,
\eeqn
where $\,V=W,Z$\,.\,  It shows that the partial width of $\,H^0\to WW,ZZ$\,
tend to be smaller than the SM values. For the mass-range of
$\,M_H^{}> M_{V'}+m_V^{}$\, with $\,V'=W',Z'$,\,
we have the new decay channels $\,H\to W'W,Z'Z$\, opened.
We derive these decay rates as,
\beqn
\label{eq:HVVpwid}
\Gamma[H\to V'V] &=&
\frac{m_V^2 M_{V'}^2 }{4\pi v^2M_H}\xi_{HVV'}^2
\Big(2+ \frac{(M_H^2-M_{V'}^2-m_V^2)^2}{4m_V^2 M_{V'}^2} \Big)
\nonumber\\[2mm]
&& \times
\sqrt{1-2\frac{m_V^2\!+\!M_{V'}^2}{M_H^2}+
       \(\!\frac{M_{V'}^2\!-\!m_V^2}{M_H^2}\!\)^2\,} \,.
\eeqn
Noting that the size of the dimensionless couplings $\,\xi_{HVV'}^{}\,$
are comparable to the $\,\xi_{HVV}^{}$,\,
this decay width (\ref{eq:HVVpwid}) are even enhanced over that of
$\,H^0\to WW,ZZ$,\, since we have,
$\,\Gamma[H \!\to\! V'V]/\Gamma[H\!\to\! VV]\propto(M_{V'}^{}/m_W^{})^2$\,.\,
It is clear that the $\,H\to V'V$\, can be the dominant decay modes
once they are kinematically allowed, as shown in Fig.\,\ref{fig:HBR400} of Sec.\,\ref{sec4.1}.
For $H^0$ decays into the SM fermions, the partial decay widths differ
from the SM Higgs values by the corresponding coupling factor as given in
Eq.\,(\ref{eq:tth/H}),
\beqn
\label{eq:Hffwid}
\frac{~\Gamma[H\to f \bar f]~~~}{~~\Gamma[H\to f \bar f]_{\rm SM}~}
& ~\simeq~ & \xi_{Hff}^2 \,\leqq\, 1\,,
\eeqn
where $\,f\,$ stands for the SM fermion.

Finally, the relevant loop-induced radiative decay channel
$\,H\to gg\,$ in the present model also receives contributions
from the six heavy vector-like partners of the SM quarks.
This partial decay width differs from the corresponding SM value,
\beqn
\label{eq:Hggwid}
&&
\frac{~\Gamma[H\to gg]~~~}{~~\Gamma[H\to gg]_{\rm SM}~} ~=~
\frac{~\displaystyle\left|
\sum_{f=t, Q} \xi_{Hff}^{} A_{1/2}^H(\tau_f^{}) \right|^2
}{\displaystyle \left| \sum_{f= t} A_{1/2}^H(\tau_f^{}) \right|^2}\,,
\eeqn
where $\,\tau_f^{}\equiv M_H^2/(4m_f^2)$\,.


\newpage

\end{document}

Title: LHC Higgs Signatures from Extended Electroweak Gauge Symmetry
\\
Authors:  Tomohiro Abe, Ning Chen, Hong-Jian He
\\
Comments: JHEP Journal Version, 48 pages, 19 Figs, (only minor refinements + extended discussions)
\\
We study LHC Higgs signatures from the extended electroweak gauge symmetry
SU(2) x SU(2) x U(1). Under this gauge structure, we present an effective
UV completion of the 3-site moose model with ideal fermion delocalization,
which contains two neutral Higgs states (h, H) and three new gauge bosons
(W', Z'). We study the unitarity, and reveal that the exact E^2 cancellation
in the longitudinal WW scattering amplitudes is achieved by the joint role of
exchanging both spin-1 new gauge bosons and spin-0 Higgs bosons.
We identify the lighter Higgs state h with mass 125GeV, and derive the
unitarity bound on the mass of heavier Higgs boson H. The parameter space of
this model is highly predictive. We study the production and decay signals of
this 125GeV Higgs boson h at the LHC. We demonstrate that the h Higgs boson
can naturally have enhanced signals in the diphoton channel $gg \to h \to\gamma\gamma$,
while the events rates in the reactions $gg \to h \to WW^*$ and $gg \to h \to ZZ^*$
are generally suppressed relative to the SM expectation. Searching the h Higgs boson
via associated productions and vector boson fusions are also discussed for our model.
We further analyze the LHC signals of the heavier Higgs boson H as a distinctive
new physics discriminator from the SM. For wide mass-ranges of H, we derive constraints
from the existing LHC searches, and study the discovery potential of H
at the LHC(8TeV) and LHC(14TeV).